\documentclass[a4paper,11pt]{article} %11pt in original
\pdfoutput=1 % if your are submitting a pdflatex (i.e. if you have
\usepackage{jheppub}
\usepackage[whole]{bxcjkjatype}  %%日本語環境

\usepackage{subfig}
\usepackage{amsmath, amssymb}
\usepackage{mathrsfs}
\usepackage{graphicx}% Include figure files
\usepackage{dcolumn}% Align table columns on decimal point
\usepackage{bm}% bold math

\usepackage{comment}
\usepackage{braket}
\usepackage{appendix}
\usepackage{color,soul}
\usepackage{amsthm}
\usepackage{bbm}
\usepackage{bbm,bm,graphicx,mathtools,color,hyperref,slashed}
\usepackage{pdflscape,afterpage}
\usepackage{textgreek}

\newcommand{\diff}{\mathrm{d}}

\newcommand{\ve}{\varepsilon}
\newcommand{\Diff}{{\mathcal{D}}}

\newcommand{\tr}{\mathrm{tr}}
\newcommand{\im}{\mathrm{i}}

\newcommand{\rmd}{\mathrm{d}}
\newcommand{\rme}{\mathrm{e}}

\newcommand{\lcm}{\operatorname{lcm}}

\title{Non-supersymmetric duality cascade of QCD(BF) via semiclassics on $\mathbb{R}^2\times T^2$ with the baryon-'t Hooft flux}

\author[]{Yui Hayashi,}
\emailAdd{yui.hayashi@yukawa.kyoto-u.ac.jp}

\author[]{Yuya Tanizaki,}
\emailAdd{yuya.tanizaki@yukawa.kyoto-u.ac.jp}

\author[]{and Hiromasa Watanabe}
\emailAdd{hiromasa.watanabe@yukawa.kyoto-u.ac.jp}

\affiliation[]{Yukawa Institute for Theoretical Physics, Kyoto University, Kitashirakawa Oiwakecho, Sakyo-ku, Kyoto 606-8502, Japan
}

\preprint{YITP-24-41}

\abstract{
We study the phase diagrams of the bifundamental QCD (QCD(BF)) of different ranks, which is the $4$d $SU(N_1) \times SU(N_2)$ gauge theory coupled with a bifundamental Dirac fermion. 
After discussing the anomaly constraints on possible vacuum structures, 
we apply a novel semiclassical approach on $\mathbb{R}^2\times T^2$ with the baryon-'t~Hooft flux to obtain the concrete dynamics. 
The $2$d effective theory is derived by the dilute gas approximation of center vortices, and it serves as the basis for determining the phase diagram of the model under the assumption of adiabatic continuity. 
As an application, we justify the non-supersymmetric duality cascade between different QCD(BF), which has been conjectured in the large-${N}$ argument. 
Combined with the semiclassics and the large-$N_{1,2}$ limit, we construct the explicit duality map from the parent theory, $SU(N_1) \times SU(N_2)$ QCD(BF), to the daughter theory, $SU(N_1) \times SU(N_2-N_1)$ QCD(BF), including the correspondence of the coupling constants. 
We numerically examine the validity of the duality also for finite $N_{1,2}$ within our semiclassics, finding a remarkable agreement of the phase diagrams between the parent and daughter sides. 
}

\begin{document}

\maketitle

\section{Introduction and Summary}

Non-Abelian gauge theories in $4$d are, typically, strongly coupled at low energies, which cause rich nonperturbative dynamics, such as color confinement and chiral symmetry breaking, and the $\theta$ angle serves as a useful tool to learn about the mechanism behind those phenomena. 
In this paper, we focus on a specific class of $4$d gauge theories, Quantum ChromoDynamics with the Bi-Fundamental fermion (QCD(BF)), which is the $4$d $SU(N_1) \times SU(N_2)$ gauge theory with a Dirac fermion $\Psi$ in the bifundamental representation $(\bm{N_1},\bm{\overline{N}_2})$. %$(\Box, \overline{\Box})$.

This theory has two vacuum angles $(\theta_1,\theta_2)$, and it is expected that there are various phase transitions in the $(\theta_1,\theta_2)$ space. 
Indeed, the phase diagram of the QCD(BF) has been investigated from the anomaly matching as well as from some calculable limits \cite{Tanizaki:2017bam, Karasik:2019bxn} (See also Refs.~\cite{tHooft:1979rat, Wen:2013oza, Kapustin:2014lwa, Kapustin:2014zva, Cho:2014jfa, Gaiotto:2017yup, Kikuchi:2017pcp, Komargodski:2017dmc, Tanizaki:2018xto, Cordova:2019jnf, Cordova:2019uob} for recent developments of anomaly matching and global inconsistency), and these studies uncover its nontrivial phase diagram. 
In particular, when $N_1\not=N_2$, the $(\theta_1,\theta_2)$ phase diagram has different features between the massless fermion limit, $m\approx 0$, and the massive fermion limit, $m\to \infty$. 
Karasik and Komargodski~\cite{Karasik:2019bxn} studied how the topological structure of the phase diagram is deformed as we change the fermion mass, and they proposed an interesting duality conjecture from their observations: 
There is an infrared (IR) duality between $SU(N_1) \times SU(N_2)$ QCD(BF) and $SU(N_1) \times SU(N_2-N_1)$ QCD(BF). 
This suggests the possibility of a ``duality cascade'' in non-supersymmetric theories.

\begin{figure}[t]
\centering
\includegraphics[scale=0.55]{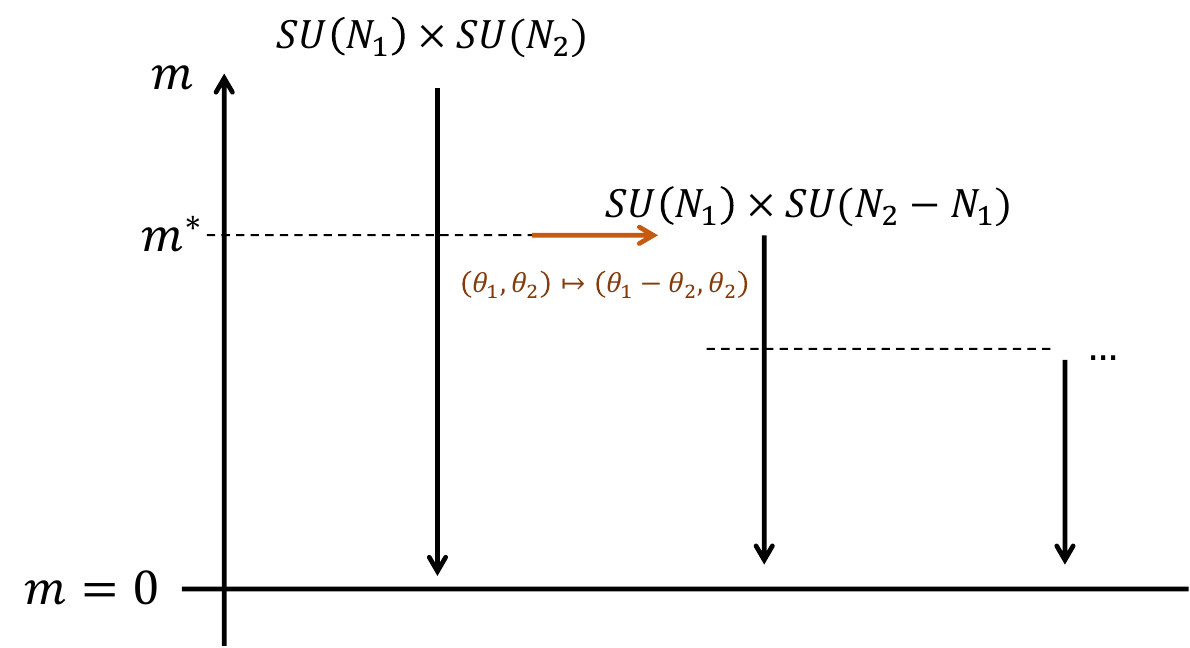}
\caption{Schematic picture of the duality cascade in QCD(BF) \cite{Karasik:2019bxn}.
As the fermion mass decreases, we encounter the topology-changing mass $m = m^*$, and it has been conjectured that the phase diagram coincides with that of $SU(N_1)\times SU(N_2-N_1) ~\mathrm{QCD(BF)~on~} (\theta_1-\theta_2,\theta_2)$ below this mass $m < m^*$ with an appropriate reparametrization.
}
\label{fig:KK_duality_conjecture}
\end{figure}

The original idea of duality cascade was developed in studying the extension of the AdS/CFT correspondence to the branes at conical singularities~\cite{Gubser:1998fp, Klebanov:1998hh, Klebanov:1999rd, Klebanov:2000nc, Klebanov:2000hb}, where one can construct the gravity dual of $\mathcal{N}=1$ supersymmetric QCD(BF). 
For $\mathcal{N} = 1$ supersymmetric $SU(N_1) \times SU(N_2)$ gauge theories, one can repeatedly use the Seiberg duality, reducing the rank of the gauge groups along with lowering the renormalization group scale~\cite{Klebanov:2000hb}.
Karasik and Komargodski~\cite{Karasik:2019bxn} conjectured the presence of a similar duality cascade\footnote{As detailed later, the meaning of duality cascade is slightly different from the original supersymmetric one.
First, the IR duality here means a correspondence between the vacuum structures after suitable reparametrization. Second, although the original supersymmetric duality cascade occurs with lowering the renormalization group scale, the non-supersymmetric duality cascade involves the change of the bifundamental fermion mass.} in non-supersymmetric $SU(N_1) \times SU(N_2)$ QCD(BF), so that one can easily understand the topology changing phenomena of the $(\theta_1,\theta_2)$ phase diagram by repeating the duality transformation $SU(N_1) \times SU(N_2) \rightarrow SU(N_1) \times SU(N_2-N_1)$ (See Figure~\ref{fig:KK_duality_conjecture}).
The equal-rank $SU(N)\times SU(N)$ QCD(BF) can then be regarded as the terminal of the duality cascade.

In our previous paper \cite{Hayashi:2023wwi}, we have developed a novel semiclassical method to reveal the vacuum structure of the equal-rank $SU(N) \times SU(N)$ QCD(BF).
Aside from the duality cascade, the equal-rank QCD(BF) is the daughter theory of the large-$N$ orbifold equivalence from the $\mathcal{N}=1$ $SU(2N)$ super Yang-Mills (SYM) theory~\cite{Kachru:1998ys, Bershadsky:1998cb, Schmaltz:1998bg, Strassler:2001fs, Dijkgraaf:2002wr, Kovtun:2003hr, Kovtun:2004bz, Armoni:2005wta, Kovtun:2005kh}. 
The nonperturbative justification of the orbifold equivalence requires uncovering the symmetry-breaking pattern of these theories~\cite{Kovtun:2003hr, Kovtun:2004bz}, and our semiclassical description gives its affirmative support by explicitly computing the vacuum phase diagram of equal-rank QCD(BF)~\cite{Hayashi:2023wwi} (see also~\cite{Shifman:2008ja}).

In this paper, extending the preceding one \cite{Hayashi:2023wwi}, we investigate the different-rank $SU(N_1) \times SU(N_2)$ QCD(BF) by employing the semiclassical approach with the 't Hooft flux $T^2$-compactification.
In this semiclassical approach, we realize the weak-coupling description of confinement in the four-dimensional gauge theories by putting them on $\mathbb{R}^2 \times T^2$ incorporating the 't Hooft flux~\cite{Tanizaki:2022ngt, Tanizaki:2022plm, Hayashi:2023wwi, Hayashi:2024qkm}. 
We can then perform the explicit computation of the confinement phenomena by using the dilute gas approximation of center vortices, so this setting gives an explicit realization for one of the prevalent understandings for the quark confinement~\cite{DelDebbio:1996lih, Faber:1997rp, Langfeld:1998cz, Kovacs:1998xm, Greensite:2011zz}.
Grounded in the adiabatic continuity conjecture, we expect that the $(\theta_1,\theta_2)$ phase diagram of the $4$d QCD(BF) has the same structure with the weakly-coupled confining theory at small $T^2$.
It is noteworthy that the application of this approach has been successful in distinctly outlining the qualitative features of various confining gauge theories~\cite{Tanizaki:2022ngt, Tanizaki:2022plm, Hayashi:2023wwi, Hayashi:2024qkm}, including $SU(N)$ Yang-Mills theory, $\mathcal{N} = 1$ $SU(N)$ SYM theory, QCD with fundamental quarks, QCD with $2$-index quarks, and equal-rank QCD(BF).

Let us emphasize that it is not straightforward to extend the previous analysis of equal-rank QCD(BF) to different-rank QCD(BF).
In different-rank QCD(BF), the naive insertion of minimal 't Hooft flux for $SU(N_1) \times SU(N_2)$ violates the single-valuedness of the bifundamental matter.
We resolve this problem by introducing $U(1)$ baryon magnetic flux at the same time as we have done for QCD with fundamental quarks~\cite{Tanizaki:2022ngt, Hayashi:2024qkm}, but then the $2$d semiclassical theory becomes the different one from what we obtained for the equal-rank QCD(BF) in \cite{Hayashi:2023wwi}.

\subsection{Summary and highlights}

In Section~\ref{sec:model}, we explain some basics of QCD(BF) and also give a review of the duality conjecture~\cite{Karasik:2019bxn}, which states that two different QCD(BF) have the same structure of the $(\theta_1,\theta_2)$ phase diagram:
\begin{align}
    SU(N_1)&\times SU(N_2) \mathrm{~QCD(BF)~on~} (\theta_1,\theta_2)~\mathrm{for}~m<m^* \notag \\
    &\longleftrightarrow  SU(N_1)\times SU(N_2-N_1) ~\mathrm{QCD(BF)~on~} (\theta_1-\theta_2,\theta_2), \tag{\ref{eq:KK_duality_conjecture}}
\end{align}
where $m^*$ denotes the (largest) fermion mass of the topology-changing point for the phase diagram.
In Ref.~\cite{Karasik:2019bxn}, this duality is proposed based on observations of the vacuum branches with the large-$\operatorname{gcd}(N_1,N_2)$ argument. This duality should be understood as a matching between the meta-stable vacuum branches after a suitable reparametrization of the fermion mass and dynamical scales.
If this is true, it causes the non-supersymmetric ``duality cascade'' as the fermion mass is lowered, which is sketched in Figure~\ref{fig:KK_duality_conjecture}.

In Section~\ref{sec:anomaly_BFQCD}, we derive kinematic constraints on phase diagrams arising from the global symmetry by extending the results of previous works~\cite{Tanizaki:2017bam, Karasik:2019bxn, Hayashi:2023wwi}. 
The vector-like symmetry of the $SU(N_1) \times SU(N_2)$ QCD(BF) is given by the baryon-number symmetry and the $1$-form center symmetry, 
\begin{equation}
    G_{\mathrm{vec}}=\frac{U(1)_V}{\mathbb{Z}_{\lcm(N_1, N_2)}} \times \mathbb{Z}^{[1]}_{\gcd(N_1,N_2)}, 
    \tag{\ref{eq:vect_sym}}
\end{equation}
and its $4$d symmetry-protected topological (SPT) states are characterized by the two discrete labels, $(\tilde{k},\tilde{\ell})\in (\mathbb{Z}_{\gcd(N_1,N_2)})^2$. 
This model in the massless limit also enjoys the discrete chiral symmetry $(\mathbb{Z}_{\gcd(N_1,N_2)})_{\chi}$, and we see that it has to be spontaneously broken in the confinement state because of the mixed anomaly.
For the massive case, we show that the vector-like symmetry $G_{\mathrm{vec}}$ has the global inconsistency with the $2 \pi$ shifts of theta angles because $\theta_1 \rightarrow \theta_1 + 2\pi$ and $\theta_2 \rightarrow \theta_2 + 2\pi$ changes the SPT labels $(\tilde{k},\tilde{\ell})$ if $\operatorname{gcd}(N_1,N_2) \neq 1$.
These results impose restrictions on the possible vacuum structures.

Next, we develop the novel semiclassical approach on $\mathbb{R}^2\times T^2$ with the baryon-'t~Hooft flux to reveal the low-energy dynamics concretely. 
Through the semiclassical analysis shown in Section~\ref{sec:QCDBF_2dEFT}, we derive the following $2$d effective theory:
\begin{itemize}
    \item %$\operatorname{gcd}(N_1,N_2) \neq 1$: 
    Let us write $
    N_1 = \operatorname{gcd}(N_1,N_2) N_1',~N_2 = \operatorname{gcd}(N_1,N_2) N_2'$, then the $2$d effective theory is described by the $2 \pi N_1' N_2'$-periodic scalar $\varphi$ and discrete vacuum labels $\Tilde{k}, \Tilde{\ell} = 0,\cdots, \operatorname{gcd}(N_1,N_2)-1$. 
    The effective potential is given by
\begin{align}
    V[\varphi; \Tilde{k},\Tilde{\ell}] &= - m \mu \cos (\varphi) - 2K^{(1)} \rme^{-S_\mathrm{I}^{(1)}/N_1} \cos \left( \frac{N_2 \varphi + 2 \pi (N'_2 \tilde{\ell} + M_2 \tilde{k}) - \theta_1}{N_1} \right) \notag \\
    &~~~~~~~~~ - 2K^{(2)} \rme^{-S_\mathrm{I}^{(2)}/N_2} \cos \left( \frac{N_1 \varphi
    + 2 \pi (N'_1 \tilde{\ell} - M_1 \tilde{k}) - \theta_2}{N_2} \right), \tag{\ref{eq:potential_2deff}}
\end{align}
    where $M_1,M_2$ is the Bezout coefficient, $M_1N'_2+M_2N'_1=1$, with the special choice satisfying $M_1/N'_2, M_2/N'_1\in \mathbb{Z}$, $\mu$ is a dimensionful scale introduced by bosonization, and $K^{(i)} \rme^{-S_\mathrm{I}^{(i)}/N_i}$ is the semiclassical strong scales for $SU(N_i)~(i=1,2)$.
\end{itemize}
The compact scalar describes the phase of the chiral condensate $\tr(\overline{\Psi}\frac{1+\gamma_5}{2}\Psi)\sim \rme^{\im \varphi}$, so its periodicity is originally given by $\varphi\sim \varphi+2\pi$ but is extended to $\varphi\sim \varphi+2\pi N'_1N'_2$ when integrating out the gauge fields. 
This extension of the periodicity occurs in the same mechanism as that of $\eta'$ periodicity in QCD with fundamental quarks \cite{Hayashi:2024qkm}.
We show how the $4$d anomaly constraints obtained in Section~\ref{sec:anomaly_BFQCD} are satisfied in the semiclassical theory in Section~\ref{sec:anomaly_in_semiclassics}, and the above discrete labels $\tilde{k},\tilde{\ell}$ in semiclassics are identified with the $4$d SPT labels completely.
Under the assumption of adiabatic continuity, we expect this $2$d effective theory predicts the vacuum structure of the original $4$d $SU(N_1) \times SU(N_2)$ QCD(BF).

Let us highlight some of our results coming out of this $2$d effective description:
\begin{itemize}
    \item The $(\theta_1,\theta_2)$ phase diagram of the different-rank QCD(BF) is concretely determined for various values of the fermion mass $m$ including the topology-changing point.
    
    As will be shown in Figure~\ref{fig:two_limits}, the phase diagrams at the massless and large-mass limits have different topologies, and 
    we explicitly demonstrate the topology-changing phenomena by applying the above semiclassical description at intermediate masses. 
    We also numerically examine phase diagrams: Section~\ref{sec:numerical_phase_diagrams_gcd1} presents the results of $SU(2) \times SU(3)$ QCD(BF) as an example of $\operatorname{gcd}(N_1,N_2) = 1$ case, and Section~\ref{sec:numerical_phase_diagrams_gcd!=1} gives that of $SU(2) \times SU(4)$ QCD(BF) as an example of $\operatorname{gcd}(N_1,N_2) \neq 1$ case.

    \item We investigate the conjectured duality (\ref{eq:KK_duality_conjecture}) between $SU(N_1) \times SU(N_2)$ QCD(BF) and $SU(N_1) \times SU(N_2 - N_1)$ QCD(BF) based on the above semiclassical framework:
% The main observations are listed as follows:
\begin{itemize}
    \item We construct the explicit duality map in the large-$N_{1,2}$ limit.  

    In the large-$N_{1,2}$ limit, the $2$d effective theory is reduced to discrete vacuum branches.
    We then concretely construct the duality map of the semiclassical parameters $(m \mu, K^{(1)} \rme^{-S_\mathrm{I}^{(1)}/N_1}, K^{(2)} \rme^{-S_\mathrm{I}^{(2)}/N_2})$, so that these vacuum branches match between the parent and daughter theories, in Sections~\ref{sec:recovering_large_N_from_semiclassics} and~\ref{sec:comment_large_N_gcd!=1}.
    
    \item 
    The duality also holds in the hierarchical limit $ \Lambda_1,m \ll \Lambda_2$ without large-$N_{1,2}$.
    
    In the limit where the dynamical scales of $SU(N_2)$ and $SU(N_2 - N_1)$ gauge are extremely high, we can forget most of the $SU(N_2)$ and $SU(N_2 - N_1)$ dynamics, and it becomes possible to match the effective $SU(N_1)$ gauge theories extracted from $SU(N_1) \times SU(N_2)$ QCD(BF) and $SU(N_1) \times SU(N_2 - N_1)$ QCD(BF).
    This matching is presented in Section~\ref{sec:duality_hierarchical}.
    
    \item The duality cascade is practically effective even away from those limits.
    
    We numerically study the phase diagrams of $SU(2) \times SU(5)$ QCD(BF) and $SU(2) \times SU(3)$ QCD(BF) in Section~\ref{sec:duality_numerics} and Appendix \ref{app:more_phase_diagrams}.
    Choosing the semiclassical parameters $(m \mu, K^{(1)} \rme^{-S_\mathrm{I}^{(1)}/N_1}, K^{(2)} \rme^{-S_\mathrm{I}^{(2)}/N_2})$ according to the dictionary in large-$N_{1,2}$ limit, we find a remarkable agreement of the phase diagrams between these theories.
    Quantitatively, the duality is violated due to finite-$N_{1,2}$ effects, but we argue it is just the $O(N_2^{-4})$ effect in Appendix~\ref{app:more_phase_diagrams}.
    
\end{itemize}

\end{itemize}

\section{Model: QCD(BF) with \texorpdfstring{$N_1 \neq N_2$}{N1<=N2}}
\label{sec:model}

In this section, we give a brief review on QCD(BF), including its symmetry and also some large-$N$ properties.
In particular, the large-$N$ argument predicts a nontrivial relation between the $SU(N_1)\times SU(N_2)$ QCD(BF) with the $SU(N_1)\times SU(N_2-N_1)$ QCD(BF), and this is called the ``duality cascade,'' proposed in Ref.~\cite{Karasik:2019bxn}.

\subsection{Model and notation}
\label{sec:QCDBF_model}

Here we introduce the $SU(N_1)\times SU(N_2)$ QCD(BF) and notations. The field contents are 
\begin{equation}
    \left\{\begin{aligned}
    a_1:& \text{ $SU(N_1)$ gauge field with the field strength }f_1=\diff a_1+\im a_1\wedge a_1, \\
    a_2:& \text{ $SU(N_2)$ gauge field with the field strength }f_2=\diff a_2+\im a_2\wedge a_2, \\
    \Psi:& \text{ $N_1\times N_2$ matrix-valued Dirac fermion in the bifundamental representation}.
\end{aligned}
\right.
\end{equation}
Under the $SU(N_1)\times SU(N_2)$ gauge transformations $(U_1, U_2)$, they transform as 
\begin{equation}
    \Psi\mapsto U_1\Psi U_2^{\dagger},\quad 
    a_i\mapsto U_i a_i U_i^{\dagger} - \im\, U_i\diff U_i^{\dagger}\quad (i=1,2),
\end{equation}
and the covariant derivative for $\Psi$, denoted $\slashed{D}\Psi$, is given by
\begin{align}
    \slashed{D}\Psi=\gamma^{\mu}(\partial_{\mu}\Psi+\im\, a_{1\mu}\Psi-\im\, \Psi a_{2\mu}). 
\label{eq:derivative_covariant_bifund}
\end{align}
The action of this model reads,
\begin{align}
S=&  \frac{1}{g_1^2} \int  |f_1|^2 +  \frac{1}{g_2^2} \int |f_2|^2 +\int \operatorname{tr} 
 \overline{\Psi}(\slashed{D}+m)\Psi \nonumber\\
&+\frac{\im\theta_1}{8\pi^2}\int \operatorname{tr}(f_1\wedge f_1)+\frac{\im\theta_2}{8\pi^2}\int \operatorname{tr} (f_2\wedge f_2), 
\label{eq:action_bifundamental}    
\end{align}
where $|f|^2 := \operatorname{tr}(f \wedge \star f)$ to represent the gauge kinetic terms. 

Without loss of generality, we can assume that the fermion mass is positive $m>0$ because the phase of $m$ only results in the redefinition of $(\theta_1,\theta_2)$ due to the chiral anomaly. 
Let us also assume $N_2>N_1$, and we will limit our consideration to theories that are asymptotically free, i.e., $N_2 < \frac{11}{2} N_1$. 
We define $\Lambda_1$ and $\Lambda_2$ as the dynamical scales for $SU(N_1)$ and $SU(N_2)$ sectors, respectively: Within the $1$-loop renormalization group, 
\begin{equation}
    \Lambda_1 = \mu \exp\left(-\frac{8\pi^2}{\left(\frac{11}{3}N_1 - \frac{2}{3}N_2\right) g_1^2(\mu)}\right), \quad 
    \Lambda_2 = \mu \exp\left(-\frac{8\pi^2}{\left(\frac{11}{3}N_2 - \frac{2}{3}N_1\right) g_2^2(\mu)}\right), 
\end{equation}
where $\mu$ is the renormalization scale. 

For later convenience, we define the coprime numbers $(N_1',N_2')$ as
\begin{align}
    N_1' := \frac{N_1}{\operatorname{gcd}(N_1,N_2)}, ~~N_2' := \frac{N_2}{\operatorname{gcd}(N_1,N_2)}.
\end{align}
Let us also introduce the Bezout coefficient $(M_1,M_2)$ that satisfies
\begin{equation}
    N_1' M_2 + N_2' M_1 = 1, \label{eq:def_K1_K2}
\end{equation}
or $N_1 M_2+ N_2 M_1 = \gcd(N_1,N_2)$. The Bezout coefficient is not unique as the relation is invariant under $(M_1,M_2)\to (M_1+N'_1, M_2-N'_2)$, so we pick one representative. 
Since $\gcd(N'_1,N'_2)=1$ implies $\gcd((N'_1)^2, (N'_2)^2)=1$, we can require that 
\begin{equation}
    M_1\in N_2' \mathbb{Z},\quad M_2\in N_1' \mathbb{Z},
\end{equation}
so we choose such a one.  
Note that $M_1, M_2$ satisfy $\operatorname{gcd}(M_1,N_1') = \operatorname{gcd}(M_2,N_2') = 1$.

\subsection{Symmetry}
\label{sec:symmetry}

At general parameters, this model enjoys the vector-like global symmetry, 
\begin{equation}
    G_{\mathrm{vec}}=\frac{U(1)_V}{\mathbb{Z}_{\lcm(N_1, N_2)}} \times \mathbb{Z}^{[1]}_{\gcd(N_1,N_2)}, 
    \label{eq:vect_sym}
\end{equation}
where $U(1)_V$ acts on the bifundamental fermion as $\Psi\mapsto \rme^{\im \alpha }\Psi, \overline{\Psi}\mapsto \rme^{-\im \alpha}\overline{\Psi}$ and $\mathbb{Z}_{\gcd(N_1,N_2)}^{[1]}$ is the $1$-form symmetry acting on Wilson loops. 
At the massless point $(m=0)$, the model acquires the discrete chiral symmetry, 
\begin{equation}
    G_{\mathrm{massless}}=G_{\mathrm{vec}}\times (\mathbb{Z}_{\gcd(N_1,N_2)})_{\chi}, 
    \label{eq:symmetry_massless}
\end{equation}
where the chiral symmetry $(\mathbb{Z}_{\gcd(N_1,N_2)})_{\chi}$ is generated by 
\begin{equation}
    \Psi\mapsto \rme^{\frac{2\pi \im}{\gcd(N_1,N_2)}\frac{1+\gamma_5}{2}}\Psi, \quad 
    \overline{\Psi} \mapsto \overline{\Psi}\rme^{-\frac{2\pi \im}{\gcd(N_1,N_2)}\frac{1-\gamma_5}{2}}. 
\end{equation}

Let us first look at the vector-like symmetry in detail. We note that $(U_1,U_2,\rme^{\im \alpha})\in SU(N_1)_{\mathrm{gauge}}\times SU(N_2)_{\mathrm{gauge}}\times U(1)_V$ acts on $\Psi$ as 
\begin{equation}
    \Psi\mapsto \rme^{\im \alpha} U_1 \Psi U_2^{\dagger}. 
\end{equation}
This action has the redundancy that is generated by $z_1=(\rme^{2\pi \im/N_1}\bm{1}_{N_1}, \bm{1}_{N_2}, \rme^{-2\pi \im/N_1})$ and $z_2=(\bm{1}_{N_1}, \rme^{2\pi \im/N_2}\bm{1}_{N_2}, \rme^{2\pi \im/N_2})$, and they form the subgroup $\mathbb{Z}_{N_1}\times \mathbb{Z}_{N_2}$. 
As a result, we can describe the structure group acting faithfully on the matter fields (including the gauge and global symmetries) as 
\begin{align}
    G^{\mathrm{gauge+global}} &= \frac{SU(N_1)_{\mathrm{gauge}} \times SU(N_2)_{\mathrm{gauge}} \times U(1)_V}{\mathbb{Z}_{N_1} \times \mathbb{Z}_{N_2}} \nonumber\\
    & = \frac{ [(SU(N_1)\times SU(N_2))_{\mathrm{gauge}}/\mathbb{Z}_{\gcd(N_1,N_2)} ] \times U(1)_V}{\mathbb{Z}_{\lcm(N_1,N_2)}}. 
    \label{eq:total_symmetry}
\end{align}
To obtain the second line, it is convenient to change the basis for $\mathbb{Z}_{N_1}\times \mathbb{Z}_{N_2}$ as 
\begin{equation}
    \mathbb{Z}_{N_1}\times \mathbb{Z}_{N_2}\simeq \mathbb{Z}_{\gcd(N_1,N_2)}\times \mathbb{Z}_{\lcm(N_1,N_2)},  
    \label{eq:basis_rotation_group}
\end{equation}
where $\mathbb{Z}_{\gcd(N_1,N_2)}$ is generated by $z_1^{N_1'}z_{2}^{N_2'}=(\rme^{2\pi \im/\gcd(N_1,N_2)}\bm{1}_{N_1}, \rme^{2\pi \im/\gcd(N_1,N_2)}\bm{1}_{N_2}, 1)$ and $\mathbb{Z}_{\lcm(N_1,N_2)}$ by $z_1^{-M_1}z_2^{M_2}=(\rme^{-2\pi \im M_1/N_1}\bm{1}_{N_1}, \rme^{2\pi \im M_2/N_2}\bm{1}_{N_2}, \rme^{2\pi \im/\lcm(N_1,N_2)})$.  
Therefore, the subgroup $\mathbb{Z}_{{\operatorname{gcd}(N_1,N_2)}}$ of the quotients $\mathbb{Z}_{N_1} \times \mathbb{Z}_{N_2}$ lives completely inside of the gauge group $SU(N_1)_{\mathrm{gauge}} \times SU(N_2)_{\mathrm{gauge}}$, and we find the second line of \eqref{eq:total_symmetry}.

To get the global symmetry of the $SU(N_1)\times SU(N_2)$ QCD(BF), we can simply neglect the gauge group in \eqref{eq:total_symmetry} for the $0$-form symmetry and we find $U(1)_V/\mathbb{Z}_{\lcm(N_1,N_2)}$. 
In addition, the $\mathbb{Z}_{\gcd(N_1,N_2)}$ subgroup of the gauge group does nothing on the matter field, and it gives the $1$-form symmetry acting on the loop operators, and we obtain \eqref{eq:vect_sym}.

In the massless case ($m = 0$), the $SU(N_1)\times SU(N_2)$ QCD(BF) is classically invariant under the $U(1)_\chi$ chiral transformation,
\begin{align}
    \Psi\mapsto \mathrm{e}^{\im \frac{1+\gamma_5}{2} \alpha}\Psi, \qquad \overline{\Psi}\mapsto \overline{\Psi} \mathrm{e}^{-\im \frac{1-\gamma_5}{2} \alpha}.
\end{align}
Due to the Adler-Bell-Jackiw (ABJ) anomaly, this $U(1)_\chi$ transformation shifts the theta angles as
\begin{align}
    \theta_1 \rightarrow \theta_1 + N_2 \alpha,~~ \theta_2 \rightarrow \theta_2 + N_1 \alpha. \label{eq:axial_transformation_theta}
\end{align}
When both shifts are quantized in the unit of $2\pi$, it becomes the well-defined symmetry, so it becomes the $\mathbb{Z}_{\operatorname{gcd}(N_1,N_2)}$ subgroup, which gives \eqref{eq:symmetry_massless}.
Note that the physics at the massless point depends only on the particular combination of theta angles,
\begin{align}
    \theta_{\mathrm{phys}} = N_1' \theta_1 -N_2' \theta_2,
\end{align}
due to the ABJ anomaly (\ref{eq:axial_transformation_theta}).

\subsection{Large-\texorpdfstring{$N$}{N} argument and duality conjecture}
\label{sec:KK_duality}

In Ref.~\cite{Karasik:2019bxn}, Karasik and Komargodski investigated phase diagrams using the large-$N$ counting with $N= \operatorname{gcd} (N_1,N_2)$ and gave a conjecture that different QCD(BF) theories are related by non-supersymmetric dualities.
In this subsection, we review some features of phase diagrams predicted by the large-$N$ argument.

At large $N$, the vacuum energy is the quadratic function of the theta angles in each vacuum branch \cite{Witten:1980sp}, so we expect:
\begin{align}
    E_{(k_1,k_2)}(\theta_1,\theta_2) &= E_{(0,0)} (\theta_1 - 2 \pi k_1, \theta_1 - 2 \pi k_2), \notag \\
    E_{(0,0)} (\theta_1 , \theta_2 ) &\simeq E_0 + a_1 \theta_1^2  + a_2 \theta_2^2 + b \theta_1 \theta_2. \label{eq:large-n_vac_str}
\end{align}
Here, $(k_1,k_2)\in \mathbb{Z}\times \mathbb{Z}$ is introduced to restore the $2\pi$-periodicities of $\theta_{1,2}$, which describes the multi-branch structure of (quasi-)stable vacua at large $N$. 
The vacuum branch $(k_1,k_2)$ gives the lowest energy at $(\theta_1,\theta_2) = (2 \pi k_1,2 \pi k_2)$. 
As the fermion determinant is positive definite when $m>0$, the Vafa-Witten argument from the QCD inequalities claims that $\theta_1=\theta_2=0$ should be the lowest-energy state, which requires the positivity of the Hessian matrix of $E_{(0,0)}(\theta_1,\theta_2)$. This gives the constraint that $a_1, a_2 >0, b^2<4a_1 a_2$ when $m>0$. 
At the massless point, one of the inequalities is saturated; $b^2=4 a_1 a_2$. 

\subsubsection{Phase diagram of massless QCD(BF)}
\begin{figure}[t]
\centering
\begin{minipage}{.45\textwidth}
\subfloat[Massless case $m=0$]{
\includegraphics[width= \textwidth]{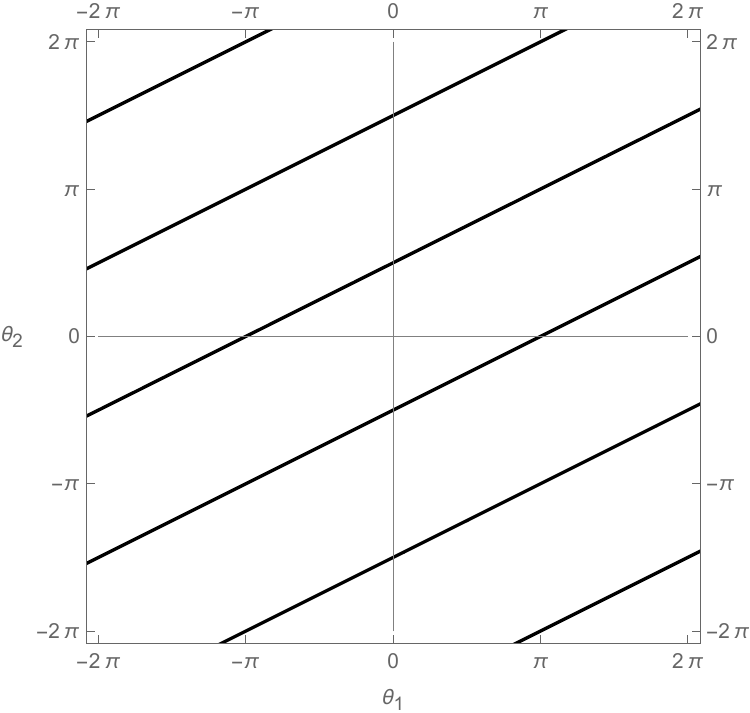}
\label{fig:BFQCDmassless}
}\end{minipage}\quad
\begin{minipage}{.45\textwidth}
\subfloat[Small-mass case $m \ll \Lambda_1,~\Lambda_2$]{\includegraphics[width= \textwidth]{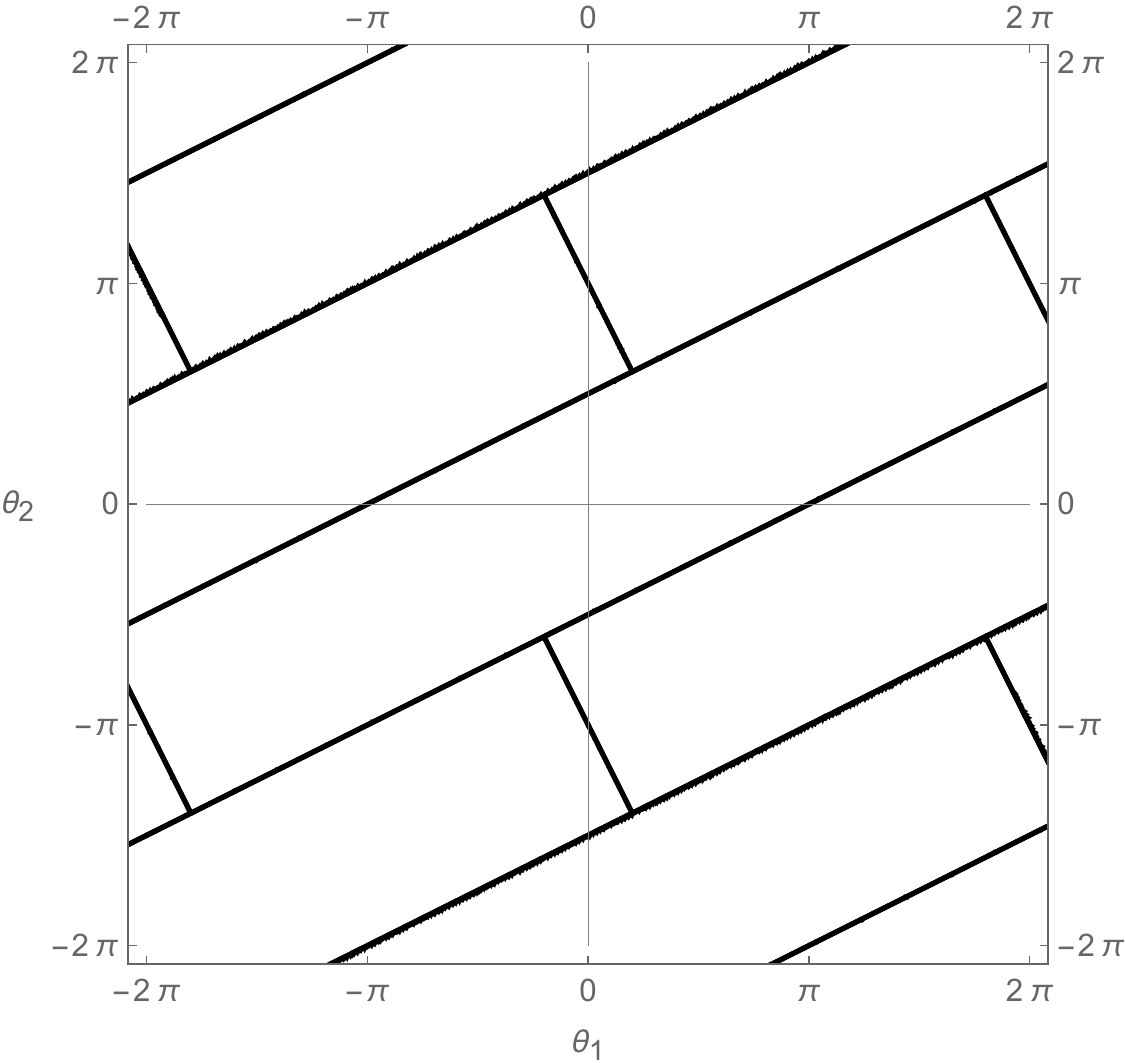}
\label{fig:BFQCDsmallmass}}
\end{minipage}
\caption{Phase diagrams suggested by the large-$N$ argument for the massless and small-mass cases.  The phase diagrams of $SU(N_1) \times SU(N_2)$ QCD(BF) with $N_1' = 1$ and $N_2' =2$ are plotted.
(\textbf{a})~Phase diagram of the massless case $m=0$, where only $\theta_{\mathrm{phys}} = N_1' \theta_1 -N_2' \theta_2$ is physical.
From the simple large-$N$ vacuum energy, the phase boundaries are located on $ \theta_{\mathrm{phys}} = \pi ~(\operatorname{mod}2\pi)$.
On each phase, the $\operatorname{gcd}(N_1,N_2)$ degenerate states are expected to appear as a consequence of the discrete chiral symmetry breaking.
(\textbf{b})~Phase diagram with infinitesimal fermion mass $m$. The mass perturbation splits the $\operatorname{gcd}(N_1,N_2)$ degenerate vacua. 
The phase boundaries due to this splitting are line segments through the halfway points $\{(N_2' \pi +2\pi k_1, N_1' \pi +2\pi k_2) \}_{k_1,k_2 \in \mathbb{Z}}$.}
\label{fig:BFQCD_massless_smallmass}
\end{figure}

Before introducing the duality conjecture, let us consider the massless case $m = 0$. 
We note that the large-$N$ limit in this section is similar to the Veneziano-type large-$N$ limit, not the 't~Hooft-type, so the anomalously-broken $U(1)_\chi$ does not restore as $N\to \infty$. 
Due to the ABJ anomaly, the vacuum energy should depend only on $\theta_{\mathrm{phys}.}=N'_1\theta_1-N'_2\theta_2$, and we obtain
\begin{align}
    E_{(k_1,k_2)} (\theta_1 , \theta_2 ) &\simeq E_0 + a \left[ N_1' \theta_1 -N_2' \theta_2 - 2\pi(N_1' k_1 - N_2' k_2) \right]^2. 
\end{align}
For $- \pi + 2 \pi n < \theta_{\mathrm{phys}} < \pi + 2 \pi n$, the branches $(k_1,k_2)$ with $N_1' k_1 - N_2' k_2 = n$ are the ground states, and the solutions can be expressed using the Bezout coefficient \eqref{eq:def_K1_K2} as
\begin{equation}
    (k_1,k_2)\in (n M_2, -n M_1)+(N'_2, N'_1)\mathbb{Z}.
\end{equation}
This implies that the system is infinitely degenerate.
For instance, the branches $(k_1,k_2) = (0,0), \pm (N_2',N_1'), \cdots$ are all the true vacua for $- \pi < \theta_{\mathrm{phys}} < \pi$.
For finite $(N_1,N_2)$, it would be natural to think that there are $N=\operatorname{gcd}(N_1,N_2)$ degenerate states on each sector from the chiral symmetry breaking.
An example of the phase diagram is shown in  Figure~\ref{fig:BFQCDmassless}.

We can also speculate the phase diagram at small fermion mass $m$.
The $\operatorname{gcd}(N_1,N_2)$ degeneracy is resolved by the mass perturbation.
So we guess that the vacuum energy is,
\begin{align}
    E_{(k_1,k_2)}(\theta_1 , \theta_2 ) &= E_{(0,0)} (\theta_1 - 2 \pi k_1, \theta_1 - 2 \pi k_2) \notag \\
    E_{(0,0)} (\theta_1 , \theta_2 ) &\simeq E_0 + a \left[ N_1' \theta_1 -N_2' \theta_2 \right]^2 + \epsilon_1 \theta_1^2  + \epsilon_2 \theta_2^2 + \epsilon_{12} \theta_1 \theta_2,
\end{align}
where $\epsilon_1,\epsilon_2,\epsilon_{12}$ are small coefficients, which vanish as $m \rightarrow 0$.
The positivity of the Hessian requires $4a_1 a_2-b^2=4a(N_1'N_2' \epsilon_{12}+(N_1')^2 \epsilon_2 +  (N_2')^2 \epsilon_1)+O(\epsilon^2)>0$ for consistency with the QCD inequalities. 

To see how the mass perturbation affects the phase diagram, let us gradually increase the theta parameters along the almost flat direction, $(\theta_1,\theta_2)\to (\theta_1+2\pi N_2',\theta_2+2\pi N'_1)$, so that $\theta_{\mathrm{phys}}$ is unchanged. 
We note that $(2\pi N'_2, 2\pi N'_1)$ is the minimal $2\pi$ periodicity along this direction because $\gcd(N'_1, N'_2)=1$, 
so let us calculate the phase boundary between $(k_1,k_2) = (0,0)$ and $(k_1,k_2) = (N_2',N_1')$ phases. 
The phase boundary between these phases is given by $E_{(N_2',N_1')} (\theta_1 , \theta_2 ) = E_{(0,0)} (\theta_1 , \theta_2 )$, which gives a linear equation
\begin{align}
   (2 \epsilon_1 N_2' + \epsilon_{12} N_1' ) \theta_1 + (2 \epsilon_2 N_1' + \epsilon_{12} N_2' ) \theta_2 = 2 \pi \left( \epsilon_1 (N_2')^2 + \epsilon_2 (N_1')^2 + \epsilon_{12} N_1'  N_2' \right).
\end{align}
This shows that the phase boundary between $(k_1,k_2)=(0,0), (N'_2, N'_1)$ is the straight line that contains the halfway point $(\theta_1,\theta_2) = (\pi N_2', \pi N_1')$. 
The other details of the phase boundary depend on the specific form of the correction $(\epsilon_1 \theta_1^2  + \epsilon_2 \theta_2^2 + \epsilon_{12} \theta_1 \theta_2)$, and
an example of the phase diagram is shown in Figure~\ref{fig:BFQCDsmallmass}.
One can see that the phase boundaries are consistent with the global inconsistency constraints, which will be discussed in Section~\ref{sec:anomaly_BFQCD}.

When $m=0$, there are $N(=\gcd(N_1,N_2))$ degenerate vacua due to the discrete chiral symmetry breaking, and the mass perturbation lifts the degeneracy. 
This new phase boundary at nonzero $m$ can be understood as the exchange of the different chiral broken vacua in the massless limit. 
Therefore, we can guess that there is no qualitative change between the massless and small-mass phase diagrams for the case $\gcd(N_1,N_2)=1$ because of the absence of discrete chiral symmetry, while this is outside the scope of the large-$N$ discussion in this subsection.

\subsubsection{Duality cascade}

Let us imagine that we dial the mass parameter $m$ from almost the massless point to almost the infinitely large one. 
When we consider the limit $m\gg \Lambda_1, \Lambda_2$, the bifundamental fermion decouples from the dynamics, and the QCD(BF) reduces to the product of $SU(N_1)$ Yang-Mills theory and $SU(N_2)$ Yang-Mills theory. 
Then, the $(\theta_1,\theta_2)$ phase diagram looks like the checkerboard, where phase boundaries are on $\theta_i = \pi~\operatorname{mod}2 \pi~(i=1,2)$. 
It is very nontrivial how the phase diagram at small mass (Figure~\ref{fig:BFQCDsmallmass}) is deformed to the phase diagram at large mass because even the topologies of phase boundaries are different.
For instance, the phases $(k_1,k_2) = (0,0)$ and $(k_1,k_2) = (0,1)$ are adjacent in the large-mass limit, which is not the case in the small-mass limit (Figure~\ref{fig:BFQCDsmallmass}). Therefore, as the fermion mass $m$ increases, a ``topology change'' is expected to occur at some point. 

First, let us consider what the phase diagram looks like at this topology-changing point.
Following Ref.~\cite{Karasik:2019bxn}, to observe the duality cascade, we here consider the phase boundary between  $(k_1,k_2) = (0,0)$ and $(k_1,k_2) = (1,1)$ phases in the example of $N_1' = 1$, $N_2' = 2$ (Figure~\ref{fig:BFQCDsmallmass}).
The phase boundary is located on the line:
\begin{align}
   E_{(0,0)} &(\theta_1, \theta_2) - E_{(1,1)} (\theta_1, \theta_2) \notag \\
   & = (4\pi a_1 + 2\pi b) \theta_1 + (4\pi a_2 + 2 \pi b) \theta_2 - 4 \pi^2 a_1  - 4 \pi^2 a_2 - 4 \pi^2 b = 0.
\end{align}
This line goes through $(\theta_1,\theta_2) = (\pi,\pi)$, and its slope depends on the parameters: $- \frac{2 a_1 + b}{2 a_2 +  b}$.
We know that the slope is $1/2$ in the small-mass limit $m \rightarrow +0$, and also that the slope becomes negative\footnote{The negative slope at $(\theta_1,\theta_2) = (\pi,\pi)$ is expected as well, from the perturbation theory at the decoupling limit \cite{Karasik:2019bxn} or semiclassical analysis (see below or Ref.~\cite{Hayashi:2023wwi}).} in the large-mass limit because $a_1 \sim O(\Lambda_1^4),~a_2 \sim O(\Lambda_2^4)$ and $b \sim 0$ for $m \rightarrow +\infty$.
Thus, the slope vanishes (or diverges) at some mass $m = m_*$, which is actually the topology-changing point.

At $m = m_*$, where $2a_1 + b = 0$ holds\footnote{We here assume that the slope vanishes $2a_1 + b = 0$ at the topology-changing point, rather than $2a_2 + b = 0$. This scenario is plausible for $N_2 > N_1$.}, the vacuum energy reads,
\begin{align}
    E_{(k_1,k_2)}(\theta_1 , \theta_2 ) &= E_{(0,0)} (\theta_1 - 2 \pi k_1, \theta_1 - 2 \pi k_2) \notag \\
    E_{(0,0)} (\theta_1 , \theta_2 ) &\simeq E_0 + a_1 \theta_1^2  + a_2 \theta_2^2 - 2 a_2 \theta_1 \theta_2 \notag \\
    &= a_1 (\theta_1 - \theta_2)^2 + (a_2-a_1) \theta_2^2,
\end{align}
which is identical to the vacuum energy of the pure product-group Yang-Mills theory at $(\theta_1',\theta_2') = (\theta_1 - \theta_2, \theta_2)$.

One can further guess the ranks of the gauge groups by matching the massless phase diagrams:
\begin{align}
    \theta_{\mathrm{phys}} &= N_1' \theta_1 - N_2' \theta_2 = N_1'(\theta_1 - \theta_2) - (N_2' - N_1') \theta_2.
\end{align}
Therefore, we have the following coincidence of the phase diagrams at the massless limit:
\begin{align}
    SU(N_1)&\times SU(N_2) \mathrm{~QCD(BF)~on~} (\theta_1,\theta_2)~\mathrm{at}~m=0 \notag \\
    &\longleftrightarrow  SU(N_1)\times SU(N_2-N_1) ~\mathrm{QCD(BF)~on~} (\theta_1-\theta_2,\theta_2)~\mathrm{at}~m=0
\end{align}

Based on these observations, Karasik and Komargodski~\cite{Karasik:2019bxn} conjectured a new infrared duality,
\begin{align}
    SU(N_1)&\times SU(N_2) \mathrm{~QCD(BF)~on~} (\theta_1,\theta_2)~\mathrm{for}~m<m^* \notag \\
    &\longleftrightarrow~  SU(N_1)\times SU(N_2-N_1) ~\mathrm{QCD(BF)~on~} (\theta_1-\theta_2,\theta_2), \label{eq:KK_duality_conjecture}
\end{align}
where the fermion mass of the dual theory, $m'$, should be a function of $m$ that satisfies $m'|_{m=0}=0$ and $m'|_{m=m^*}=\infty$. Here, the duality means the matching of the phase diagrams and meta-stable vacua between these theories.

Now, let us assume that the fermion mass $m$ of the parent theory is set to be sufficiently small, then $m'$ of the daughter theory would also be small enough and it is below the topology-changing point of $SU(N_1)\times SU(N_2-N_1)$ QCD(BF). 
If so, we can further apply the duality~\eqref{eq:KK_duality_conjecture} to obtain another daughter QCD(BF), and we can repeat this procedure, which is called the duality cascade in Ref.~\cite{Karasik:2019bxn} (see Figure~\ref{fig:KK_duality_conjecture}). 
Although this duality cascade involves changes in the fermion mass $m$, its structure resembles the duality cascade of the $\mathcal{N}=1$ supersymmetric bifundamental gauge theories~\cite{Klebanov:2000hb}.
Understanding this observation is quite valuable for extending non-perturbative techniques towards non-supersymmetric theories.

As an additional remark, it is worth mentioning that the asymptotic freedom of the daughter theory requires an additional constraint.
Indeed, an asymptotically free parent theory satisfies $N_1 < N_2 < \frac{11}{2}N_1$.
Then, the asymptotic freedom for the $SU(N_1)$ gauge, $N_2 - N_1 < \frac{11}{2}N_1$, automatically holds, but the other condition $N_1 < \frac{11}{2} (N_2 - N_1)$, i.e. $N_1 < \frac{11}{13} N_2$ gives an additional constraint.

\section{Anomaly and global inconsistency}
\label{sec:anomaly_BFQCD}

In this section, we extend the anomaly and global inconsistency analysis for QCD(BF) \cite{Tanizaki:2017bam, Karasik:2019bxn, Hayashi:2023wwi} to the different-rank QCD(BF).
We first discuss the background gauge fields for the vector-like symmetry \eqref{eq:vect_sym} and classify its possible symmetry-protected topological (SPT) states. 
Using this knowledge, we derive the 't~Hooft anomaly and global inconsistency for the theta periodicities and the discrete chiral symmetry.

\subsection{Background gauging of the vector-like symmetry and its SPT classification}

As discussed in Section~\ref{sec:symmetry}, the vector-like symmetry $G_{\mathrm{vec}}$ of QCD(BF) consists of the baryon-number symmetry $U(1)_V/\mathbb{Z}_{\lcm(N_1,N_2)}$ and the $1$-form symmetry $\mathbb{Z}_{\gcd(N_1,N_2)}^{[1]}$, and let us introduce its background gauge field. 
We have seen that the origins of the quotient in the baryon-number symmetry and of the $1$-form symmetry both come from the quotient of the structure group $G^{\mathrm{gauge+global}}$ in \eqref{eq:total_symmetry}. 
This indicates that the background gauge fields for $G_\mathrm{vec}$ consist of 
\begin{equation}
    \left\{
    \begin{aligned}
    A:& \text{ $U(1)_V$ $1$-form gauge field},\\
    B_1:& \text{ $\mathbb{Z}_{N_1}$ $2$-form gauge field}, \\
    B_2:& \text{ $\mathbb{Z}_{N_2}$ $2$-form gauge field}. 
    \end{aligned}
    \right.
\end{equation}
Following Ref.~\cite{Kapustin:2014gua}, we describe the $\mathbb{Z}_{N_{1,2}}$ $2$-form gauge fields as pairs of $U(1)$ $2$-form and $1$-form gauge fields $(B_i,C_i)$ satisfying
\begin{equation}
    N_i B_i=\diff C_i,
\end{equation}
for $i = 1,2$.
To reduce the pair of $U(1)$ $2$-form and $1$-form gauge fields into the discrete $2$-form gauge field, we postulate the $1$-form gauge invariance:
\begin{equation}
    B_i\mapsto B_i+\diff \Lambda_i,\quad C_i \mapsto C_i+ N_i \Lambda_i, 
\end{equation}
where the $1$-form gauge parameters $\Lambda_{1,2}$ are $U(1)$ gauge fields. 

To represent the $SU(N_1) \times SU(N_2)$ gauge fields with the $\mathbb{Z}_{N_1} \times \mathbb{Z}_{N_2}$ backgrounds, we locally promote the $SU(N_1) \times SU(N_2)$ gauge fields $(a_1,a_2)$ to the $U(N_1) \times U(N_2)$ gauge fields $(\widetilde{a}_1,\widetilde{a}_2)$, 
\begin{equation}
    \widetilde{a}_i=a_i+\frac{1}{N_i} C_i \bm{1}_{N_i} \quad (\text{locally}). 
\end{equation}
The $U(N_i)$ field strength is defined by $\widetilde{f}_i=\diff \widetilde{a}_i+\im \widetilde{a}_i\wedge \widetilde{a}_i$, and its $1$-form gauge-invariant combination is given by $f_i = \widetilde{f}_i - B_i$.
To implement the quotient structure, we note that the $(a_1 + A)$ combination (resp.~$a_2 - A$ combination) is kept under the $\mathbb{Z}_{N_1}$ background (resp.~$\mathbb{Z}_{N_2}$ background) gauge transformations.
Hence, we locally modify the $U(1)_V$ background gauge field $A$ as,
\begin{equation}
    \widetilde{A} =  A - \frac{1}{N_1} C_1 + \frac{1}{N_2} C_2 \quad (\text{locally}),
    \label{eq:localdef_U1withtwist}
\end{equation}
and regard $\widetilde{A}$ as a proper $U(1)$ gauge field (satisfying the cocycle condition).
Then, the combination $f_1 + \diff A$ (resp.~$f_2 - \diff A$) does not have a fractional $\mathbb{Z}_{N_1}$ (resp.~$\mathbb{Z}_{N_2}$) flux.
Incidentally, the covariant derivative is not affected by the $\mathbb{Z}_{N_1} \times \mathbb{Z}_{N_2}$ two-form backgrounds,
\begin{equation}
    \slashed{D} \Psi \Rightarrow \widetilde{\slashed{D}}[A]\Psi=\gamma^{\mu}\left(\partial_\mu \Psi+\im\, \widetilde{a}_1 \Psi-\im\,\Psi \widetilde{a}_2+\im\,\widetilde{A}\Psi\right). 
    \label{eq:gaugedDirac}
\end{equation}
To sum up, the introduction of the background gauge fields replaces the field strengths as
\begin{align}
    & f_1 \Rightarrow \widetilde{f}_1 - B_1 \bm{1}_{N_1}, \notag \\
    & f_2 \Rightarrow \widetilde{f}_2 - B_2 \bm{1}_{N_2},  \notag \\
    & \diff A \Rightarrow  \diff \widetilde{A} + B_1-B_2,
    \label{eq:couplingsBackgroundFields}
\end{align}
where $\widetilde{f}_1, \widetilde{f}_2,$ and $\diff \widetilde{A}$ obey the standard Dirac quantization, and the backgrounds $(B_1,B_2)$ introduce fractional fluxes.

For later purposes, it is convenient to rotate the basis of $(B_1,B_2)$ in order to separate the $\mathbb{Z}_{\operatorname{lcm}(N_1,N_2)}$ ``quotient part'' $B_A$ for the $U(1)_V/\mathbb{Z}_{\lcm(N_1,N_2)}$ symmetry and the $\mathbb{Z}_{{\operatorname{gcd}(N_1,N_2)}}^{[1]}$-symmetry background $B$ \cite{Sulejmanpasic:2020zfs}: The translation rule turns out to be 
\begin{align}
\left\{ \,
    \begin{aligned}
    & B_A = -B_1 + B_2 , \\
    & B = - (M_2N_1' B_1 + M_1 N_2' B_2),
    \end{aligned}
\right.
\label{eq:def_B_BA}
\end{align}
where $M_1,M_2$ are Bezout coefficient~\eqref{eq:def_K1_K2}, and the inverse relation is given by
\begin{align}
\left\{ \,
    \begin{aligned}
    & B_1 = -  M_1 N_2' B_A - B , \\
    & B_2 = M_2N_1' B_A - B. \label{eq:transformation_formula}
    \end{aligned}
\right.
\end{align} 
Let us derive this result. The ``quotient part'' $B_A$ can be easily found from the minimal coupling $\diff A \Rightarrow  \diff \widetilde{A} + B_1-B_2 =: \diff \widetilde{A} - B_A$, and 
\begin{align}
    B_A = - B_1 + B_2 = \frac{-N_2' \diff C_1 + N_1' \diff C_2}{\operatorname{lcm}(N_1,N_2)},
\end{align}
is indeed a 2-form $\mathbb{Z}_{\operatorname{lcm}(N_1,N_2)}$ background field since $N_1'$ and $N_2'$ are coprime integers.
In terms of a pair of $U(1)$ gauge fields $(C_A,B_A),$ the $\mathbb{Z}_{\operatorname{lcm}(N_1,N_2)}$ background is
\begin{align}
    \operatorname{lcm}(N_1,N_2) B_A = \diff C_A,~~~
    C_A = -N_2' C_1 + N_1'  C_2 ,
\end{align}
with the 1-form gauge invariance
\begin{align}
    C_A \sim C_A + \operatorname{lcm}(N_1,N_2) (- \Lambda_1 + \Lambda_2). \label{eq:gauge-inv-quotient}
\end{align}
Next, we shall extract the $\mathbb{Z}_{{\operatorname{gcd}(N_1,N_2)}}^{[1]}$ symmetry background $B$ from $(B_1,B_2)$, which should satisfy $\gcd(N_1, N_2) B=\diff C$ with some $U(1)$ gauge field $C$. 
According to the basis rotation of the group \eqref{eq:basis_rotation_group}, this can be achieved as
\begin{align}
    \begin{pmatrix}
C_A & \\
C & \\
\end{pmatrix}
=
\begin{pmatrix}
-N_2' & N_1' \\
-M_2 & -M_1 \\
\end{pmatrix}
\begin{pmatrix}
C_1 \\
C_2 \\
\end{pmatrix}. 
\end{align}
This transformation is invertible within integer coefficients as the transformation matrix has a unit determinant.
For $C = -(M_2 C_1 + M_1 C_2)$, the 1-form gauge invariance becomes
\begin{align}
    C \sim C - \operatorname{gcd}(N_1,N_2) ( M_2 N_1' \Lambda_1 + M_1 N_2' \Lambda_2). \label{eq:gauge-inv-1-form}
\end{align}
Since the transformation matrix is invertible with integer coefficients, we can reparametrize the 1-form gauge transformations (\ref{eq:gauge-inv-quotient}) and (\ref{eq:gauge-inv-1-form}) as,
\begin{align}
    C_A \sim C_A + \operatorname{lcm}(N_1,N_2) \Lambda_A,~~~C \sim C + \operatorname{gcd}(N_1,N_2) \Lambda, 
\end{align}
where $\Lambda$ and $\Lambda_A$ are canonically normalized $U(1)$ gauge fields.
Now, we can see the $\mathbb{Z}_{{\operatorname{gcd}(N_1,N_2)}}^{[1]}$ symmetry background is expressed by the following pair of $(B,C)$,
\begin{align}
    \operatorname{gcd}(N_1,N_2) B = \diff C,~~C= -(M_2 C_1 + M_1 C_2).
\end{align}
To summarize, after the basis rotation, the $1$-form gauge transformations for the background gauge fields are given by 
\begin{equation}
    \left\{
    \begin{aligned}
        \widetilde{A}\mapsto& \widetilde{A}+\Lambda_A, \\\
        B_A\mapsto& B_A +\diff \Lambda_A,&& \text{with}~  C_A\mapsto C_A + \lcm(N_1,N_2)\Lambda_A,\\
        B \mapsto& B + \diff \Lambda,&& \text{with}~ C\mapsto C+\gcd(N_1,N_2)\Lambda.
    \end{aligned}
    \right.
\end{equation}
We note that ``$\lcm(N_1,N_2)(\diff \tilde{A}-B_A)$'' gives the canonically quantized $U(1)$ gauge field, which can be understood as the baryon-number gauge field~\cite{Tanizaki:2018wtg}. 

The gauge-invariant local topological action ($\bmod \, 2\pi$) of the background gauge fields is then given by 
\begin{align} 
    S^{k,\ell,\phi}_{\mathrm{SPT}} [\widetilde{A}, B_A, B] &= \frac{\im \operatorname{gcd}(N_1,N_2) k}{4\pi}\int B \wedge B + \frac{\im \operatorname{lcm}(N_1,N_2)  \ell}{2\pi}\int B \wedge (\diff \widetilde{A} - B_A) \notag \\
    &~~+\frac{\im (\operatorname{lcm}(N_1,N_2))^2 \phi}{8\pi^2} \int (\diff \widetilde{A} - B_A)^2, 
    \label{eq:BFQCD-SPT}
\end{align}
where $k, \ell\in \mathbb{Z}_{\operatorname{gcd}(N_1,N_2)}$ are discrete labels, and $\phi$ is continuous parameter with $2\pi$ periodicity. 
When the partition functions of the gapped vacua belong to different discrete labels $(k,\ell)$, there is no continuous deformation of the theory that connects those vacua. 
This means that they are discriminated as SPT states with the $G_{\mathrm{vec}}$ symmetry, and those states must be separated by some quantum phase transitions.

\subsection{Anomaly for the \texorpdfstring{$\theta_{1,2}$}{theta1,2} periodicity and for the chiral symmetry \texorpdfstring{$(\mathbb{Z}_{\operatorname{gcd}(N_1,N_2)})_{\chi}$}{(Z{gcd(N1,N2)})chiral}}

To find an 't~Hooft anomaly of QCD(BF),  we define the partition function equipped with the gauge background $\widetilde{A}, B_A, B$,
\begin{align}
    Z_{\theta_1,\theta_2} [\widetilde{A}, B_A, B] := \int \mathcal{D}\widetilde{a}_1 \mathcal{D}\widetilde{a}_2 \mathcal{D} \overline{\Psi} \mathcal{D} \Psi ~\rme^{-S[\widetilde{A}, B_A, B]} ,
\end{align}
where the gauged action is given by
\begin{align}
    S[\widetilde{A}, B_A, B] &:=   \frac{1}{g_1^2} \int  |\widetilde{f}_1-B_1|^2 +  \frac{1}{g_2^2} \int |\widetilde{f}_2-B_2|^2 +\int \operatorname{tr} \overline{\Psi}(\widetilde{\slashed{D}}[A]+m)\Psi \nonumber\\
&+\frac{\im\theta_1}{8\pi^2}\int \operatorname{tr}\left[(\widetilde{f}_1-B_1)^2\right]+\frac{\im\theta_2}{8\pi^2}\int \operatorname{tr} \left[(\widetilde{f}_2-B_2)^2\right].
\label{eq:action_with_background}  
\end{align}
Here, $(B_1,B_2)$ are given by (\ref{eq:transformation_formula}) in terms of $(B_A, B)$.

Let us first discuss the generalized anomaly (or global inconsistency) related to the $\theta$ periodicities.
This partition function violates the usual $\theta$ angle periodicity, and it acquires the local counterterm; 
\begin{align}
    Z_{\theta_1+2\pi,\theta_2} [\widetilde{A}, B_A, B] &= \exp\left( \frac{\im N_1}{4 \pi} \int B_1 \wedge B_1 \right) Z_{\theta_1,\theta_2} [A, B_A, B] , \notag \\
    Z_{\theta_1,\theta_2+2\pi} [\widetilde{A}, B_A, B] &=  \exp\left( \frac{\im N_2}{4 \pi} \int B_2 \wedge B_2  \right) Z_{\theta_1,\theta_2} [A, B_A, B] .
    \label{eq:anomalyQCDBF}
\end{align}
We consider the implications of these global inconsistencies.
Let us suppose that the system has a trivial vacuum for some $(\theta_1,\theta_2)$, 
then the partition function is described by the local topological action of the background fields in the low-energy limit:
\begin{align}
    Z_{\theta_1,\theta_2}[\widetilde{A}, B_A, B] & \xrightarrow{\text{RG flow}} \exp \left( S^{k,\ell,\phi}_{\mathrm{SPT}} [\widetilde{A}, B_A, B] \right), 
\end{align}
which has discrete labels $k,\ell \in \mathbb{Z}_{\gcd(N_1,N_2)}$. 
For general $2\pi$ shifts of $\theta$ angles $(\theta_1,\theta_2) \mapsto (\theta_1+2\pi n_1,\theta_2+2\pi n_2)$, the relation \eqref{eq:anomalyQCDBF} with \eqref{eq:transformation_formula} shows that 
the discrete labels of the effective action~\eqref{eq:BFQCD-SPT} must jump as
\begin{align}
    \begin{pmatrix}
        k\\
        \ell
    \end{pmatrix}
    \mapsto 
    \begin{pmatrix}
        k\\
        \ell
    \end{pmatrix}
    +
    \begin{pmatrix}
    N_1' & N_2' \\
    -M_1 & M_2 \\
    \end{pmatrix}
    \begin{pmatrix}
    n_1 \\
    n_2 \\
    \end{pmatrix}.
    % (k,\ell)\mapsto (k + N_1' n_1 +  N_2' n_2, \ell - M_1 n_1 +  M_2 n_2). 
\end{align}
As long as these shifts are nontrivial in $\mathbb{Z}_{\gcd(N_1,N_2)}\times \mathbb{Z}_{\gcd(N_1,N_2)}$, the two points $(\theta_1,\theta_2)$ and $(\theta_1+2\pi n_1,\theta_2+2\pi n_2)$ can be distinguished as SPT states.
Note that the above $2\times 2$ matrix has a unit determinant, so it is invertible with integer coefficients.
Therefore, there is a global inconsistency for the shift $(\theta_1,\theta_2) \mapsto (\theta_1+2\pi n_1,\theta_2+2\pi n_2)$ unless both $n_1$ and $n_2$ are multiples of $\operatorname{gcd}(N_1,N_2)$.
This global inconsistency rules out a trivially gapped phase connecting the two points $(\theta_1,\theta_2)$ and $ (\theta_1+2\pi n_1,\theta_2+2\pi n_2)$.
One can interpret this obstruction as an anomaly between $U(1)_V/\mathbb{Z}_{\operatorname{lcm}(N_1,N_2)}$, $\mathbb{Z}_{\operatorname{gcd}(N_1,N_2)}^{[1]}$, and the $\theta_1$ or $\theta_2$ periodicity (see Refs.~\cite{Shimizu:2017asf, Gaiotto:2017tne, Tanizaki:2017mtm, Tanizaki:2018wtg, Yonekura:2019vyz, Anber:2019nze, Morikawa:2022liz} for related anomalies), and one can check the large-$N$ phase diagram (e.g. Figure~\ref{fig:BFQCDsmallmass}) satisfies this constraint.

In the massless case $m=0$, the $SU(N_1) \times SU(N_2)$ QCD(BF) additionally has the discrete chiral symmetry $(\mathbb{Z}_{\operatorname{gcd}(N_1,N_2)})_{\chi}$, and let us discuss its mixed 't~Hooft anomaly with $G_{\mathrm{vec}}$.
Since the axial transformation shifts the theta angles, the calculation is parallel to that of global inconsistency (see also Refs.~\cite{Sulejmanpasic:2020zfs,Hayashi:2023wwi}).
The discrete chiral transformation $(\mathbb{Z}_{ \operatorname{gcd}(N_1,N_2)})_\chi$ changes the fermion integration measure, 
\begin{align}
    \Diff \overline{\Psi} \Diff \Psi & \mapsto \exp\left(\frac{2\pi \im}{\operatorname{gcd}(N_1,N_2)}\mathrm{ind}(\widetilde{\slashed{D}}[A])\right) \Diff \overline{\Psi}\Diff \Psi, 
    \label{eq:chiralanomaly}
\end{align}
so we have to evaluate the index of the Dirac operator~\eqref{eq:gaugedDirac},
\begin{align}
    \mathrm{ind}(\widetilde{\slashed{D}}[A])
    &=\int\frac{1}{8\pi^2}\tr\left[(\widetilde{f}_1\otimes \bm{1}_{N_2}-\bm{1}_{N_1}\otimes\widetilde{f}_2+\diff \widetilde{A} ~\bm{1}_{N_1}\otimes \bm{1}_{N_2})^2\right] \notag\\
    &=\int \frac{1}{8\pi^2}\tr\left[\left((\widetilde{f}_1-B_1)\otimes \bm{1}_{N_2}-\bm{1}_{N_1} \otimes(\widetilde{f}_2- B_2)+(\diff \widetilde{A} - B_A)\bm{1}_{N_1} \otimes \bm{1}_{N_2} \right)^2\right]\notag\\
    &=\int \frac{1}{8\pi^2}\left(N_2 \tr(\widetilde{f}_1-B_1)^2 + N_1 \tr(\widetilde{f}_2-B_2)^2+N_1N_2(\diff \widetilde{A} - B_A)^2\right).
\end{align}
Note that the anomaly~\eqref{eq:chiralanomaly} depends only on $\mathrm{ind}(\widetilde{\slashed{D}}[A])$ in mod $\gcd(N_1,N_2)$. 
This modulo-$\operatorname{gcd}(N_1,N_2)$ index can be calculated as follows\footnote{We added $\diff \widetilde{A}$, which gives no contribution, in order to make the gauge-invariant combination $\diff \widetilde{A} - B_A$ and align the expressions with (\ref{eq:BFQCD-SPT}).}
\begin{align}
    \mathrm{ind}(\widetilde{\slashed{D}}[A])
    &= \frac{N_1 N_2}{8\pi^2} \int (-B_1 \wedge B_1 - B_2 \wedge B_2 +  B_A \wedge B_A) ~~[\operatorname{mod}~\operatorname{gcd}(N_1,N_2)] \notag \\
    &= -\frac{2 N_1 N_2}{8\pi^2} \int B \wedge B + \frac{2 N_1 N_2 (M_1 N_2' - M_2 N_1' ) }{8\pi^2} \int  B \wedge (\diff \widetilde{A} - B_A) \notag \\
    & ~~~ + \frac{2 M_1 M_2 (\operatorname{lcm} (N_1,N_2))^2 }{8\pi^2} \int  (\diff \widetilde{A} - B_A)^2 ~~[\operatorname{mod}~\operatorname{gcd}(N_1,N_2)].
\end{align}
Here we have used $N_1 N_2 (1- (M_1 N_2')^2 - (M_2 N_1')^2 ) = 2 M_1 M_2 (\operatorname{lcm} (N_1,N_2))^2$, which follows from the definition of $(M_1,M_2)$, (\ref{eq:def_K1_K2}).

This anomaly provides a constraint on possible phases, which rules out the trivially gapped phase. 
One of the minimal possibilities to satisfy the anomaly matching is to spontaneously break the discrete chiral symmetry completely. 
To illustrate this, we note that the labels of the topological action \eqref{eq:BFQCD-SPT} are shifted by the discrete chiral transformation as follows,
\begin{equation}
    (k,\ell,\phi)\mapsto \left(k-2 N_1'N_2', \ell + M_1N_2' - M_2 N_1', \phi + 2\pi\frac{2 M_1 M_2}{\operatorname{gcd}(N_1,N_2)} \right). \label{eq:shifts_chiral}
\end{equation}
We can see that all the nontrivial discrete chiral transformations $(\mathbb{Z}_{\operatorname{gcd}(N_1,N_2)})_\chi$ change the discrete labels. 
To this end, let us rewrite the above transformation of the discrete labels as 
\begin{equation}
    \begin{pmatrix}
        k\\
        \ell
    \end{pmatrix}
    \mapsto
    \begin{pmatrix}
        k\\
        \ell
    \end{pmatrix}
    -
    \begin{pmatrix}
        N_2' & N_1'\\
        M_2  & -M_1
    \end{pmatrix}
    \begin{pmatrix}
        N_1'\\
        N_2'
    \end{pmatrix},
\end{equation}
which suggests that it is convenient to change the $(k,\ell)$ basis to another basis defined by 
\begin{equation}
    \begin{pmatrix}
        \tilde{k}\\
        \tilde{\ell}
    \end{pmatrix}
    =
    \begin{pmatrix}
        -N_2' & N_1'\\
        M_2  & M_1
    \end{pmatrix}
    \begin{pmatrix}
        N_2' & N_1'\\
        M_2  & -M_1
    \end{pmatrix}^{-1}
    \begin{pmatrix}
        k\\
        \ell
    \end{pmatrix}
    =
    \begin{pmatrix}
        -N_2' & N_1'\\
        M_2  & M_1
    \end{pmatrix}
    \begin{pmatrix}
        M_1 & N_1'\\
        M_2 & -N_2'
    \end{pmatrix}
    \begin{pmatrix}
        k\\
        \ell
    \end{pmatrix}.
    \label{eq:newBasis_discreteLabels}
\end{equation}
Then, the discrete chiral transformation acts as $(\tilde{k},\tilde{\ell})\mapsto (\tilde{k},\tilde{\ell}-1)$, and it is now evident that all $(\mathbb{Z}_{ \operatorname{gcd}(N_1,N_2)})_{\chi}$ are anomalous under the $G_{\mathrm{vec}}$ background gauge field. 
Somewhat surprisingly, the discrete label $(\tilde{k},\tilde{\ell})$ naturally appears in the semiclassics as we shall discuss in Section~\ref{sec:center_vortex_residual_N1N2}. 
As a summary, the anomaly matching condition requires the complete chiral symmetry breaking, 
\begin{equation}
(\mathbb{Z}_{\operatorname{gcd}(N_1,N_2)})_{\mathrm{\chi}}\xrightarrow{\text{SSB}} 1, 
\end{equation}
under the following assumptions: (1) The system is confining so that $\mathbb{Z}_{\operatorname{gcd}(N_1,N_2)}^{[1]}$ is unbroken, (2) the vector-like $U(1)_V/\mathbb{Z}_{\operatorname{lcm}(N_1,N_2)}$ symmetry is unbroken, and (3) there is no accidental massless mode or topological order in the low-energy theory.

\section{Semiclassics through \texorpdfstring{$T^2$}{T2} compactification with the baryon-'t Hooft flux}
\label{sec:QCDBF_2dEFT}

In this section, we develop a semiclassical framework for studying the dynamics of the different-rank QCD(BF). 
It is based on the novel semiclassical approach to the $4$d gauge theories through the $T^2$-compactification with the 't~Hooft flux~\cite{Tanizaki:2022ngt}, which preserves the $4$d 't~Hooft anomalies in $2$d effective theories and explains the confinement by the proliferation of center vortices or fractional instantons. 
The effective description of this $T^2$-compactified setup captures qualitative features of strongly-coupled $4$d confining gauge theories, including pure $SU(N)$ Yang-Mills theory, $\mathcal{N}=1$ $SU(N)$ SYM, QCD with fundamental quarks, $2$-index quarks, and equal-rank QCD(BF)~\cite{Tanizaki:2022ngt, Tanizaki:2022plm, Hayashi:2023wwi, Hayashi:2024qkm} (see also Refs.~\cite{Yamazaki:2017ulc, Cox:2021vsa, Poppitz:2022rxv} for related studies on $\mathbb{R}\times T^3$). 

Section \ref{sec:T2compactification} describes the detailed setup of the $T^2$ compactification for the different-rank QCD(BF).
In Section \ref{sec:derive_2dEFT}, we derive the two-dimensional semiclassical effective theory, which serves as a foundational framework for the exploration of phase diagrams in subsequent sections under the assumption of the adiabatic continuity.

\subsection{\texorpdfstring{$T^2$}{T2} compactification with baryon-'t Hooft flux}
\label{sec:T2compactification}

We put the QCD(BF) on $\mathbb{R}^2\times T^2$ with the nontrivial background gauge fields on $T^2$ so that the minimal 't~Hooft fluxes are inserted for both $SU(N_1)$ and $SU(N_2)$ gauge groups: 
\begin{equation}
    \int_{T^2} B_1=\frac{2\pi}{N_1}, \quad \int_{T^2} B_2=\frac{2\pi}{N_2}. 
\end{equation}
For this purpose, we translate these conditions into those of the background fields for the global symmetry, $U(1)_V/\mathbb{Z}_{\lcm(N_1,N_2)}\times \mathbb{Z}_{\gcd(N_1,N_2)}^{[1]}$, according to \eqref{eq:def_B_BA} (with setting $\widetilde{A}=0$), 
\begin{equation}
    \int_{T^2} \lcm(N_1,N_2)(\diff \widetilde{A}-B_A)=2\pi \left(N_2'-N_1'\right), \quad 
    \int_{T^2} B = 2\pi \frac{-(M_1+ M_2)}{\gcd(N_1, N_2)}. 
    \label{eq:baryon_tHooft}
\end{equation}
This is the baryon-'t~Hooft flux in our $T^2$-compactification. 

To obtain the $2$d semiclassical description, it is convenient to realize the baryon-'t~Hooft flux in terms of the transition functions. 
Let us denote the coordinates as $x = (x_1,x_2,x_3,x_4) = (\Vec{x}, x_3, x_4) \in \mathbb{R}^2 \times T^2$, and we identify $(x_3,x_4)$ with both $(x_3+L,x_4)$ and $(x_3,x_4+L)$ to construct the torus $T^2$.
We call the transition functions on $T^2$ for $SU(N_i)$ gauge group $g_3^{(i)}(x_4)$ and $g_4^{(i)}(x_3)$~($i = 1,2$).
The 't~Hooft fluxes require that the transition functions satisfy
\begin{align}
    g_3^{(i)}(L)^\dagger g_4^{(i)}(0)^\dagger g_3^{(i)}(0) g_4^{(i)}(L) = \rme^{\frac{2 \pi \im}{N_i}}\bm{1}_{N_i}. \label{eq:tHooft_flux_transition_fct}
\end{align}
Using the gauge transformations, we get the coordinate-independent transition functions,
\begin{align}
    g_3^{(i)}(x_4) = S_i,~~~~g_4^{(i)}(x_3) = C_i~~~(i=1,2), 
\end{align}
where $C_i$ and $S_i$ are the $SU(N_i)$ clock and shift matrices: $C_i \propto \operatorname{diag}(1,\rme^{\frac{2 \pi \im}{N_i}}, \cdots,\rme^{\frac{2 \pi \im (N_i-1)}{N_i}})$ and $(S)_{mn} \propto \delta_{m+1,n}$.
For $N_1 \neq N_2$, the wavefunction of the bifundamental field cannot be single-valued with these 't~Hooft fluxes, and we also introduce the fractional $U(1)_V$ flux in the torus $T^2$ to compensate it:
\begin{align}
    A = \left( \frac{2 \pi}{N_1 L^2} - \frac{2 \pi}{N_2 L^2} \right) x_3 \diff x_4
\end{align}
Note that this is equivalent to \eqref{eq:baryon_tHooft} as $\diff A = \diff \widetilde{A}-B_A$ according to the local expression~\eqref{eq:localdef_U1withtwist}. 
Consequently, the boundary conditions for the bifundamental fermion $\Psi(x)$ are written as
\begin{align}
    \Psi(\Vec{x}, x_3+L, x_4) &= \rme^{\im\left( \frac{2 \pi}{N_1 L} - \frac{2 \pi}{N_2 L} \right) x_4} g_3^{(1)}(x_4)^\dagger \psi(\Vec{x}, x_3, x_4) g_3^{(2)}(x_4)  \notag \\
    \Psi(\Vec{x}, x_3, x_4+L) &=  g_4^{(1)}(x_3)^\dagger \Psi(\Vec{x}, x_3, x_4) g_3^{(2)}(x_4).
\end{align}
Now, these twisted boundary conditions are consistent with the single-valuedness of the bifundamental fermion $\Psi(\Vec{x}, x_3, x_4)$. 

\subsection{2d effective description}
\label{sec:derive_2dEFT}

The adiabatic continuity conjecture implies that the weakly-coupled theory on $\mathbb{R}^2 \times T^2$ at small $T^2$ can provide insight into the qualitative aspects of the original strongly-coupled theory on $\mathbb{R}^4$. 
With the compactification described above, let us derive the 2d effective theory for the weakly-coupled theory on $\mathbb{R}^2 \times T^2$ at small $T^2$.

Our approach involves the following two steps: 
First, we identify low-energy modes that remain perturbatively massless after the small $T^2$ compactification.
Subsequently, we incorporate center-vortex contributions together with the residual gauge symmetry after the Higgsing by the Polyakov loops along $T^2$.
 
\subsubsection{Perturbative analysis}

To construct the 2d effective theory, let us first identify the low-energy modes of the $2$d effective theory.
For $SU(N_1)$ and $SU(N_2)$ gluons, the boundary conditions admit no constant modes, and all excitations have $O(1/NL)$ mass on the small torus $T^2$.
This can be understood as the adjoint Higgsing, $SU(N_1)\times SU(N_2) \rightarrow \mathbb{Z}_{N_1} \times \mathbb{Z}_{N_2}$ (see Ref.~\cite{Tanizaki:2022ngt} for details). 
The residual discrete gauge fields turn out to play an important role beyond perturbation theory, but we may neglect it for a while. 

For the bifundamental fermion, the low-energy modes can be found by solving the $2$d zero-mode equation on $T^2$,
\begin{align}
    \left[ \gamma_3 \partial_3 + \gamma_4 \left( \partial_4 + \im A_{4} \right) \right] \Psi(x_3,x_4) = 0,
\end{align}
with the boundary conditions
\begin{align}
    \Psi(x_3+L,x_4) &= \rme^{\im\left( \frac{2 \pi}{N_1 L} - \frac{2 \pi}{N_2 L} \right) x_4} S_1^\dagger \Psi(x_3,x_4) S_2, \\
    \Psi(x_3,x_4+L) &= C_1^\dagger \Psi(x_3,x_4) C_2.  
\end{align}
We can easily count the number of such zero-modes by using the 2d index theorem,
\begin{align}
    \operatorname{ind} (D) = \operatorname{dim} (\mathrm{Rep}) \int_{T^2} \frac{\diff A}{2 \pi}=N_2-N_1, \label{eq:low_energy_modes_indextheorem}
\end{align}
where the dimension of the representation is $\operatorname{dim} (\mathrm{Rep}) = N_1N_2$, and the flux is $\int_{T^2} \frac{F_{U(1)}}{2 \pi} = \frac{1}{N_1} - \frac{1}{N_2}$. 
As a result, there are $(N_2 - N_1)$ of 2d massless Dirac fermions at the classical level.
Of course, we can obtain the same result by directly solving the zero-mode equation with the above boundary condition, and we demonstrate it in Appendix \ref{app:counting_BFQCD}.

We note that most of these zero-modes are accidental. Recall that QCD(BF) does not have continuous chiral symmetries that can force massless fermions. 
To be more precise, we are now restricting ourselves to the perturbative analysis, so there still exists the $U(1)_\chi$ symmetry. 
This requires the presence of ``one'' massless fermion, but the other $N_2-N_1-1$ massless fermions are supposed to be gapped out by considering the interaction effect within the perturbation theory. 
To see it more clearly, let us perform the Abelian bosonization and we have $N_2-N_1$ compact bosonz, $\varphi_1,\ldots, \varphi_{N_2-N_1}$. 
The $U(1)_\chi$ acts on them as the shift symmetry, $\varphi_i\mapsto \varphi_i + \alpha$. 
As the difference $\varphi_i-\varphi_j$ is invariant under $U(1)_\chi$, its mass term can be generated by the perturbative diagrams, in particular by the $1$-loop diagrams~\cite{Tanizaki:2022plm}:
\begin{equation}
    \Delta S
    =
    \begin{gathered}
        \includegraphics[scale=0.35]{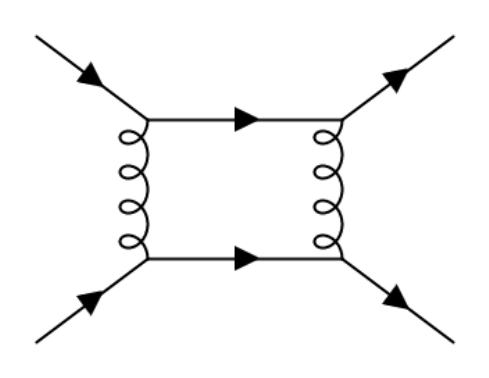}
    \end{gathered}
    =\int d^2 x \sum_{i\neq j} J_{ij}\cos(\varphi_i-\varphi_j).
    \label{eq:4fermi_bosonized}
\end{equation}
Note that this mass term for $\varphi_i-\varphi_j$ is perturbatively generated: $J_{ij} \sim O(g^4/(NL)^2)$, which is far larger than the (nonperturbatively small) dynamical scale $\rme^{-\sharp/g^2}$.
Therefore, the genuine low-energy mode for small $T^2$ is described by the diagonal one,
\begin{align}
    \varphi := \varphi_1 = \varphi_2 = \cdots = \varphi_{N_2-N_1}.
\end{align}
In what follows, we consider the 2d low-energy effective theory in terms of $\varphi$.
Since we obtain 2d free Dirac fermions, the action for the 2d effective theory in the lowest order would be,
\begin{align}
    S_\mathrm{2d}^{(0)}[\varphi] =  \int \diff^2x \left[ \frac{N_2-N_1}{8\pi} (\partial_\mu \varphi)^2 - m \mu \cos \varphi \right],
\end{align}
with a (scheme-dependent) dimensionful constant $\mu$.

\subsubsection{Center vortex and the residual \texorpdfstring{$\mathbb{Z}_{N_1} \times \mathbb{Z}_{N_2}$}{ZN1xZN2} gauge field}
\label{sec:center_vortex_residual_N1N2}

The perturbatively massless mode $\varphi$, analogous to the $\eta'$ particle, acquires a nonperturbatively small mass induced by center vortices.
The second step to derive the 2d effective theory is to include contributions from center vortices.

Prior to moving on to the calculation, it would be pertinent to mention some aspects of center vortices in this setup.
Virtually, let us further compactify $\mathbb{R}^2$ into another $T^2$ with nontrivial 't Hooft flux $n_{12}^{(1)} = 1$ for $SU(N_1)$.
Then, there exists a classical fractional instanton with
\begin{align}
    (Q_{\mathrm{top}}^{(1)}, Q_{\mathrm{top}}^{(2)}) = (1/N_1,0),~~ 
\end{align}
where $Q_{\mathrm{top}}^{(i)}$ denotes the instanton number for the $SU(N_i)$ gauge field. 
Because of the perturbative gap, this classical solution should behave as a local vortex in the decompactified limit $T^2 \times T^2 \rightarrow \mathbb{R}^2 \times T^2$. 
As suggested by numerical studies~\cite{Gonzalez-Arroyo:1998hjb, Montero:1999by, Montero:2000pb} (see also Refs.~\cite{Anber:2022qsz, Anber:2023sjn} for analytic studies), we further assume that this vortex solution saturates the BPS bound and the action is given by $S = S_\mathrm{I}^{(1)}/N_1$, with the instanton action $S_\mathrm{I}^{(1)} = \frac{8 \pi^2}{g^2_1}$. 
In parallel, there also exists $(Q_{\mathrm{top}}^{(1)}, Q_{\mathrm{top}}^{(2)}) = (0,1/N_2)$ fractional instanton with the action $S = S_\mathrm{I}^{(2)}/N = \frac{8 \pi^2}{N_2 g^2_2}$ as a local vortex solution.
Within the framework of 2d effective theory, we can identify these fractional instantons as so-called center vortices.

Let us include the contributions from center vortices by the dilute gas approximation.
Using the $U(1)_{\mathrm{axial}}$ spurious symmetry,
\begin{align}
    \varphi &\rightarrow \varphi + 2 \alpha, ~~\theta_1 \rightarrow \theta_1 + 2 N_2\alpha, ~~\theta_2 \rightarrow \theta_2 + 2 N_1\alpha, \label{eq:axial_bosonized}
\end{align}
the vertices of center vortices for $SU(N_1)$ and $SU(N_2)$ can be uniquely fixed:
\begin{align}
(\pm 1/N_1,0)~\mathrm{vortex}~~
\begin{cases}
    \mathcal{V}_1(\Vec{x}) = K^{(1)} \rme^{-S_\mathrm{I}^{(1)}/N_1 + \im \theta_1 /N_1} \rme^{- \im N_2 \varphi /N_1} \\
    \mathcal{V}_1^* (\Vec{x}) = K^{(1)} \rme^{-S_\mathrm{I}^{(1)}/N_1 - \im \theta_1 /N_1} \rme^{ \im N_2 \varphi /N_1}, 
\end{cases} \notag \\
(0,\pm 1/N_2)~\mathrm{vortex}~~
\begin{cases}
    \mathcal{V}_2(\Vec{x}) = K^{(2)} \rme^{-S_\mathrm{I}^{(2)}/N_2 + \im \theta_2 /N_2} \rme^{- \im N_1 \varphi /N_2}\\
    \mathcal{V}_2^* (\Vec{x}) = K^{(2)} \rme^{-S_\mathrm{I}^{(2)}/N_2 - \im \theta_2 /N_2} \rme^{ \im N_1 \varphi /N_2},  \label{eq:BFQCD_vertex_oper}
\end{cases}
\end{align}
where $K^{(1)}$ and $K^{(2)}$ are some dimensionful positive constants of $O(1/L^2)$. Using the $1$-loop renormalization group, the magnitudes of these vertices can be roughly estimated as
\begin{equation}
    K^{(1)}\rme^{-S^{(1)}_{\mathrm{I}}/N_1}\simeq \frac{1}{L^2}(\Lambda_1 L)^{(11-2 N_2/N_1)/3},\quad 
    K^{(2)}\rme^{-S^{(2)}_{\mathrm{I}}/N_2}\simeq \frac{1}{L^2}(\Lambda_2 L)^{(11-2 N_1/N_2)/3}. 
\end{equation}

The operators $ \rme^{- \im N_2 \varphi /N_1}$ and $\rme^{- \im N_1 \varphi /N_2}$ do not respect the $2\pi$ periodicity of $\varphi$ and demand further refinement.
We should notice that there are residual $\mathbb{Z}_{N_1} \times \mathbb{Z}_{N_2}$ gauge fields $(\Tilde{a}_1,\Tilde{a}_2)$, which magnetically couple to $\varphi$:
\begin{align}
    \frac{\im}{2\pi}\int (\theta_1-N_2 \varphi)\diff \Tilde{a}_1+\frac{\im}{2\pi}\int (\theta_2-N_1\varphi)\diff \Tilde{a}_2. 
    % \frac{i N_2}{2 \pi } \int \Tilde{a}_1 \wedge \diff \varphi,~~ \frac{i N_1}{2 \pi } \int \Tilde{a}_2 \wedge \diff \varphi 
    \label{eq:BFQCD_residual_coupling}
\end{align}
One can easily check consistency with the center-vortex vertices~(\ref{eq:BFQCD_vertex_oper}).\footnote{Roughly speaking, the center vortices are the defects with fractional fluxes: $\diff \Tilde{a}_1 \sim 2\pi/N_1$ and $\diff \Tilde{a}_2 \sim 2\pi/N_2$. The above magnetic coupling (\ref{eq:BFQCD_residual_coupling}) reproduces the vertices (\ref{eq:BFQCD_vertex_oper})}
The integration over the gauge field $(\Tilde{a}_1,\Tilde{a}_2)$ restrict the compact boson to satisfy $\int \diff \varphi \in 2 \pi N_1'$ and $\int \diff \varphi \in 2 \pi N_2'$, which extends the periodicity of $\varphi$ as $\varphi \sim \varphi + 2 \pi N_1' N_2'$, and $\rme^{- \im N_2 \varphi /N_1}$ and $\rme^{- \im N_1 \varphi /N_2}$ become well-defined.
Here, we still have $N_1' N_2'$ ambiguity to lift the $2\pi$-periodic scalar to a $2 \pi N_1' N_2'$-periodic one, and this requires the discussion on the discrete vacuum labels. 

The discrete vacuum labels appear because the appropriate boundary condition for $\mathbb{R}^2$ constrain the total topological charges $(Q_{\mathrm{top}}^{(1)}, Q_{\mathrm{top}}^{(2)})$ to be integers.
Indeed, the dilute gas approximation of fractional instantons with this constraint yields
\begin{align}
    Z &= \int \mathcal{D}\varphi~\rme^{-S_\mathrm{2d}^{(0)}[\varphi] } \notag \\
    &~~~~\times \left[ \sum_{n_1, \bar{n}_1 \geq 0} \frac{1}{n_1! \bar{n}_1 !} \left( \int \rmd^2\Vec{x} ~\mathcal{V}_1(\Vec{x}) \right)^{n_1} \left( \int \rmd^2\Vec{x} ~\mathcal{V}_1^*(\Vec{x}) \right)^{\bar{n}_1} \delta_{n_1-\bar{n}_1 \in N_1 \mathbb{Z}}\right] \notag \\
    &~~~~\times \left[ \sum_{n_2, \bar{n}_2 \geq 0} \frac{1}{n_2! \bar{n}_2 !} \left( \int \rmd^2\Vec{x} ~\mathcal{V}_2(\Vec{x}) \right)^{n_2} \left( \int \rmd^2\Vec{x} ~\mathcal{V}_2^*(\Vec{x}) \right)^{\bar{n}_2} \delta_{n_2-\bar{n}_2 \in N_2 \mathbb{Z}}\right]
    \notag\\
    &= \frac{1}{N_1 N_2} \sum_{k_1 =0}^{N_1-1} \sum_{k_2 = 0}^{N_2 -1} \int \mathcal{D}\varphi~\rme^{-S_\mathrm{2d}^{(0)}[\varphi] - S_{\mathrm{center~vortex}}^{(k_1,k_2)}[\varphi]}, 
    \label{eq:dilute_gas_paritition_fct}
\end{align}
where $k_1\in \mathbb{Z}_{N_1}$ and $k_2\in \mathbb{Z}_{N_2}$ are the discrete vacuum labels and 
\begin{align}
    S_{\mathrm{center~vortex}}^{(k_1,k_2)}[\varphi] &= \int \diff^2\Vec{x}~ \left\{- 2K^{(1)} \rme^{-S_\mathrm{I}^{(1)}/N_1} \cos \left( \frac{N_2 \varphi + 2 \pi k_1 - \theta_1}{N_1} \right)\right. \notag \\
    &~~~ \left. - 2K^{(2)} \rme^{-S_\mathrm{I}^{(2)}/N_2} \cos \left( \frac{N_1 \varphi  + 2 \pi k_2 - \theta_2}{N_2} \right)\right\}.
\end{align}
The periodicity of discrete labels, $k_1\sim k_1+N_1$ and $k_2\sim k_2+N_2$, are trivially realized, but the $2\pi$ shift of $\varphi$ requires a nontrivial identification, 
\begin{equation}
    (\varphi+2\pi, k_1, k_2) \sim (\varphi, k_1+N_2, k_2+N_1).
    \label{eq:identification_phi_k1_k2}
\end{equation}
As $\mathbb{Z}_{N_1}\times \mathbb{Z}_{N_2}\simeq \mathbb{Z}_{\gcd(N_1,N_2)}\times \mathbb{Z}_{\lcm(N_1,N_2)}$, we change the basis for the discrete labels as\footnote{Recall that our representative of the Bezout coefficient~\eqref{eq:def_K1_K2} satisfies $M_1\in N'_2\mathbb{Z}$ and $M_2\in N'_1 \mathbb{Z}$, and thus $\ell=M_1 k_1+M_2 k_2$ is well-defined in $\mathbb{Z}_{\lcm(N_1,N_2)}$. } 
\begin{equation}
    \begin{pmatrix}
        \tilde{k}\\
        \tilde{\ell}
    \end{pmatrix}
    =\begin{pmatrix}
        N_1' & -N_2'\\
        M_1 & M_2
    \end{pmatrix}
    \begin{pmatrix}
        k_1 \\
        k_2
    \end{pmatrix}, \label{eq:change_of_discrete_labels}
\end{equation}
so that $\tilde{k}\sim \tilde{k}+\gcd(N_1, N_2)$ and $\tilde{\ell} \sim \tilde{\ell}+\lcm(N_1,N_2)$. In this basis, \eqref{eq:identification_phi_k1_k2} becomes  
\begin{equation}
    (\varphi+2\pi, \tilde{k}, \tilde{\ell})\sim (\varphi, \tilde{k}, \tilde{\ell}+\gcd(N_1,N_2)). 
    \label{eq:identification_phi_k_l}
\end{equation}
Since $\lcm(N_1,N_2)=\gcd(N_1,N_2)N_1'N_2'$, this shows that $\tilde{\ell}$ also becomes the $\mathbb{Z}_{\gcd(N_1,N_2)}$ label by extending the periodicity of $\varphi$ as $\varphi\sim \varphi+2\pi N_1' N_2'$. 
Equivalently, we may regard $k_1,k_2$ are both the $\mathbb{Z}_{\gcd(N_1,N_2)}$ labels once $\varphi$ is extended as $\varphi\sim \varphi+2\pi N_1' N_2'$.

Consequently, the low-energy effective theory is described by the extended periodic scalar $\varphi$, $\varphi \sim \varphi + 2\pi N_1' N_2'$, and the $(\mathbb{Z}_{\gcd(N_1,N_2)})^2$ vacuum labels $(\tilde{k},\tilde{\ell})$ (or $(k_1,k_2)$), and its potential is given by 
\begin{align}
    V[\varphi; \tilde{k},\tilde{\ell}] &= - m \mu \cos (\varphi) - 2K^{(1)} \rme^{-S_\mathrm{I}^{(1)}/N_1} \cos \left( \frac{N_2 \varphi + 2 \pi (N'_2 \tilde{\ell} + M_2 \tilde{k}) - \theta_1}{N_1} \right) \notag \\
    &~~~~~~~~~ - 2K^{(2)} \rme^{-S_\mathrm{I}^{(2)}/N_2} \cos \left( \frac{N_1 \varphi
    + 2 \pi (N'_1 \tilde{\ell} - M_1 \tilde{k}) - \theta_2}{N_2} \right), \label{eq:potential_2deff}
\end{align}
where $\mu$ is some dimensionful constant introduced in the Abelian bosonization.\footnote{In Sections~\ref{sec:analysis_gcd1} and~\ref{sec:analysis_gcd!=1}, we are going to identify the vacua by finding the minima of this potential, which corresponds to the classical analysis. Strictly speaking, the $2$d effective theory still has the renormalization-group (RG) flow, and it changes the coefficient of the effective potential. We note that \eqref{eq:potential_2deff} contains the relevant perturbation, and thus the RG flow stops at a certain energy scale. Therefore, the classical analysis in Sections~\ref{sec:analysis_gcd1} and~\ref{sec:analysis_gcd!=1} is going to be justified by a suitable rescaling of these parameters. }

\subsection{Realization of anomaly constraints in semiclassics}
\label{sec:anomaly_in_semiclassics}

Here, we shall see how the anomaly constraints are realized within our semiclassical framework. 
In this $T^2$-compactification, we have introduced the background gauge fields~\eqref{eq:baryon_tHooft} along the compactified direction to preserve $4$d anomalies in the $2$d effective theory~\cite{Tanizaki:2017qhf, Yamazaki:2017dra}.
As we have discussed in Section~\ref{sec:anomaly_BFQCD}, there are the following constraints in the $4$d setup:
\begin{itemize}
    \item Global inconsistency

    In the $(\theta_1,\theta_2)$ plane, it is impossible to connect $(\theta_1,\theta_2)$ and $(\theta_1 + 2 \pi n_1,\theta_2 + 2 \pi n_2)$ with a trivially gapped phase, unless both $n_1$ and $n_2$ are multiples of $\operatorname{gcd}(N_1,N_2)$.

    \item Mixed anomaly between $(\mathbb{Z}_{\operatorname{gcd}(N_1, N_2)})_{\chi}$ and $U(1)_V/\mathbb{Z}_{\operatorname{lcm}(N_1,N_2)} \times \mathbb{Z}_{\operatorname{gcd}(N_1, N_2)}^{[1]}$
    
    In the massless case $m = 0$, the anomaly requires the complete spontaneous breaking of discrete chiral symmetry.
\end{itemize}

Let us first discuss the anomaly for the discrete chiral symmetry $(\mathbb{Z}_{\gcd(N_1,N_2)})_{\chi}$. 
As the compact boson is related to the chiral condensate as $\tr(\overline{\Psi}\frac{1+\gamma_5}{2}\Psi)\sim \rme^{\im \varphi}$, the discrete chiral symmetry acts as the shift symmetry of $\varphi$ in the $2$d effective description. 
Indeed, the potential~\eqref{eq:potential_2deff} at the massless point $m=0$ is invariant under the following transformation, 
\begin{align}
    \begin{pmatrix}
        \varphi\\
        \tilde{k}\\
        \tilde{\ell}
    \end{pmatrix}
    \mapsto
    \begin{pmatrix}
        \varphi+\frac{2\pi}{\gcd(N_1,N_2)}\\
        \tilde{k}\\
        \tilde{\ell}-1
    \end{pmatrix}
    \mapsto \cdots 
    \mapsto
    \begin{pmatrix}
        \varphi+2\pi\\
        \tilde{k}\\
        \tilde{\ell}-\gcd(N_1,N_2)
    \end{pmatrix}
    \sim 
    \begin{pmatrix}
        \varphi\\
        \tilde{k}\\
        \tilde{\ell}
    \end{pmatrix},
    % \varphi &\mapsto \varphi + \frac{2 \pi \tilde{k}}{\operatorname{gcd}(N_1, N_2)} + 2 \pi (N_1' \tilde{k}_2  + N_2' \tilde{k}_1), \notag \\
    % k_1'' &\mapsto k_1'' + \tilde{k}_1''~~~~~   k_2'' \mapsto k_2'' + \tilde{k}_2'' 
    \label{eq:chiral_transf_semiclassics}
\end{align}
where \eqref{eq:identification_phi_k_l} is used to see that this transformation forms $\mathbb{Z}_{\gcd(N_1,N_2)}$. 
Notably, the discrete chiral transformation shifts the discrete label, $\tilde{\ell} \mapsto \tilde{\ell}-1$, so we can immediately conclude the $\gcd(N_1,N_2)$ degenerate vacua associated with the chiral symmetry breaking. 
Therefore, the requirement of $4$d 't~Hooft anomaly is correctly realized in the $2$d semiclassical framework.

Similarly, we can understand the global inconsistency constraint when $m\not =0$ from the $2$d viewpoint. 
The $2\pi$ shifts of $\theta$ angles, $(\theta_1,\theta_2)\mapsto (\theta_1+2\pi n_1, \theta_2+2\pi n_2)$, can be absorbed by the shifts of discrete labels $(k_1,k_2)\mapsto (k_1+n_1, k_2+n_2)$, which is equivalent to 
\begin{equation}
    \begin{pmatrix}
        \tilde{k}\\
        \tilde{\ell}
    \end{pmatrix}
    \mapsto 
    \begin{pmatrix}
        \tilde{k}\\
        \tilde{\ell}
    \end{pmatrix}
    + 
    \begin{pmatrix}
        N'_1 & -N'_2 \\
        M_1 & M_2
    \end{pmatrix}
    \begin{pmatrix}
        n_1\\
        n_2
    \end{pmatrix}. 
    \label{eq:theta_shifts_semiclassics}
\end{equation}
We should note that both $\tilde{k},\tilde{\ell}$ are the $\mathbb{Z}_{\gcd(N_1,N_2)}$ labels after the periodicity extension, $\varphi\sim \varphi+2\pi N'_1N'_2$. 
Since the discrete labels are non-trivially shifted unless both $n_1$ and $n_2$ are multiples of $\operatorname{gcd}(N_1,N_2)$, the unique and trivially gapped vacuum is impossible to connect $(\theta_1,\theta_2)$ and $(\theta_1+2\pi n_1, \theta_2+2\pi n_2)$ without encountering quantum phase transitions. 
This is exactly the statement of the global inconsistency in $4$d.

Lastly, we point out that these transformation properties~\eqref{eq:chiral_transf_semiclassics} and~\eqref{eq:theta_shifts_semiclassics} of $(\tilde{k},\tilde{\ell})$ turns out to be exactly the same with the ones for the label $(\tilde{k},\tilde{\ell})$ of \eqref{eq:newBasis_discreteLabels}, so it would be quite natural to identify them. Then, $(\tilde{k},\tilde{\ell})$ in the semiclassics carries the information of the $4$d SPT action completely even after the $T^2$ compactification with the baryon-'t~Hooft flux, and this is a nontrivial evidence for the adiabatic continuity.

\section{Phase diagrams of QCD(BF) for \texorpdfstring{$\operatorname{gcd}(N_1,N_2) = 1$}{gcd(N1,N2)=1}}
\label{sec:analysis_gcd1}

In this section, we explore the consequences of the semiclassical description for different-rank QCD(BF) with $\operatorname{gcd}(N_1,N_2) = 1$. In this case, the discrete labels $k_1, k_2$ (or $\tilde{k},\tilde{\ell}$) are completely absorbed by the periodicity extension, $\varphi\sim \varphi+2\pi N_1 N_2$, and the potential~\eqref{eq:potential_2deff} becomes 
\begin{align}
    V[\varphi] &= - m \mu \cos (\varphi) - 2K^{(1)} \rme^{-S_\mathrm{I}^{(1)}/N_1} \cos \left( \frac{N_2 \varphi - \theta_1}{N_1} \right) - 2K^{(2)} \rme^{-S_\mathrm{I}^{(2)}/N_2} \cos \left( \frac{N_1 \varphi - \theta_2}{N_2} \right). 
    \label{eq:potential_GCD1_BFQCD}
\end{align}
In Section~\ref{sec:phase_diagrams_gcd1}, we study the phase diagrams in the large-mass and small-mass limits by finding the minima of this effective potential, and we also provide a numerical demonstration of how these two limits are connected via the topology change. 
In Section~\ref{sec:duality_gcd1}, we establish the duality conjecture by Karasik and Komargodski~\cite{Karasik:2019bxn} in the hierarchical and large-$N$ limits within our semiclassical effective theory. Moreover, by constructing the explicit duality map as suggested by the combination of semiclassics and the large-$N$ limit, we numerically justify its validity even for finite parameters. 

% \subsection{The large-mass and massless limits and the topology change}
\subsection{Phase diagrams for \texorpdfstring{$\gcd(N_1,N_2)=1$}{gcd(N1,N2)=1} with semiclassics}
\label{sec:phase_diagrams_gcd1}

We can analytically study the small-mass and large-mass limits $m \rightarrow +0,+\infty$. 
Let us consider the large-mass limit first, and then go to the small-mass limit.
We shall see that the semiclassical description supports the natural scenarios for both the small-mass and large-mass limits.
We then numerically explore how the intermediate-mass range connects these different limits and observe the topology change of the $(\theta_1,\theta_2)$ phase diagram. 

\begin{figure}[t]
\centering
\begin{minipage}{.48 \textwidth}
\includegraphics[width = 0.9\textwidth]{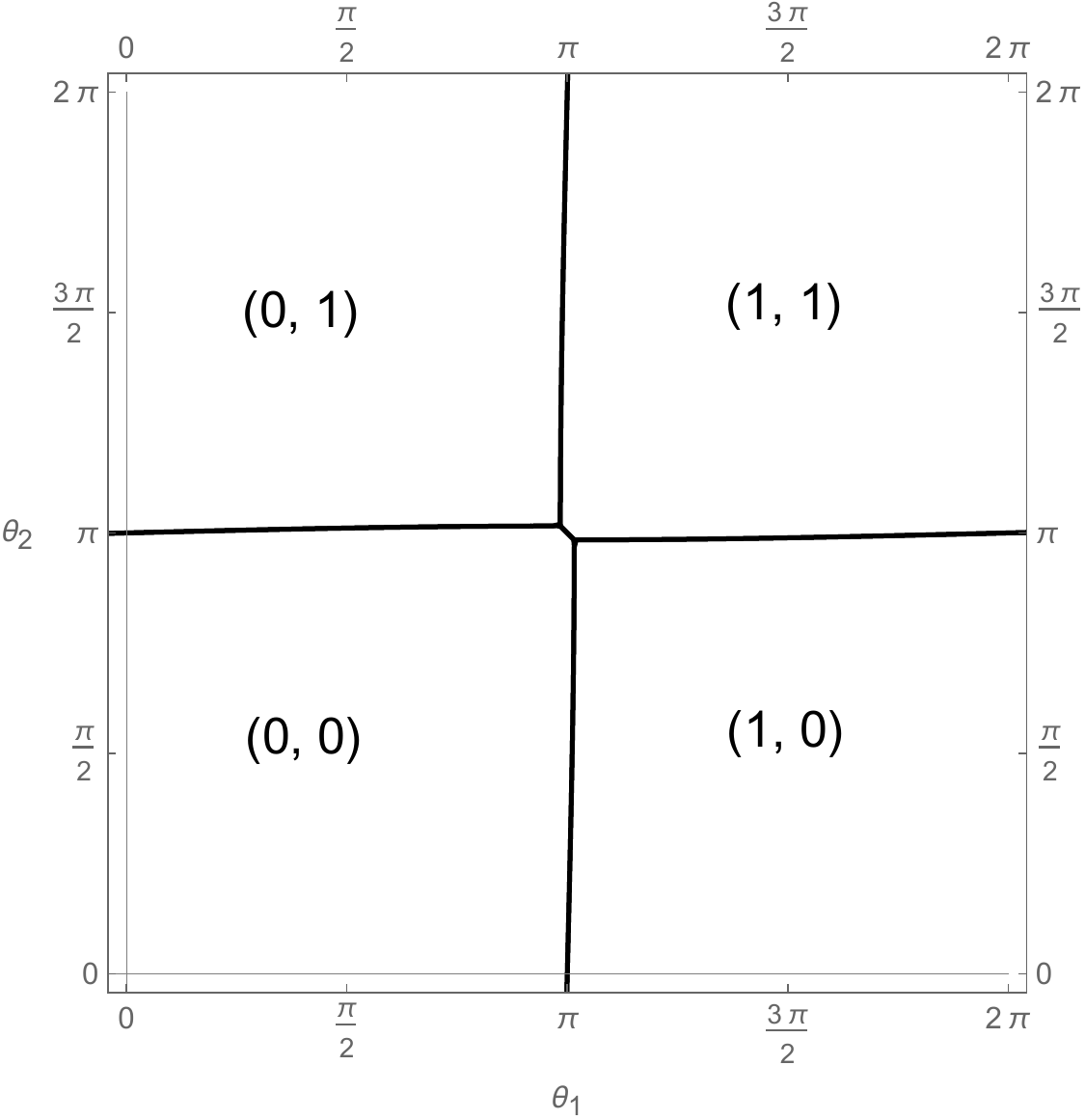}
\label{fig:conjectured_massive}
\end{minipage}\quad
\begin{minipage}{.48 \textwidth}
\includegraphics[width = 0.9\textwidth]{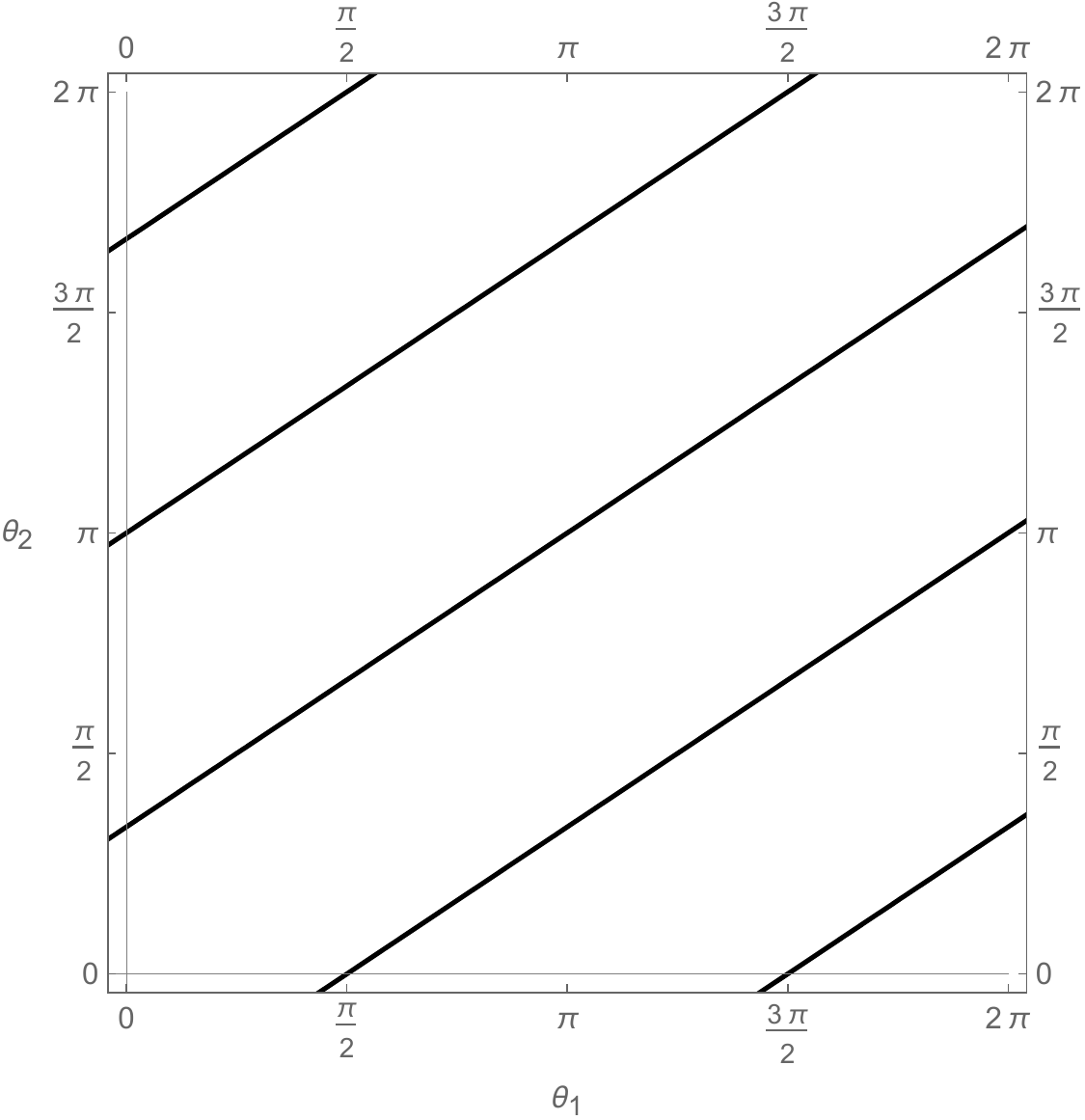}
\label{fig:conjectured_massless}
\end{minipage}
\caption{Two limits of phase diagrams of QCD(BF) on the $(\theta_1, \theta_2)$ plane.
The plots are made only on the fundamental domain $0<\theta_1 < 2\pi$ and $0<\theta_2 < 2\pi$. 
(\textbf{Left panel})~The phase diagram for the large-mass limit $\Lambda_1, \Lambda_2 \ll m (\ll 1/(NL))$. The labels show the discrete vacuum label $(k_1, k_2) \in \mathbb{Z}_{N_1} \times \mathbb{Z}_{N_2}$, which is related to the $2\pi N_1 N_2$-periodic scalar field as $\varphi=2\pi(M_1 k_1+M_2 k_2)$. 
(\textbf{Right panel})~The phase diagram for the massless or small-mass limit, and this figure is drawn for the case $N_1 = 2,~N_2 = 3$. The phase transition lines are given by $\theta_{\mathrm{phys}}=N_1\theta_1-N_2\theta_2=\pi$ (mod $2\pi$) in the massless limit. Since the discrete chiral symmetry is absent for $\gcd(N_1,N_2)=1$, the small mass perturbation does not introduce new phase transition lines unlike the case of $\gcd(N_1,N_2)\not=1$. }
\label{fig:two_limits}
\end{figure}

% \begin{itemize}
%     \item Large-mass limit $m \rightarrow +\infty$

\subsubsection{Large-mass limit \texorpdfstring{$m\to \infty$}{m->infty}}
\label{sec:large-mass_gcd=1}

In the limit $m\to \infty$, we first need to minimize $-m \mu \cos(\varphi)$ in \eqref{eq:potential_GCD1_BFQCD}, and the candidates of minima are given by $\varphi=2\pi \hat{\ell}$ with $\hat{\ell}\sim \hat{\ell}+N_1N_2$ at order $O(m)$. 
    % $\varphi \in 2 \pi \mathbb{Z}$ at order $O(m)$, and we can write $\varphi = 2 \pi (N_1\ell_2 + N_2\ell_1)~~(\operatorname{mod}N_1N_2)$. 
We can decompose $\hat{\ell}=M_1 k_1+M_2 k_2$ with $k_1\sim k_1+N_1$ and $k_2\sim k_2+N_2$ as $\gcd(N_1,N_2)=1$. Since $N_2 M_1+N_1 M_2=1$ and $M_1/N_2, M_2/N_1\in \mathbb{Z}$, the potential~\eqref{eq:potential_GCD1_BFQCD} at $O(m^0)$ becomes
\begin{align}
    &V[\varphi = 2 \pi (M_1k_1 + M_2 k_2)] \notag\\
    &= - 2K^{(1)} \rme^{-S_\mathrm{I}^{(1)}/N_1} \cos \left( \frac{2 \pi N_2 M_1 k_1 - \theta_1}{N_1} \right)
    - 2K^{(2)} \rme^{-S_\mathrm{I}^{(2)}/N_2} \cos \left( \frac{2 \pi N_1 M_2 k_2 - \theta_2}{N_2} \right) \notag\\
    &=- 2K^{(1)} \rme^{-S_\mathrm{I}^{(1)}/N_1} \cos \left( \frac{2 \pi k_1 - \theta_1}{N_1} \right) 
    - 2K^{(2)} \rme^{-S_\mathrm{I}^{(2)}/N_2} \cos \left( \frac{2 \pi k_2 - \theta_2}{N_2} \right).
\end{align}
This reproduces the phase diagram of the $SU(N_1)\times SU(N_2)$ pure Yang-Mills theory, where the phase boundaries are located on $\theta_1 = \pi ~(\operatorname{mod}2\pi)$ and $\theta_2 = \pi ~(\operatorname{mod}2\pi)$, and $(k_1,k_2)\in \mathbb{Z}_{N_1}\times \mathbb{Z}_{N_2}$ describes the discrete vacuum label (see the left panel of Figure~\ref{fig:two_limits}).

In the leading order, there are four-fold degenerate vacua at $(\theta_1, \theta_2) = (\pi,\pi)$, but this degeneracy is partially resolved by including the next-to-leading-order corrections.
To see this, we add the small fluctuation $\varphi = 2 \pi (M_1 k_1 + M_2 k_2) + \delta \varphi~~(\operatorname{mod}~2\pi N_1N_2) $, where $\delta \varphi$ is supposed to be the $O(m^{-1})$ quantity. 
The order $O(m^{-1})$ corrections to the potential can be evaluated as follows:
\begin{align}
    V& [2 \pi (M_1k_1 + M_2 k_2) + \delta \varphi] - V[2 \pi (M_1k_1 + M_2 k_2)] \notag \\
%    &\simeq \frac{m \mu }{2} (\delta \varphi)^2  - 2K^{(1)} \rme^{-S_\mathrm{I}^{(1)}/N_1} \cos \left( \frac{2 \pi k_1 - \theta_1}{N_1} + \frac{N_2}{N_1} \delta \varphi \right) \notag \\
%    &~~ - 2K^{(2)} \rme^{-S_\mathrm{I}^{(2)}/N_2} \cos \left( \frac{2 \pi k_2 - \theta_2}{N_2} + \frac{N_1}{N_2} \delta \varphi \right)\notag\\
    &\simeq \frac{m \mu }{2} (\delta \varphi)^2  + \left[ \frac{2K^{(1)} \rme^{-S_\mathrm{I}^{(1)}/N_1}}{N_1/N_2} \sin \left( \frac{2 \pi k_1 - \theta_1}{N_1} \right) + \frac{2K^{(2)} \rme^{-S_\mathrm{I}^{(2)}/N_2}}{N_2/N_1} \sin \left( \frac{2 \pi k_2 - \theta_2}{N_2}  \right) \right] \delta \varphi \notag\\
    &= \frac{m \mu }{2}\left(\delta \varphi+O(m^{-1})\right)^2 \notag\\
    &- \frac{1}{2 m \mu} \left[ \frac{2K^{(1)} \rme^{-S_\mathrm{I}^{(1)}/N_1}}{N_1/N_2} \sin \left( \frac{2 \pi k_1 - \theta_1}{N_1} \right) + \frac{2K^{(2)} \rme^{-S_\mathrm{I}^{(2)}/N_2}}{N_2/N_1} \sin \left( \frac{2 \pi k_2 - \theta_2}{N_2}  \right) \right]^2.
\end{align}
Therefore, the four-fold degeneracy at $(\theta_1, \theta_2) = (\pi,\pi)$ is lifted to the two-fold degeneracy, and the lowest ground states $\varphi$ are those of $(k_1, k_2) = (0,0), (1,1)$. 
The phase boundary has a negative slope (Left panel of Figure~\ref{fig:two_limits}), as also expected in the large-$N$ argument.
This two-fold degeneracy at $(\theta_1,\theta_2)=(\pi,\pi)$ is protected by the CP symmetry, which interchanges $(k_1, k_2) = (0,0)$ and $(k_1, k_2) = (1,1)$.

\subsubsection{Massless limit \texorpdfstring{$m\to 0$}{m->0} and small-mass perturbation}
    % \item Massless limit
Next, let us consider the massless limit, $m=0$. 
For $m=0$, the potential~\eqref{eq:potential_GCD1_BFQCD} depends only on $\theta_{\mathrm{phys}}=N_1\theta_1-N_2\theta_2$, and we can see it explicitly by shifting $\varphi$ to $\varphi+\theta_2/N_1$:
\begin{align}
    V\left[\varphi+\frac{\theta_2}{N_1}\right] 
    &=  - 2K^{(1)} \rme^{-S_\mathrm{I}^{(1)}/N_1} \cos \left( \frac{N_2 \varphi + \frac{N_2}{N_1}\theta_2 - \theta_1}{N_1} \right) - 2K^{(2)} \rme^{-S_\mathrm{I}^{(2)}/N_2} \cos \left( \frac{N_1 \varphi}{N_2} \right)\notag\\
    &=- 2K^{(1)} \rme^{-S_\mathrm{I}^{(1)}/N_1} \cos \left( \frac{N_2 \varphi - \frac{\theta_{\mathrm{phys}}}{N_1}}{N_1} \right) - 2K^{(2)} \rme^{-S_\mathrm{I}^{(2)}/N_2} \cos \left( \frac{N_1 \varphi}{N_2} \right). 
    \label{eq:potential_GCD1_massless}
\end{align}
This potential exhibits a unique vacuum for $\varphi\sim \varphi+2\pi N_1 N_2$ at generic values of $\theta_{\mathrm{phys}}$, and a two-fold degeneracy occurs when $\theta_{\mathrm{phys}}=\pi~(\operatorname{mod} 2 \pi)$.\footnote{Here, we would like to note that this phase transition at $\theta_{\mathrm{phys}}=\pi$ is ``not'' required by any kind of anomaly, as the anomaly and global inconsistency are absent for $\operatorname{gcd}(N_1,N_2) = 1$.
Using the center-vortex vertices $\rme^{-\im \frac{N_2 \varphi - \theta_1}{N_1}}$ and $\rme^{-\im \frac{N_1 \varphi - \theta_2}{N_2}}$, we can construct an operator
$\left( \rme^{-\im \frac{N_2 \varphi - \theta_1}{N_1}} \right)^{M_1/N_2} \left( \rme^{-\im \frac{N_1 \varphi - \theta_2}{N_2}} \right)^{M_2/N_1}=\rme^{-\frac{\im}{N_1N_2}(\varphi-(M_1\theta_1+M_2\theta_2)}$. 
By shifting $\varphi\to \varphi+\frac{\theta_2}{N_1}$, this operator can be regarded as the following symmetric perturbation, 
$\cos \left( \frac{\varphi - \frac{M_1}{N_1}\theta_{\mathrm{phys}}}{N_1 N_2} \right)$, to the potential. When this perturbation is added and dominates $V[\varphi]$, the ground state is always unique and behaves continuously as a function of $\theta_{\mathrm{phys}}$. }
We note that the discrete chiral symmetry is absent for $\gcd(N_1,N_2)=1$, so the unique vacuum state is allowed.  
The two-fold degeneracy of $\theta_{\mathrm{phys}}=\pi$ comes out of the $2\pi$ periodicity of $\theta_{\mathrm{phys}}$, which follows from the equivalence, $\theta_{\mathrm{phys}}\to \theta_{\mathrm{phys}}+2\pi~\Longleftrightarrow~(\theta_1, \theta_2)\to (\theta_1+2\pi M_2, \theta_2-2\pi M_1)$. 
Indeed, the above potential is invariant under $\theta_{\mathrm{phys}}\to \theta_{\mathrm{phys}}+2\pi$ associated with $\varphi\to \varphi+2\pi M_1/N_1$. 
As an example, the massless phase diagram with $N_1 = 2$ and $N_2 = 3$ is plotted in the right panel of Figure~\ref{fig:two_limits}.
We note that the small mass perturbation does not introduce new transition lines when $\gcd(N_1,N_2)=1$, unlike Figure~\ref{fig:BFQCDsmallmass}. 

Let us add some details to understand the structure of the massless phase diagram. 
First, we can readily confirm that the above potential at $\theta_{\mathrm{phys}}=2\pi n$ has the unique vacuum at $\varphi_{*(n)}=\frac{2\pi M_1}{N_1}n~(\bmod~2\pi N_1N_2)$. 
When we change the theta parameter gradually as $\theta_{\mathrm{phys}}=2\pi n+\delta \theta_{\mathrm{phys}}$ with $-\pi<\delta\theta_{\mathrm{phys}}<\pi$, the vacuum configuration $\varphi_{*(n)}$ also changes continuously and it lives inside 
\begin{equation}
    \operatorname{min}\left(\frac{2\pi M_1}{N_1}n, \frac{2\pi M_1}{N_1}n+\frac{\delta \theta_{\mathrm{phys}}}{N_1N_2}\right)
    <\varphi_{*(n)}<
    \operatorname{max}\left(\frac{2\pi M_1}{N_1}n, \frac{2\pi M_1}{N_1}n+\frac{\delta \theta_{\mathrm{phys}}}{N_1N_2}\right).
\end{equation}
This is because $\varphi=\frac{2\pi M_1}{N_1}n+\frac{\delta \theta_{\mathrm{phys}}}{N_1N_2}$ minimizes the first term of \eqref{eq:potential_GCD1_massless} and $\varphi=\frac{2\pi M_1}{N_1}n$ minimizes the second one, and the true minimum should be in between them.
For $\theta_{\mathrm{phys}}=\pi(\,=2\pi-\pi)$, there are two distinct configurations, $\varphi_{*(0)}$ and $\varphi_{*(1)}$, giving the same energy: The fact $\varphi_{*(0)}\not = \varphi_{*(1)}$ follows from the observation that they live in non-overlapping regions, 
\begin{equation}
    0<\varphi_{*(0)}<\frac{\pi}{N_1N_2},\quad \frac{2\pi M_1}{N_1}-\frac{\pi}{N_1N_2}<\varphi_{*(1)}<\frac{2\pi M_1}{N_1}. 
\end{equation}
Therefore, the phase transition occurs at $\theta_{\mathrm{phys}} = \pi$ for the semiclassical potential~\eqref{eq:potential_GCD1_massless}.

\subsubsection{Topology change of the phase diagrams as a function of \texorpdfstring{$m$}{fermion mass}}
\label{sec:numerical_phase_diagrams_gcd1}

\begin{figure}[t]
\centering
  \begin{minipage}[b]{0.4\linewidth}
    \centering \subfloat[$m \mu = 0.05$]{
\includegraphics[scale=0.3]{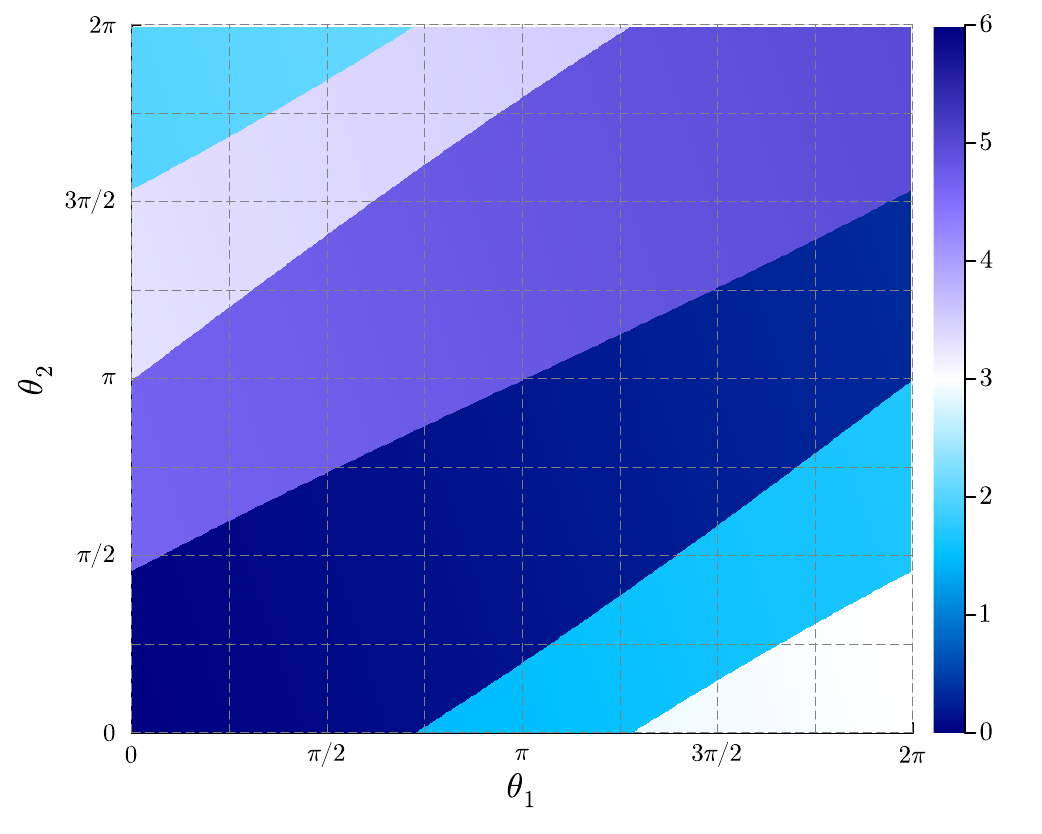}
  \label{fig:msq005} 
}
    \vspace{2ex}
  \end{minipage}%%
  \begin{minipage}[b]{0.4\linewidth}
    \centering
    \subfloat[$m \mu = 0.2$]{
\includegraphics[scale=0.3]{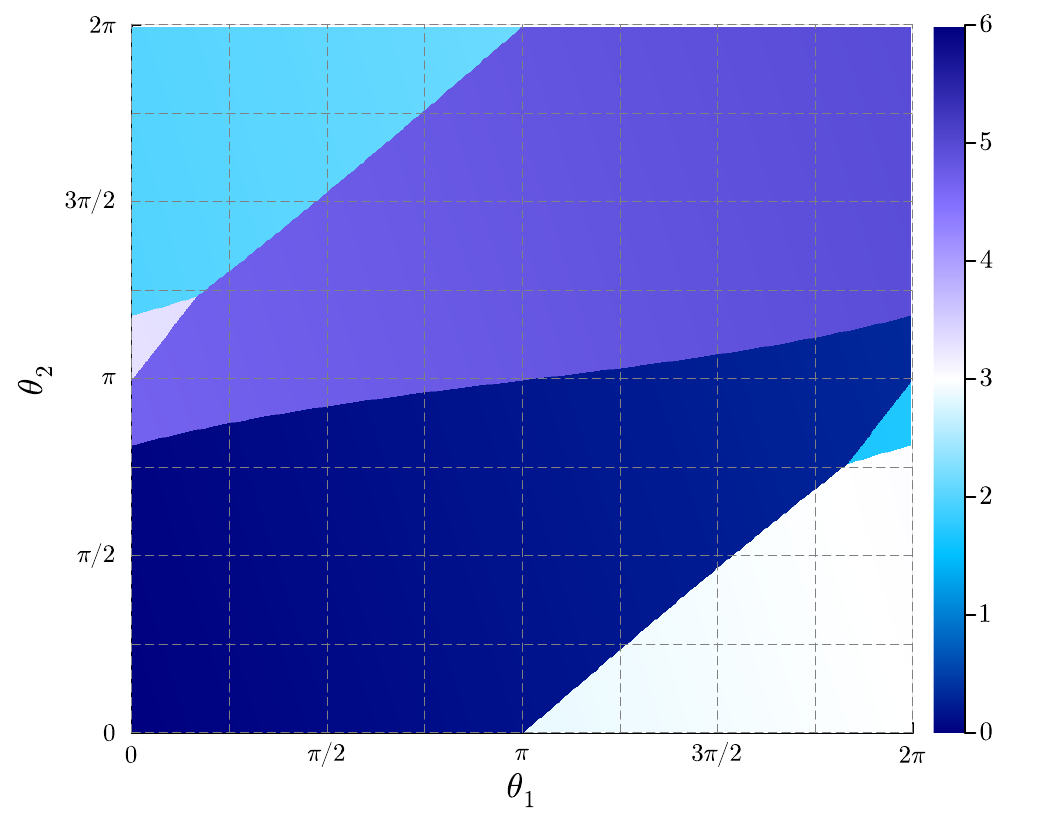}
  \label{fig:msq02} 
}
    \vspace{2ex}
  \end{minipage}
  \\
  \begin{minipage}[b]{0.4\linewidth}
    \centering
    \subfloat[$m \mu = 0.3$]{
\includegraphics[scale=0.3]{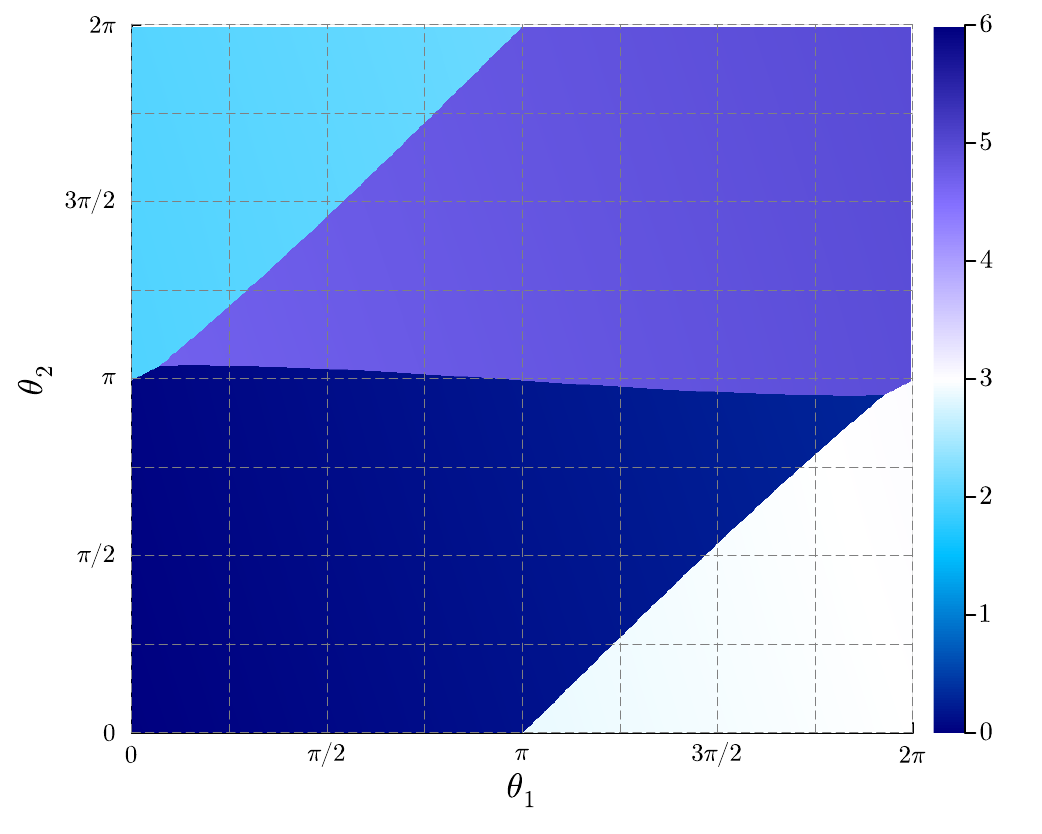}
  \label{fig:msq03} 
}
    \vspace{0ex}
  \end{minipage}%% 
  \begin{minipage}[b]{0.4\linewidth}
    \centering
    \subfloat[$m \mu = 0.5$]{
\includegraphics[scale=0.3]{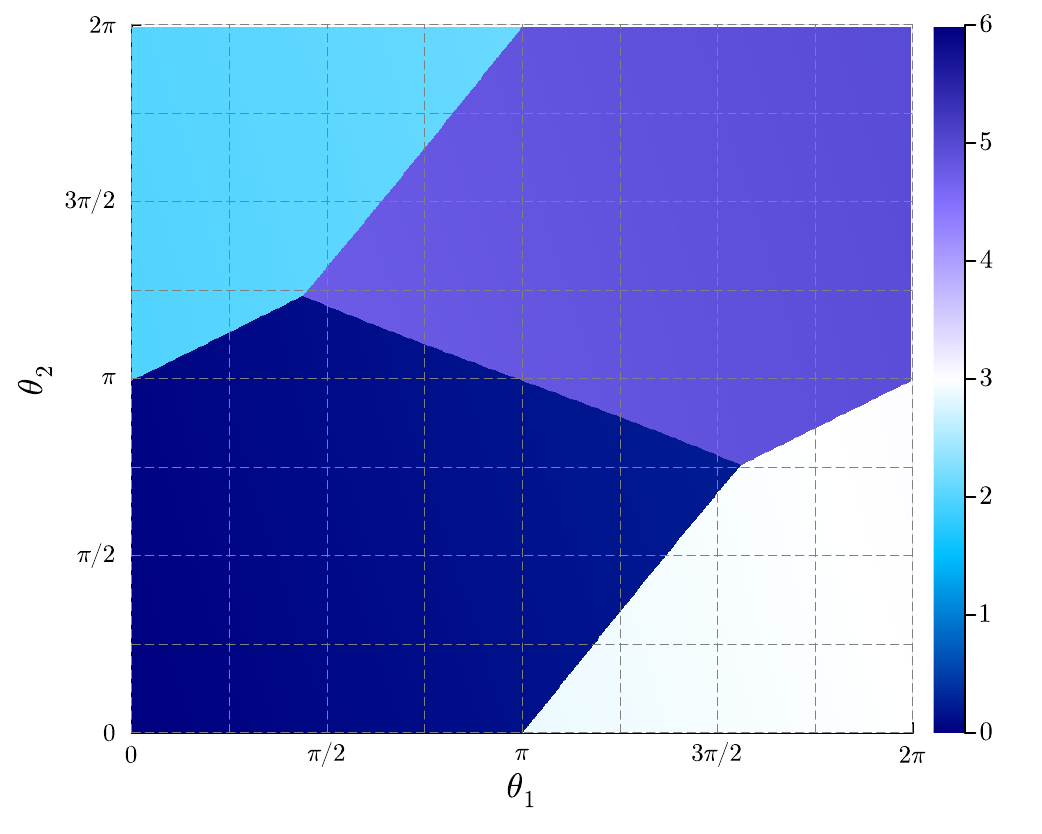}
\label{fig:msq05}
}
    \vspace{0ex}
  \end{minipage} 
  \caption{
  Phase diagrams of $(N_1,N_2) = (2,3)$ at several $m \mu$ when we set $2K^{(1)} \rme^{-S_\mathrm{I}^{(1)}/N_1} = 2K^{(2)} \rme^{-S_\mathrm{I}^{(2)}/N_2} = 1$. 
  The values of $\frac{\varphi}{2\pi}\in [0,N_1N_2)$ at the global minimum of $V[\varphi]$ are depicted with a color gradient.
  As the fermion mass is varied, we can observe the topology change of the phase diagram in the $(\theta_1,\theta_2)$ space as conjectured in the previous subsection. 
  }
  \label{fig:app-some-examples} 
\end{figure}

We have observed that the large-mass and small-mass limits yield strikingly different phase diagrams in the $(\theta_1,\theta_2)$ space, and even their topological structures are distinct. 
It is quite nontrivial how the intermediate-mass range bridges these two limits. 
To obtain an overall picture, we numerically draw phase diagrams by varying the fermion mass $m\mu$ in the following setup; $N_1 = 2,~N_2 = 3$, with $2K^{(1)} \rme^{-S_\mathrm{I}^{(1)}/N_1} = 2K^{(2)} \rme^{-S_\mathrm{I}^{(2)}/N_2}(\,=1)$. We choose the Bezout coefficient as $(M_1,M_2)=(3,-4)$ so that $M_1N_2+M_2N_1=1$ and $M_1/N_2, M_2/N_1\in \mathbb{Z}$. 
The results of the phase diagrams at several $m\mu$ are shown in Figure~\ref{fig:app-some-examples}. 

To illustrate the phase diagram, we plot the values of the global minimum $\varphi$ of the potential $V[\varphi]$ given in (\ref{eq:potential_GCD1_BFQCD}).
In this figure, the minimum values $\varphi/2\pi \in [0,N_1N_2)=[0,6)$ are illustrated through a color gradient.
This allows us to clearly observe the phase transition, which is indicated by a sudden change in color.
Let us explain how the phase diagram changes as we increase the mass parameter $m\mu$. 

\begin{enumerate}
    \item The phase diagram at $m\mu=0.05$ is shown in Figure~\ref{fig:msq005}. When the mass is sufficiently small, the phase diagram is qualitatively similar to that of the massless limit (right panel of Figure~\ref{fig:two_limits}). Moreover, the jump of the $\varphi$ values in neighboring phases is roughly given by $\frac{\Delta\varphi}{2\pi}\simeq \frac{M_1}{N_1}=\frac{3}{2}$, which is consistent with our analysis of the massless limit. 
    We note that phase boundaries are no longer parallel for nonzero mass. 

    \item As we increase the mass, some of the non-parallel phase boundaries start to merge, and triple points (three-fold degenerate vacua) appear (see Figure~\ref{fig:msq02} for the phase diagram at $m \mu = 0.2$). This is also a type of topology change, but it is a different one from what we have discussed in Section~\ref{sec:KK_duality}. 

    \item As we increase the mass further, some phases are pushed out completely from the region $0<\theta_1 < 2\pi$ and $0<\theta_2 < 2\pi$.
    During this process, the ``topology change'' discussed in Section~\ref{sec:KK_duality} happens in the vicinity of $(\theta_1,\theta_2)=(0, \pi)$.
    Figure~\ref{fig:msq03} depicts the phase diagram at $m \mu = 0.3$, slightly above the topology-changing mass.
    
    \item After that, the phase diagram converges to the one in the large-mass limit. The phase diagram at $m\mu=0.4$ is shown in Figure~\ref{fig:msq05}, and it is already qualitatively the same as the left panel of Figure~\ref{fig:two_limits}. Also quantitatively, the vacuum values of $\varphi$ are approximately given by $\frac{\varphi_{(k_1,k_2)}}{2\pi}=3k_1-4k_2$, which is consistent with our large-mass analysis. 
\end{enumerate}

\subsection{On the duality cascade conjecture}
\label{sec:duality_gcd1}

As reviewed in Section~\ref{sec:KK_duality}, Karasik and Komargodski~\cite{Karasik:2019bxn} proposed the infrared dualities between two QCD(BF)s in the large-$N_{1,2}$ limits: The duality claims that the following two theories,
\begin{align}
    SU(N_1)&\times SU(N_2) \mathrm{~QCD(BF)~on~} (\theta_1,\theta_2)~\mathrm{for}~m<m^* \notag \\
    &\longleftrightarrow~  SU(N_1)\times SU(N_2-N_1) ~\mathrm{QCD(BF)~on~} (\theta_1-\theta_2,\theta_2),  \tag{\ref{eq:KK_duality_conjecture}}
\end{align}
have the same vacuum structures in the sense of congruence in both phase diagrams and meta-stable vacua. 
This is called the duality cascade, as we can repeatedly apply these dualities when the fermion mass is sufficiently small. 
The purpose of this subsection is to understand this duality within the semiclassical framework.

We first discuss two distinct limits, at which the above duality can be analytically shown within our semiclassical framework: 
\begin{itemize}
    \item The hierarchical limit $m,\Lambda_1\ll \Lambda_2$ (Section~\ref{sec:duality_hierarchical}), and 
    \item The large-${N_{1,2}}$ limit (Section~\ref{sec:recovering_large_N_from_semiclassics}). 
\end{itemize}
Away from these limits, it is nontrivial to what extent the duality is effective. 
Therefore, we construct the explicit duality map by combining the semiclassics and the large-$N$ analysis, and 
we investigate the phase diagrams numerically using the corresponding parameters for $(N_1,N_2)=(2,3)$ and $(2,5)$ in Section~\ref{sec:duality_numerics}. 
We find that the duality map works quite effectively, and it only has a very tiny quantitative violation. 
Hence, the proposed duality is useful to understand the topology-changing structure of the phase diagram even when neither of the above limits are not strictly taken.

Again, we note that the asymptotic freedom of the $SU(N_1)\times SU(N_2-N_1)$ QCD(BF) imposes a constraint $N_1 < \frac{11}{13}N_2$ for the duality (\ref{eq:KK_duality_conjecture}).
Similarly, because of the adiabatic continuity conjecture, we restrict our scope to cases where both parent and daughter theories have the confining QCD-like vacuum, e.g., we assume that the theory is not in the conformal window. 
This also imposes a further constraint on the possible numbers of colors $(N_1,N_2)$.

\subsubsection{Duality in the hierarchical limit \texorpdfstring{$m, \Lambda_1 \ll \Lambda_2$}{m,Lambda1<<Lambda2}}
\label{sec:duality_hierarchical}

In this subsection, let us consider the limit where the strong scale $\Lambda_2$ of $SU(N_2)$ and of $SU(N_2-N_1)$ is much larger than other scales in both theories:  $m, \Lambda_1 \ll \Lambda_2$. 
As we have chosen $N_2>N_1$, we should discriminate $\Lambda_1\ll \Lambda_2$ and $\Lambda_1\gg \Lambda_2$, and we consider the first case. 

If we put an extra assumption, $N_2-N_1\gtrsim N_1$, 
the duality in this hierarchical limit can be shown directly in $\mathbb{R}^4$ by using the chiral Lagrangian. Let us have a quick look at it before working on the semiclassics. 
When $\Lambda_1\ll \Lambda_2$, the $SU(N_1)$ gauge field can be regarded as a background when discussing the dynamics of the $SU(N_2)$ or $SU(N_2-N_1)$ gauge field. 
Then, the theory enjoys the approximate $SU(N_1)_L\times SU(N_1)_R$ chiral symmetry as $m\ll \Lambda_2$, and it is likely to be spontaneously broken in both sides of the dualities as $N_2>N_1$ and we have also assumed $N_2-N_1\gtrsim N_1$. 
Then, the low-energy effective theory is described by the $SU(N_1)$-valued scalar field $U$ coupled to the $SU(N_1)$ gauge field $a_1$, and the $SU(N_1)$ gauge transformation is given by $U\to gUg^{-1}$ and $a_1\to g a_1 g^{-1}-\im g\diff g^{-1}$. 
Let us now obtain the effective action for $SU(N_1)\times SU(N_2)$ QCD(BF) at $(\theta_1,\theta_2)$. To incorporate the information of $\theta_2$ in the chiral Lagrangian, we perform the anomalous chiral transformation to put it into the phase of the fermion mass, $m_1\to m_1 \rme^{-\im \theta_2/N_1}$. This anomalous transformation also affects $\theta_1$ as 
\begin{equation}
    \theta_1\to \theta_1-\frac{N_2}{N_1}\theta_2=\frac{\theta_{\mathrm{phys}}}{N_1}, 
\end{equation}
and we obtain the following effective Lagrangian, 
\begin{align}
    S_{\mathrm{chiral}} \sim \int ~|D U|^2 - (m \rme^{- \im\theta_2/N_1} \operatorname{tr} U + \mathrm{c.c.})  + \frac{1}{g_1^2} |f_1|^2 + \frac{\im \theta_{\mathrm{phys}}/N_1}{8 \pi^2} \operatorname{tr} f_1 \wedge f_1.
    \label{eq:chiral_lagrangian_hierarchy}
\end{align}
When repeating the same discussion for $SU(N_1)\times SU(N_2-N_1)$ QCD(BF) at $(\theta_1',\theta_2)$, all the discussion is parallel except that the anomalous transformation on $\theta_1'$ is given by $\theta_1'\to \theta_1'-\frac{N_2-N_1}{N_1}\theta_2$. 
Therefore, we can find the same low-energy $\theta$ angle for $SU(N_1)$ gauge group by setting $\theta_1'=\theta_1-\theta_2$, and this suggests
\begin{align}
    SU(N_1)&\times SU(N_2) \mathrm{~QCD(BF)~on~} (\theta_1,\theta_2) \notag \\
    &\longleftrightarrow ~ SU(N_1)\times SU(N_2-N_1) ~\mathrm{QCD(BF)~on~} (\theta_1-\theta_2,\theta_2) \notag 
\end{align}
seems to hold in the limit $m, \Lambda_1 \ll \Lambda_2$.
However, this analysis with the chiral Lagrangian~\eqref{eq:chiral_lagrangian_hierarchy} is actually \textit{incomplete}.
For example, this chiral Lagrangian at $m=0$ has the apparent $(\mathbb{Z}_{N_1})_{\mathrm{chiral}}$ symmetry, $U \rightarrow \rme^{-\frac{2 \pi \im}{N_1}} U$, while the correct chiral symmetry should be $\mathbb{Z}_{\gcd(N_1,N_2)}$. 
Such a subtle $N_2$ dependence is absent in the usual chiral Lagrangian, and we must incorporate the $\eta'$ field with the extended periodicity, $\eta'\sim \eta'+2\pi N_2$~\cite{Hayashi:2024qkm}. 
Let us see how our semiclassical description takes care of this subtlety.

For semiclassical parameters, the hierarchical limit corresponds to $m, K^{(1)}\rme^{-S^{(1)}_{\mathrm{I}}/N_1}\ll K^{(2)}\rme^{-S^{(2)}_{\mathrm{I}}/N_2}$, and thus the semiclassical vacua at the leading order minimizes the last term of \eqref{eq:potential_GCD1_BFQCD}. 
On the $SU(N_1) \times SU(N_2) $ QCD(BF) side, this limit reduces the $2 \pi N_1 N_2$-periodic scalar $\varphi$ to the discrete one
\begin{align}
    \varphi = \frac{\theta_2 + 2 \pi N_2 \ell}{N_1},
\end{align}
with $\ell \in \mathbb{Z}_{N_1^2}$.
Then, there are $N_1^2$ candidates of vacua, and the energy of the $\ell$-th vacuum at the subleading order $O(\Lambda_2^0)$ becomes
\begin{align}
    V_\ell &= V \left[\varphi= \frac{\theta_2 + 2 \pi N_2 \ell}{N_1} \right] \notag \\
    &= - m \mu \cos \left(\frac{\theta_2 + 2 \pi N_2 \ell}{N_1} \right) - 2K^{(1)} \rme^{-S_\mathrm{I}^{(1)}/N_1} \cos \left( \frac{2 \pi N_2^2 \ell - \theta_{\mathrm{phys}}}{N_1^2} \right).
\end{align}
Let us now discuss the dual side. For the $SU(N_1) \times SU(N_2 - N_1) $ QCD(BF), the same limit limit reduces the $2\pi N_1(N_2-N_1)$-periodic field $\varphi$ to the discrete one, 
\begin{align}
    \varphi = \frac{\theta_2' + 2 \pi (N_2 - N_1) \ell'}{N_1},
\end{align}
with $\ell' \in \mathbb{Z}_{N_1^2}$. 
We write the theta angles in the dual theories by $(\theta_1', \theta_2')$, which are going to be set $(\theta_1'=\theta_1 - \theta_2, \theta_2' = \theta_2)$ according to the conjecture.
The energy of the $\ell'$-th vacuum at the subleading order is,
\begin{align}
    V_{\ell'}' &= - m \mu \cos \left(\frac{\theta_2' + 2 \pi (N_2-N_1) \ell'}{N_1} \right) - 2K^{(1)} \rme^{-S_\mathrm{I}^{(1)}/N_1} \cos \left( \frac{2 \pi (N_2-N_1)^2 \ell' - \theta_{\mathrm{phys}}}{N_1^2} \right) \notag \\
    &= - m \mu \cos \left(\frac{\theta_2 + 2 \pi (N_2-N_1) \ell'}{N_1} \right) - 2K^{(1)} \rme^{-S_\mathrm{I}^{(1)}/N_1} \cos \left( \frac{2 \pi (N_2-N_1)^2 \ell' - \theta_{\mathrm{phys}}}{N_1^2} \right).
\end{align}

Here, let us consider the following one-to-one correspondence between $\mathbb{Z}_{N_1^2}$ labels:
\begin{align}
    (N_2-N_1)^2 \ell' = N_2^2 \ell.
\end{align}
This is one-to-one because $\operatorname{gcd}(N_1^2, (N_2-N_1)^2) = \operatorname{gcd}(N_1^2, N_2^2) = 1$.
Here we note that the Bezout coefficient $M_1$ satisfies both $N_2 M_1 = 1 ~(\operatorname{mod} N_1)$ and $(N_2-N_1) M_1 = 1 ~(\operatorname{mod} N_1)$.
With the identification ($(N_2-N_1)^2 \ell' = N_2^2 \ell$), the vacuum energy matches as follows,
\begin{align}
    V_{\ell'}' &= - m \mu \cos \left(\frac{\theta_2 + 2 \pi M_2 (N_2-N_1)^2 \ell'}{N_1} \right) - 2K^{(1)} \rme^{-S_\mathrm{I}^{(1)}/N_1} \cos \left( \frac{2 \pi (N_2-N_1)^2 \ell' - \theta_{\mathrm{phys}}}{N_1^2} \right) \notag \\
    &= - m \mu \cos \left(\frac{\theta_2 + 2 \pi M_2 N_2^2 \ell}{N_1} \right) - 2K^{(1)} \rme^{-S_\mathrm{I}^{(1)}/N_1} \cos \left( \frac{2 \pi N_2^2 \ell - \theta_{\mathrm{phys}}}{N_1^2} \right) = V_{\ell}.
\end{align}
This establishes the duality (\ref{eq:KK_duality_conjecture}) in the hierarchical limit $m, \Lambda_1 \ll \Lambda_2$: there is an exact one-to-one correspondence between vacua of the two theories.
The generalization to the cases of $\operatorname{gcd}(N_1,N_2) \neq 1$ is straightforward.

\subsubsection{Large-\texorpdfstring{$N_{1,2}$}{N1,2} limit and the duality in semiclassics}
\label{sec:recovering_large_N_from_semiclassics}

In this section, we study the case when $N_1,N_2\gg 1$ with our semiclassical effective theory. 
We have to note that the perturbative mass gap for gluons is given by $O(1/N_i L)$, and our semiclassical effective theory is derived with the assumption that this perturbative gap is well separated from the strong scale, $N_i L \Lambda_i \ll 1$. 
Therefore, we have to make sure that the torus size $L$ is sufficiently small to satisfy this criterion especially when $N_i\gg 1$.

\begin{figure}[t]
\centering
\includegraphics[scale=0.52]{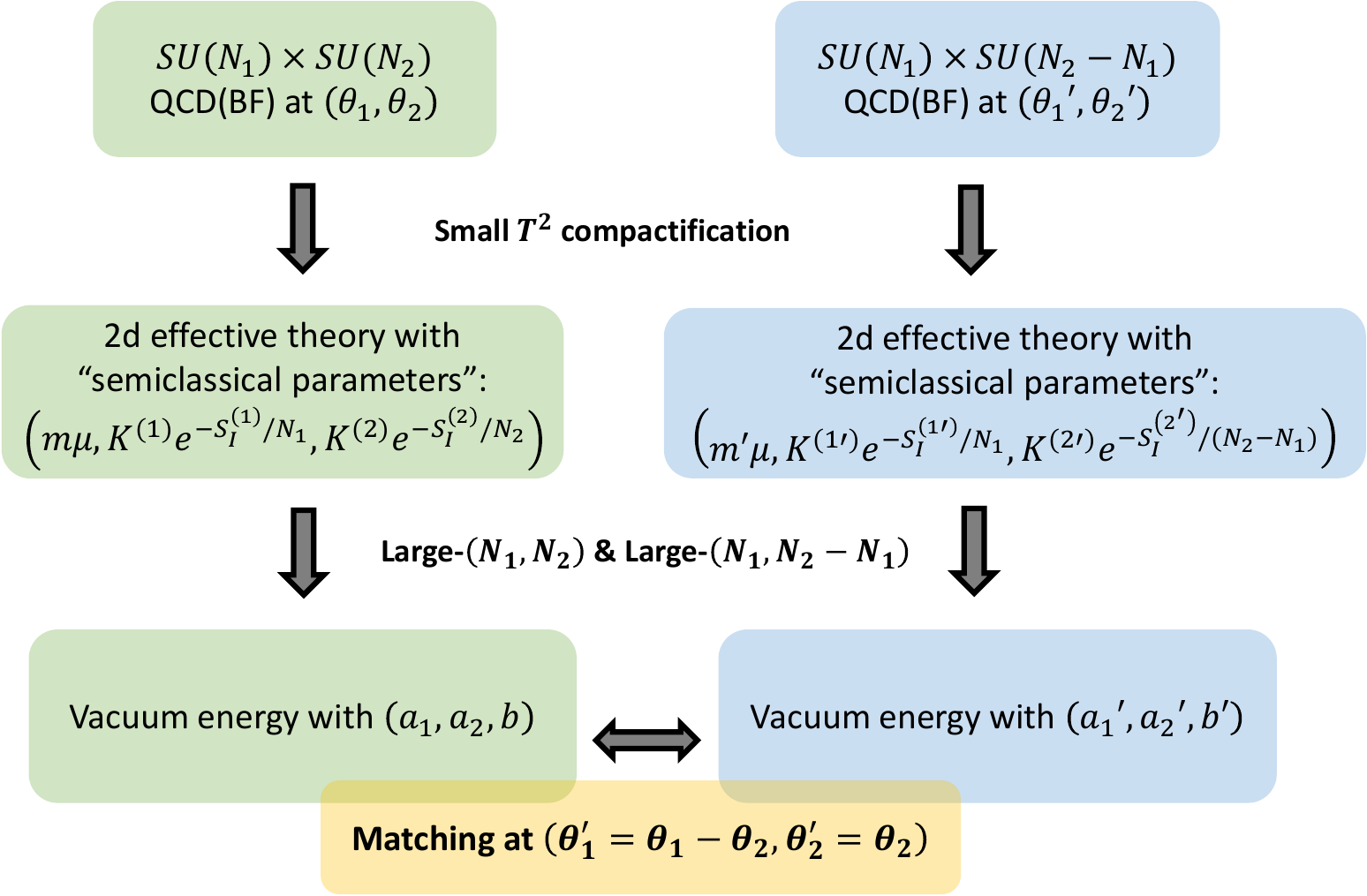}
\caption{Schematic picture for the strategy to support the large-$(N_1,N_2)$ duality.
We write the large-$N_{1,2}$ vacuum energy formulae for both $SU(N_1)\times SU(N_2)$ QCD(BF) and $SU(N_1)\times SU(N_2-N_1)$ QCD(BF) from the semiclassical descriptions, and match them at $(\theta_1' = \theta_1 - \theta_2, \theta_2' = \theta_2)$.
We can explicitly derive the correspondence of the semiclassical parameters, shown in (\ref{eq:duality_semiclssical_parameters}).
}
\label{fig:Large_N_duality}
\end{figure}

In Section~\ref{sec:KK_duality}, we discussed the vacuum structure described by (\ref{eq:large-n_vac_str}), which comes from the large-${N}$ counting argument: The vacuum structures in the large-$N_{1,2}$ limit have the multi-branch structure labeled by two integers $k_1,k_2\in \mathbb{Z}$, and the ground-state energy of each branch is a quadratic function of $\theta_1,\theta_2$. 
By combining the large-$N_{1,2}$ limit with the semiclassics, we give not only the explicit derivation of this vacuum structure but also the concrete form of the coefficients, $a_1,a_2, b$, in the formula~\eqref{eq:large-n_vac_str}. 
By applying this technique on both sides of the duality, we will show the concrete correspondence (see Figure~\ref{fig:Large_N_duality}).

We first study the $SU(N_1)\times SU(N_2)$ QCD(BF) at $(\theta_1,\theta_2)$. 
Let us rewrite the $2\pi N_1N_2$-periodic field as
$\varphi = 2 \pi (M_1 k_1 + M_2 k_2) + \delta \varphi~~(\operatorname{mod} 2 \pi N_1N_2)$, where $-\pi < \delta \varphi < \pi$, $k_1 \in \mathbb{Z}_{N_1}$, and $k_2 \in \mathbb{Z}_{N_2}$.
With this parametrization, the potential~\eqref{eq:potential_GCD1_BFQCD} reads,
\begin{align}
    V[\varphi] &= - m \mu \cos (\delta \varphi)  - 2K^{(1)} \rme^{-S_\mathrm{I}^{(1)}/N_1} \cos \left( \frac{2 \pi k_1 - \theta_1}{N_1} + \frac{N_2}{N_1} \delta \varphi \right) \notag \\
    &~~ - 2K^{(2)} \rme^{-S_\mathrm{I}^{(2)}/N_2} \cos \left( \frac{2 \pi k_2 - \theta_2}{N_2} + \frac{N_1}{N_2} \delta \varphi \right).
\end{align}
In the large-$(N_1, N_2)$ limit, the minimum point of $\delta \varphi$ is around zero $\delta \varphi \approx 0$ for ``vacuum branches'' $(k_1, k_2)$ that satisfy $\left| \frac{\theta_1 - 2 \pi k_1 }{N_1} \right| \ll 1$ and $\left| \frac{\theta_2 - 2 \pi k_2 }{N_2} \right| \ll 1$. 
We are going to consider the case $\theta_1,\theta_2\sim O(1)$, and thus these constraints on the vacuum labels become $|k_1/N_1|, |k_2/N_2|\ll 1$. 
Due to this constraint, we may neglect the information of the periodicity and regard the vacuum labels as $(k_1, k_2) \in \mathbb{Z} \times \mathbb{Z}$ in the large-$(N_1, N_2)$ limit. 
Since the vacuum expectation value of $\delta \varphi$ becomes $O(1/N_{1,2})$, we can solve for $\delta \varphi$ with the quadratic expansion. Substituting the result, the ground-state energy of each branch becomes
\begin{align}
    &\mathcal{E}_{(k_1, k_2)} (\theta_1, \theta_2)\notag\\
    &= \mathcal{E}_{(0, 0)} (\theta_1 -  2 \pi k_1, \theta_2 -  2 \pi k_2) \notag \\
    &\simeq \mathcal{E}_0 + a_1 \left( \frac{\theta_1 - 2 \pi k_1 }{N_1} \right)^2 + a_2 \left( \frac{\theta_2 - 2 \pi k_2 }{N_2} \right)^2 + b \left( \frac{\theta_1 - 2 \pi k_1 }{N_1} \right) \left( \frac{\theta_2 - 2 \pi k_2 }{N_2} \right), \label{eq:large_N_energy}
\end{align}
and each coefficient can be explicitly determined as
\begin{align}
    \mathcal{E}_0 &= - m \mu - 2K^{(1)} \rme^{-S_\mathrm{I}^{(1)}/N_1} - 2K^{(2)} \rme^{-S_\mathrm{I}^{(2)}/N_2}, \notag\\
    a_1 &= \frac{\left(m \mu + \frac{N_1^2}{N_2^2} 2K^{(2)} \rme^{-S_\mathrm{I}^{(2)}/N_2} \right) K^{(1)} \rme^{-S_\mathrm{I}^{(1)}/N_1}}{m \mu + \frac{N_2^2}{N_1^2} 2 K^{(1)} \rme^{-S_\mathrm{I}^{(1)}/N_1} + \frac{N_1^2}{N_2^2} 2 K^{(2)} \rme^{-S_\mathrm{I}^{(2)}/N_2}}, \notag \\
    a_2 &= \frac{\left(m \mu + \frac{N_2^2}{N_1^2} 2K^{(1)} \rme^{-S_\mathrm{I}^{(1)}/N_1} \right) K^{(2)} \rme^{-S_\mathrm{I}^{(2)}/N_2}}{m \mu + \frac{N_2^2}{N_1^2} 2 K^{(1)} \rme^{-S_\mathrm{I}^{(1)}/N_1} + \frac{N_1^2}{N_2^2} 2 K^{(2)} \rme^{-S_\mathrm{I}^{(2)}/N_2}}, \notag \\
    b &= - \frac{4 K^{(1)} K^{(2)}\rme^{-S_\mathrm{I}^{(1)}/N_1-S_\mathrm{I}^{(2)}/N_2}}{m \mu + \frac{N_2^2}{N_1^2} 2 K^{(1)} \rme^{-S_\mathrm{I}^{(1)}/N_1} + \frac{N_1^2}{N_2^2} 2 K^{(2)} \rme^{-S_\mathrm{I}^{(2)}/N_2}}. \label{eq:from_semiclassics_to_large_N}
\end{align}
We have the following consistency check of this result in the large-mass and massless limits:
\begin{itemize}
    \item In the large-mass limit $m \rightarrow +\infty$, the coefficients are $a_1 \simeq K^{(1)} \rme^{-S_\mathrm{I}^{(1)}/N_1}$, $a_2 \simeq K^{(2)} \rme^{-S_\mathrm{I}^{(2)}/N_2}$, and $b \rightarrow - 0$ as expected from the $SU(N_1)\times SU(N_2)$ pure YM theory.
    \item In the massless limit $m\to 0$, we find $\frac{N_2^2}{N_1^2}a_1=\frac{N_1^2}{N_2^2}a_2=-b/2(\,=: N_1^2 N_2^2 a)$, and then the vacuum energy can be written as
\begin{align}
    \mathcal{E}_{(0, 0)} (\theta_1, \theta_2) &\simeq \mathcal{E}_0 + a \left( N_1 \theta_1 - N_2 \theta_2 \right)^2 . 
\end{align}
This is consistent with the fact that physics only depends on $\theta_{\mathrm{phys}}$ at $m=0$.
\end{itemize}

We next study the dual side, $SU(N_1)\times SU(N_2-N_1)$ QCD(BF) at $(\theta'_1,\theta'_2)=(\theta_1-\theta_2,\theta_2)$ with the fermion mass $m'$. 
As above, since both $N_1$ and $N_2-N_1$ are large, we can effectively presume that the vacuum branches are labeled by integers $(k_1', k_2') \in \mathbb{Z} \times \mathbb{Z}$.
The energy density of each branch behaves as 
\begin{align}
    &\mathcal{E}_{(k_1', k_2')}' (\theta_1 - \theta_2, \theta_2)\notag\\
    % &= \mathcal{E}_{(0, 0)}' (\theta_1 - \theta_2  -  2 \pi \hat{\ell}_1, \theta_2 -  2 \pi \hat{\ell}_2) \notag \\
    &\simeq \mathcal{E}_0' + a_1' \left( \frac{\theta_1 - \theta_2 - 2 \pi k_1' }{N_1} \right)^2 + a_2' \left( \frac{\theta_2 - 2 \pi k_2' }{N_2-N_1} \right)^2 + b' \left( \frac{\theta_1 - \theta_2 - 2 \pi k_1' }{N_1} \right) \left( \frac{\theta_2 - 2 \pi k_2' }{N_2-N_1} \right), 
\end{align}
where $\mathcal{E}_0'$, $a_1'$, $a_2'$, and $b'$ are counterparts of the coefficients $\mathcal{E}_0$, $a_1$, $a_2$, and $b$ in the dual theory.
Explicitly, these coefficients are given by
\begin{align}
     a_1' &= \frac{\left(m' \mu + \frac{N_1^2}{(N_2-N_1)^2} 2K^{(2')} \rme^{-S_\mathrm{I}^{(2')}/(N_2-N_1)} \right) K^{(1')} \rme^{-S_\mathrm{I}^{(1')}/N_1}}{m' \mu + \frac{(N_2-N_1)^2}{N_1^2} 2 K^{(1')} \rme^{-S_\mathrm{I}^{(1')}/N_1} + \frac{N_1^2}{(N_2-N_1)^2} 2 K^{(2')} \rme^{-S_\mathrm{I}^{(2')}/(N_2-N_1)}}, \notag \\
    a_2' &= \frac{\left(m' \mu + \frac{(N_2-N_1)^2}{N_1^2} 2K^{(1')} \rme^{-S_\mathrm{I}^{(1')}/N_1} \right) K^{(2')} \rme^{-S_\mathrm{I}^{(2')}/(N_2-N_1)}}{m' \mu + \frac{(N_2-N_1)^2}{N_1^2} 2 K^{(1')} \rme^{-S_\mathrm{I}^{(1')}/N_1} + \frac{N_1^2}{(N_2-N_1)^2} 2 K^{(2')} \rme^{-S_\mathrm{I}^{(2')}/(N_2-N_1)}}, \notag \\
    b' &= - \frac{4 K^{(1')} K^{(2')}\rme^{-S_\mathrm{I}^{(1')}/N_1-S_\mathrm{I}^{(2')}/(N_2-N_1)}}{m' \mu + \frac{(N_2-N_1)^2}{N_1^2} 2 K^{(1')} \rme^{-S_\mathrm{I}^{(1')}/N_1} + \frac{N_1^2}{(N_2-N_1)^2} 2 K^{(2')} \rme^{-S_\mathrm{I}^{(2')}/(N_2-N_1)}},
\end{align}
where $K^{(1')} \rme^{-S_\mathrm{I}^{(1')}/N_1}$ (resp.~$K^{(2')} \rme^{-S_\mathrm{I}^{(2')}/(N_2-N_1)}$) is the center-vortex weight for $SU(N_1)$ (resp.~$SU(N_2-N_1)$) gauge field.

For the duality, we must have $\mathcal{E}_{(0, 0)} (\theta_1, \theta_2) = \mathcal{E}_{(0, 0)}' (\theta_1 - \theta_2, \theta_2)+\mathrm{const.}$, which requires
\begin{align}
     a_1 &= a_1', \notag \\
    a_2 &= N_2^2 \left( \frac{a_2'}{(N_2-N_1)^2} + \frac{a_1'}{N_1^2} - \frac{b'}{N_1 (N_2-N_1)} \right) , \notag \\
    b &= N_2 \left( -  \frac{2 a_1'}{N_1} + \frac{b'}{N_2-N_1} \right) .
\end{align}
By a tedious calculation, the parameters $(m\mu,K^{(1)} \rme^{-S_\mathrm{I}^{(1)}/N_1}, K^{(2)} \rme^{-S_\mathrm{I}^{(2)}/N_2})$ of the original theory are related to the ones of the dual theory as
\begin{align}
     m\mu &= \frac{m' \mu}{  1 + \frac{(N_2 - N_1)^2}{2 N_1 N_2 K^{(2')} \rme^{-S_\mathrm{I}^{(2')}/(N_2-N_1)}}m' \mu} , \notag \\
     K^{(1)} \rme^{-S_\mathrm{I}^{(1)}/N_1} &= \frac{K^{(2')} \rme^{-S_\mathrm{I}^{(2')}/(N_2-N_1)} N_1^3}{K^{(2')} \rme^{-S_\mathrm{I}^{(2')}/(N_2-N_1)} N_1^3 - K^{(1')} \rme^{-S_\mathrm{I}^{(1')}/N_1} (N_2 - N_1)^3} K^{(1')} \rme^{-S_\mathrm{I}^{(1')}/N_1}, \notag \\
     K^{(2)} \rme^{-S_\mathrm{I}^{(2)}/N_2} &= \frac{N_2^3}{(N_2 - N_1)^3} K^{(2')} \rme^{-S_\mathrm{I}^{(2')}/(N_2-N_1)} . 
     \label{eq:duality_semiclssical_parameters}
\end{align}
This establishes the conjectured duality~(\ref{eq:KK_duality_conjecture}) in the large-$(N_1,N_2)$ limit between the $ SU(N_1)\times SU(N_2)$ QCD(BF) on $ (\theta_1,\theta_2)$ and the $ SU(N_1)\times SU(N_2-N_1)$ QCD(BF) on $ (\theta_1-\theta_2,\theta_2)$. 
The above relation~\eqref{eq:duality_semiclssical_parameters} defines the duality map in the physical parameter range if the fermion mass of the original theory satisfies 
\begin{equation}
    m\mu\le m^*\mu:= \frac{N_1 (N_2-N_1) }{N_2^2}\cdot 2 K^{(2)} \rme^{-S_\mathrm{I}^{(2)}/N_2}. 
    \label{eq:largeN_criticalmass}
\end{equation}
In particular, the relationship of the fermion mass explicitly explains the duality at the massless point $m=0 \Leftrightarrow m' = 0$ and at the topology-changing point $m=m^* \Leftrightarrow m' = +\infty$. 

We note that the positivity of $K^{(1)} \rme^{-S_\mathrm{I}^{(1)}/N_1}$ in \eqref{eq:duality_semiclssical_parameters} puts the constraint on the strong scales of the dual theory as 
\begin{equation}
    K^{(2')} \rme^{-S_\mathrm{I}^{(2')}/(N_2-N_1)} 
    >\frac{(N_2 - N_1)^3 }{N_1^3} K^{(1')} \rme^{-S_\mathrm{I}^{(1')}/N_1}. 
    \label{eq:constraint_KKimage}
\end{equation}
Thus, the duality relation~\eqref{eq:KK_duality_conjecture} does not actually give the bijective relation but instead gives a map from the parent theory, $SU(N_1)\times SU(N_2)$ QCD(BF) with $m<m^*$, to the daughter theory, $SU(N_1)\times SU(N_2-N_1)$ QCD(BF), which is not surjective and the inverse map does not necessarily exists: 
\begin{align}
    SU(N_1)&\times SU(N_2) \mathrm{~QCD(BF)~on~} (\theta_1,\theta_2)~\mathrm{for}~m<m^* \notag \\
    &``\longrightarrow"~  SU(N_1)\times SU(N_2-N_1) ~\mathrm{QCD(BF)~on~} (\theta_1-\theta_2,\theta_2). 
    \label{eq:KK_duality_onedirection}
\end{align}
Let us emphasize that we can always find the semiclassical parameters for the daughter theory if the parent theory satisfies $m\le m^*$, so this constraint is not a problem for applying the duality cascade.

We also note that the correspondence of parameters (\ref{eq:duality_semiclssical_parameters}) is consistent with the observation at the hierarchical limit $\Lambda_2 \rightarrow \infty$ discussed in  Section~\ref{sec:duality_hierarchical}.
In our large-$N$ formula~\eqref{eq:duality_semiclssical_parameters}, we can see that $K^{(2)} \rme^{-S_\mathrm{I}^{(2)}/N_2} \rightarrow \infty$ is equivalent to $K^{(2')} \rme^{-S_\mathrm{I}^{(2')}/(N_2-N_1)} \rightarrow \infty$, so the hierarchical limit is taken on both the parent and daughter sides of the duality. Moreover, \eqref{eq:duality_semiclssical_parameters} in this limit shows that the other parameters are kept unchanged, $m = m'$ and $K^{(1')} \rme^{-S_\mathrm{I}^{(1')}/N_1} = K^{(1)} \rme^{-S_\mathrm{I}^{(1)}/N_1}$, and this is exactly what we observed in Section~\ref{sec:duality_hierarchical}.

\subsubsection{Numerical check of the duality conjecture}
\label{sec:duality_numerics}

In this subsection, we numerically investigate the phase diagrams of $SU(N_1) \times SU(N_2)$ QCD(BF) and $SU(N_1) \times SU(N_2-N_1)$ QCD(BF) with $(N_1,N_2) = (2,5)$, and examine the validity of duality when neither hierarchical nor large-$N_{1,2}$ limits are strictly taken. 
For the parent side, $SU(2)\times SU(5)$ QCD(BF), we set the strong scales as 
\begin{equation}
    2K^{(1)} \rme^{-S_\mathrm{I}^{(1)}/N_1} = 2K^{(2)} \rme^{-S_\mathrm{I}^{(2)}/N_2}=1. 
\end{equation}
Numerical search of the topology changing point $m\mu=m^*\mu$ gives $m^* \mu \approx 0.263$, and it is comparable with the large-$N_{1,2}$ formula \eqref{eq:largeN_criticalmass}, which suggests $\left.m^*\mu\right|_{\text{large-}N}=\frac{2\cdot 3}{5^2}=0.24$. 

When computing the daughter side, $SU(2)\times SU(3)$ QCD(BF), we need to set the strong scales; however, there is no clear criterion for their setting. 
Therefore, we shall use their values suggested by the large-$N_{1,2}$ formula~\eqref{eq:duality_semiclssical_parameters}, 
\begin{align}
    2K^{(1')}\rme^{-S^{(1')}_{\mathrm{I}}/N_1}
    &=\frac{2K^{(1)}\rme^{-S^{(1)}_{\mathrm{I}}/N_1}}{1+\frac{N_2^3}{N_1^3}\frac{2K^{(1)}\rme^{-S^{(1)}_{\mathrm{I}}/N_1}}{2K^{(2)}\rme^{-S^{(2)}_{\mathrm{I}}/N_2}}}=\frac{8}{133},
    \notag\\
    2K^{(2')}\rme^{-S^{(2')}_{\mathrm{I}}/(N_2-N_1)}&=\frac{(N_2-N_1)^3}{N_2^3}2K^{(2)}\rme^{-S^{(2)}_{\mathrm{I}}/N_2}=\frac{27}{125}.
    \label{eq:Kexp_relation_daughter}
\end{align}
We are going to compare the phase diagrams with several fermion masses, and we choose the fermion mass $m'\mu$ of the daughter side again from the large-$N_{1,2}$ formula~\eqref{eq:duality_semiclssical_parameters}: When $m\mu<m^*\mu$, we shall set $m'\mu$ by
\begin{equation}
    m'\mu= \frac{m\mu}{1-\frac{m\mu}{m^*\mu}}=\frac{m\mu}{1-\frac{m\mu}{0.24}}. 
    \label{eq:mass_relation_daughter}
\end{equation}
We show the results of the phase diagrams in Figure~\ref{fig:duality_comparison}. 

\afterpage{
\clearpage
\begin{landscape}
\begin{figure}[ht]
    \begin{minipage}[b]{0.196\linewidth}
        \centering \subfloat[$m \mu = 0.02$]{
        \includegraphics[scale=0.265]{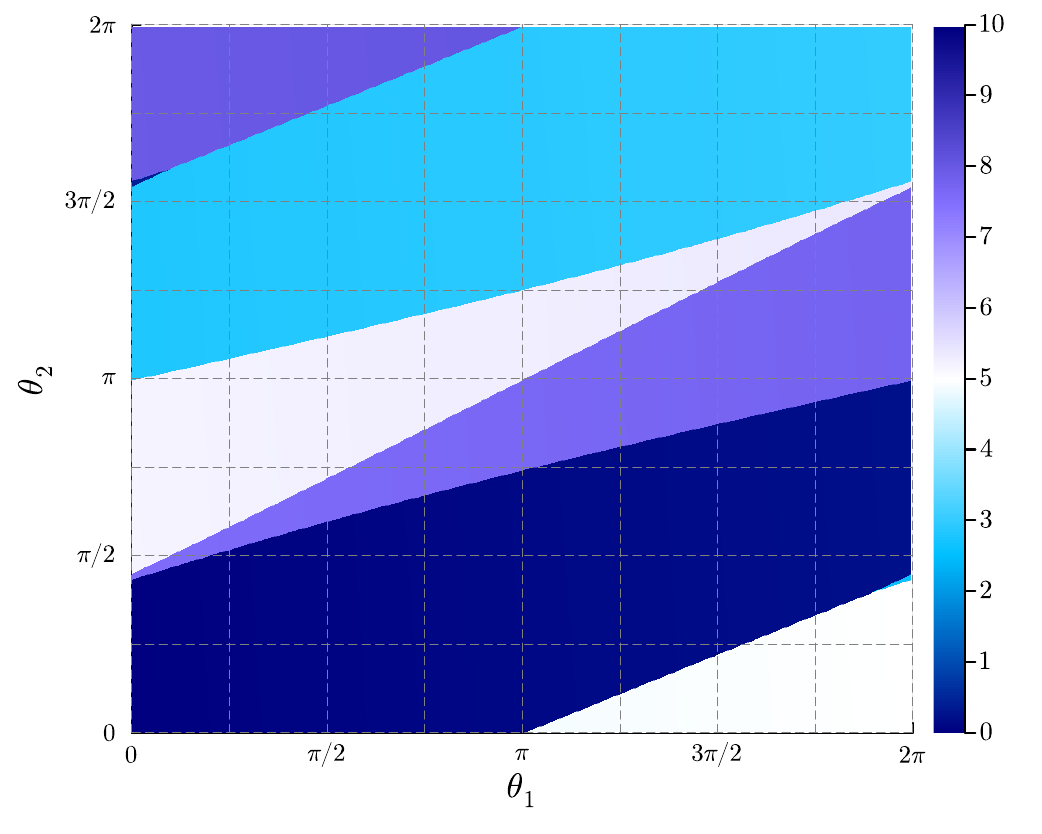}
        \label{fig:2-5_msq=0.02_c} 
        }
        \vspace{2ex}
    \end{minipage}
    \begin{minipage}[b]{0.196\linewidth}
        \centering \subfloat[$m \mu = 0.06$]{
        \includegraphics[scale=0.265]{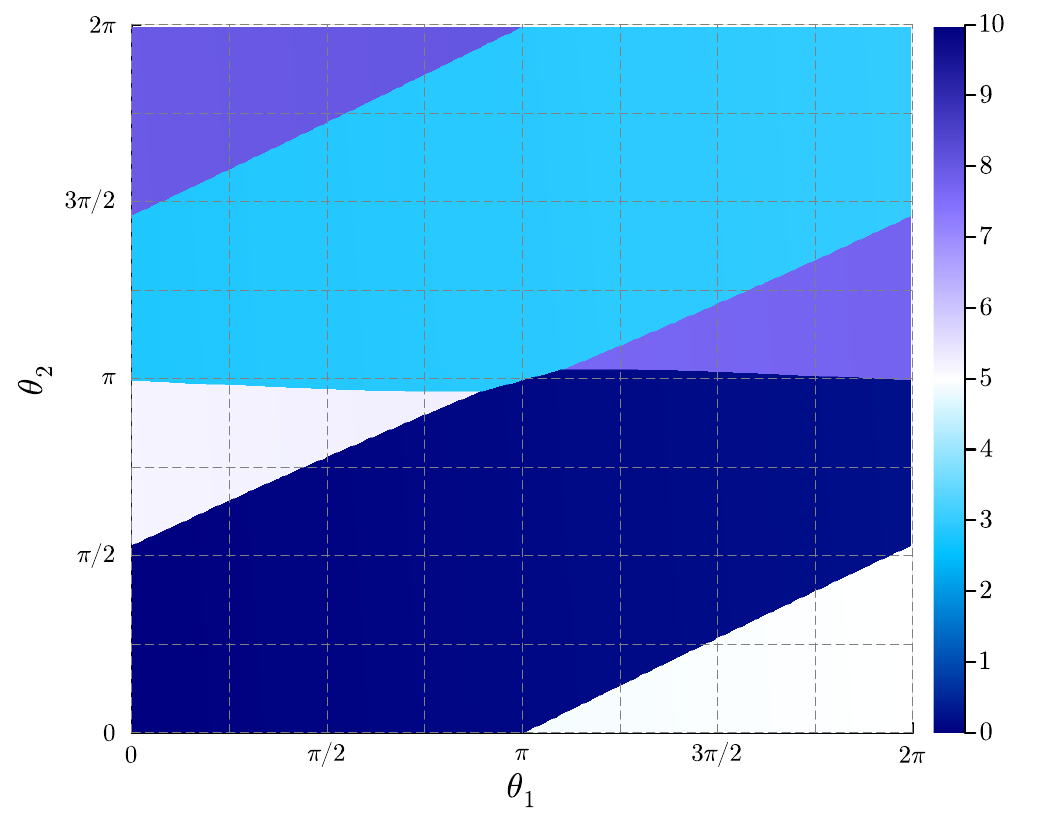}
        \label{fig:2-5_msq=0.06} 
        }
        \vspace{2ex}
    \end{minipage}
    \begin{minipage}[b]{0.196\linewidth}
        \centering \subfloat[$m \mu = 0.2$]{
        \includegraphics[scale=0.265]{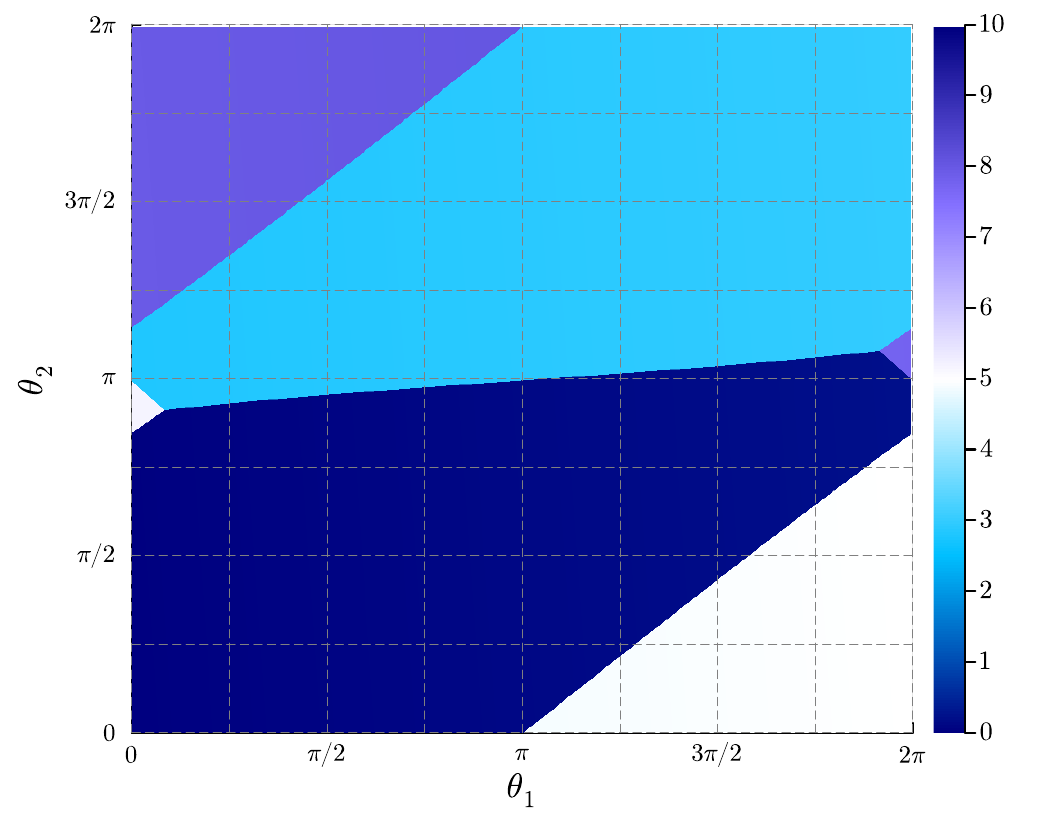}
        \label{fig:2-5_msq=0.2_c} 
        }
        \vspace{2ex}
    \end{minipage}
    \begin{minipage}[b]{0.196\linewidth}
        \centering \subfloat[$m \mu = 0.263\approx m^*\mu$]{
        \includegraphics[scale=0.265]{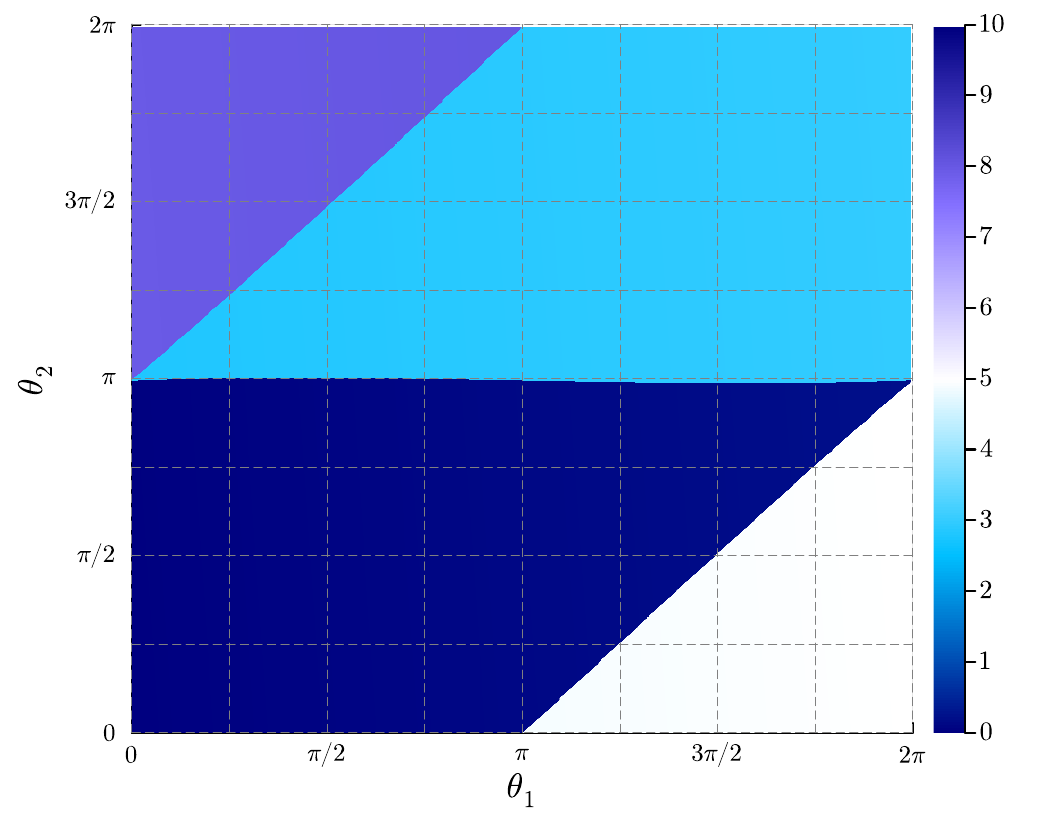}
        \label{fig:2-5_msq_crit} 
        }
        \vspace{2ex}
    \end{minipage}
    \begin{minipage}[b]{0.196\linewidth}
        \centering \subfloat[$m \mu = 0.3$]{
        \includegraphics[scale=0.265]{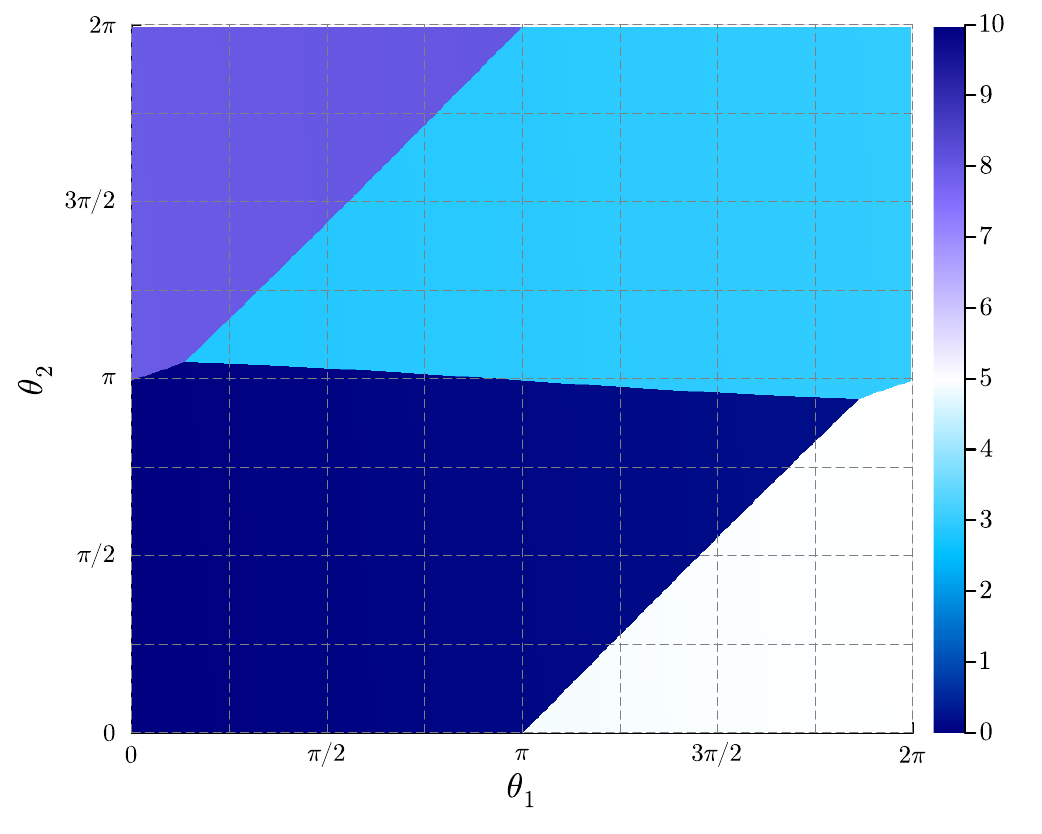}
        \label{fig:2-5_msq=0.3_c} 
        }
        \vspace{2ex}
    \end{minipage}
    \\    
    \begin{minipage}[b]{0.196\linewidth}
        \centering \subfloat[$m' \mu \approx 0.02182$]{
        \includegraphics[scale=0.265]{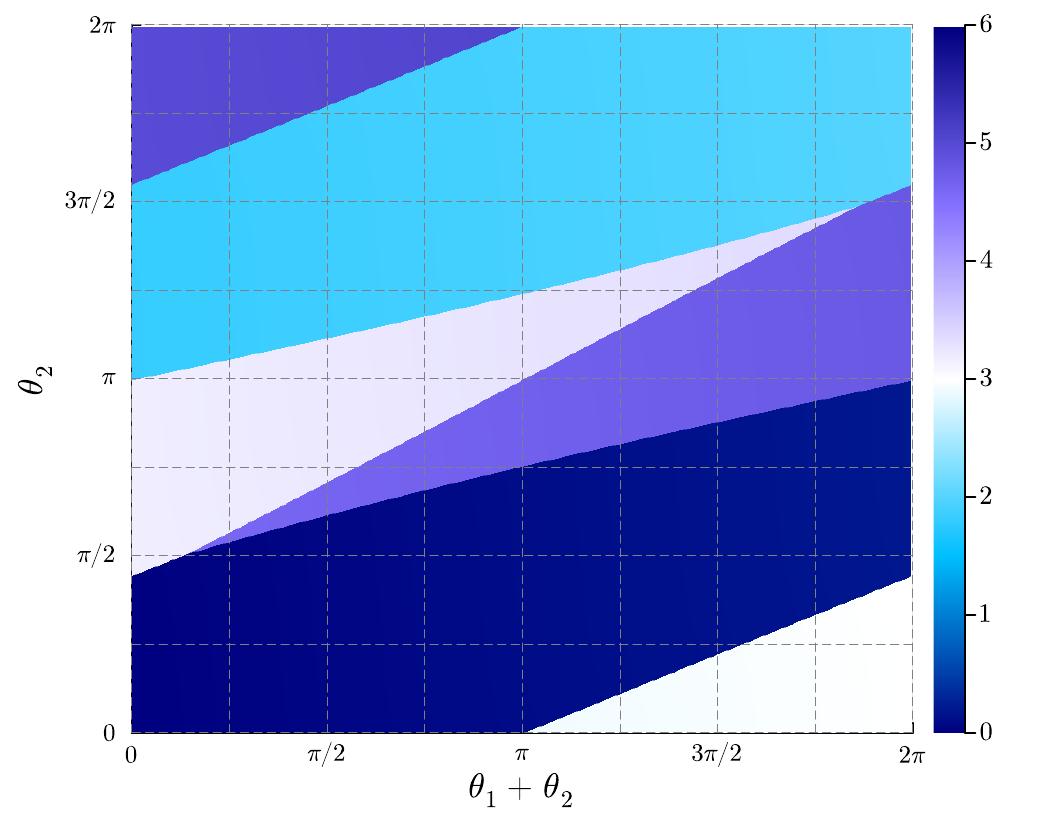}
        \label{fig:2-3_mp=0.02} 
        }
        \vspace{2ex}
    \end{minipage}
    \begin{minipage}[b]{0.196\linewidth}
        \centering \subfloat[$m' \mu = 0.08$]{
        \includegraphics[scale=0.265]{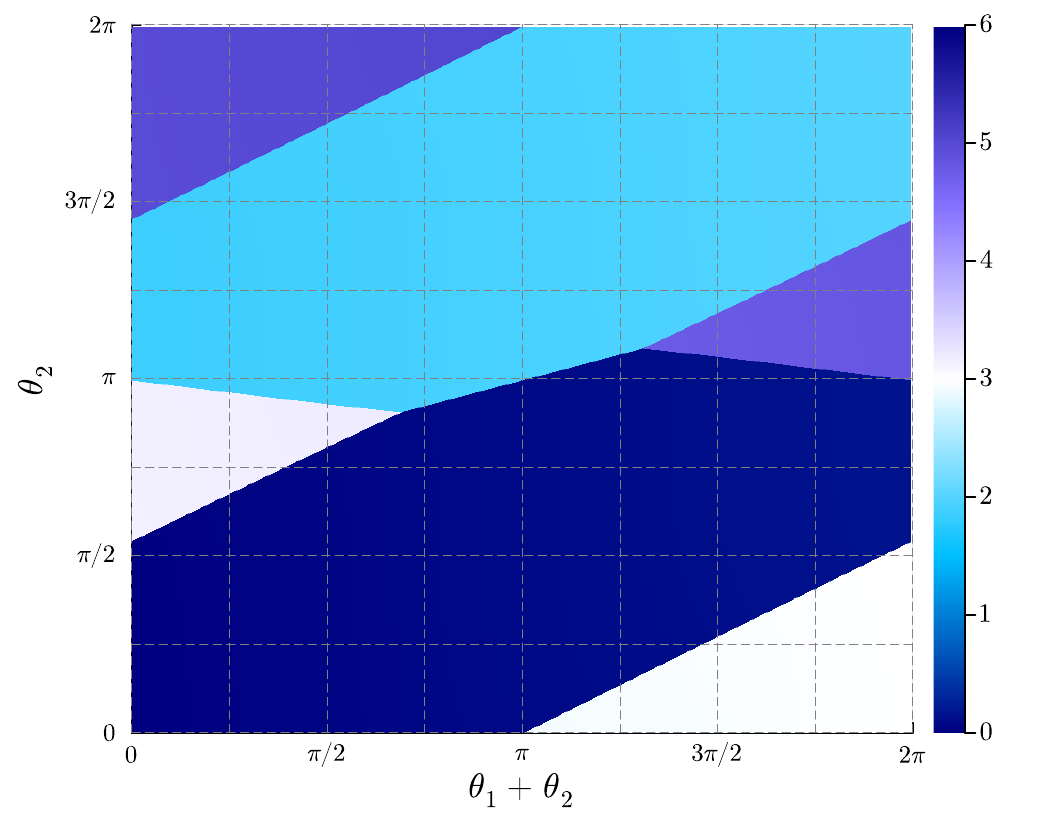}
        \label{fig:2-3_mp=0.08} 
        }
        \vspace{2ex}
    \end{minipage}
    \begin{minipage}[b]{0.196\linewidth}
        \centering \subfloat[$m' \mu = 1.2$]{
        \includegraphics[scale=0.265]{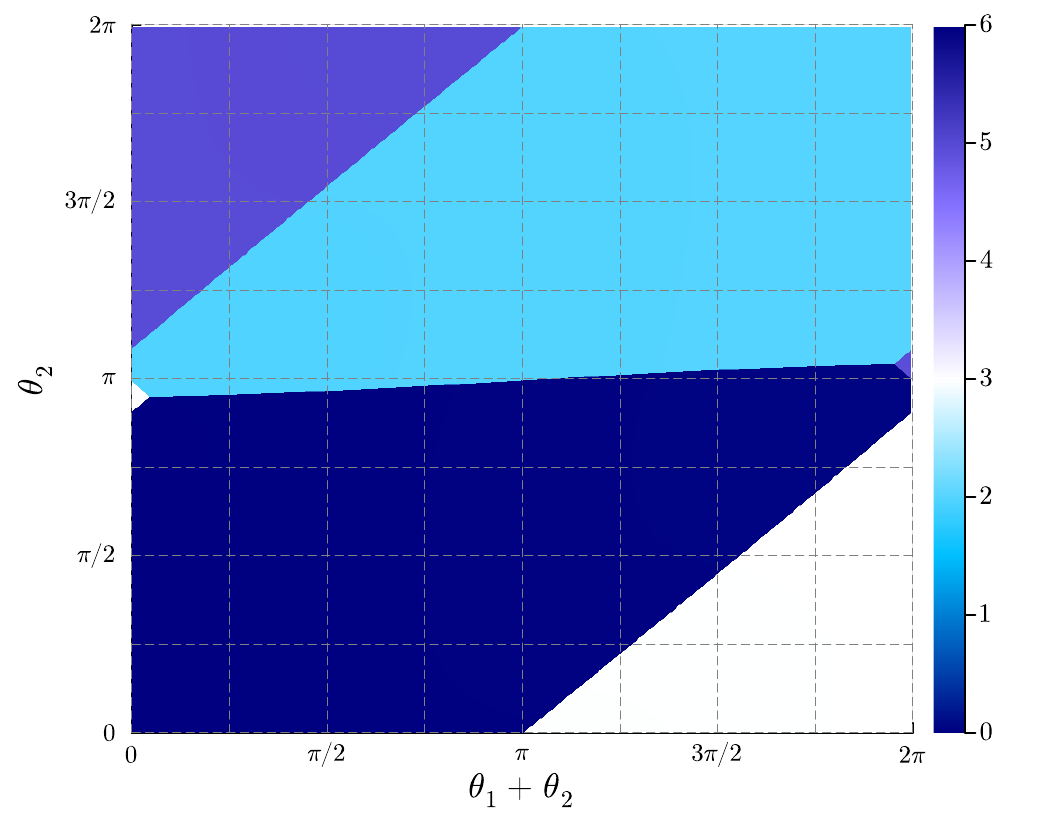}
        \label{fig:2-3_mp=1.2} 
        }
        \vspace{2ex}
    \end{minipage}
    \begin{minipage}[b]{0.196\linewidth}
        \centering \subfloat[$m' \mu = 100$]{
        \includegraphics[scale=0.265]{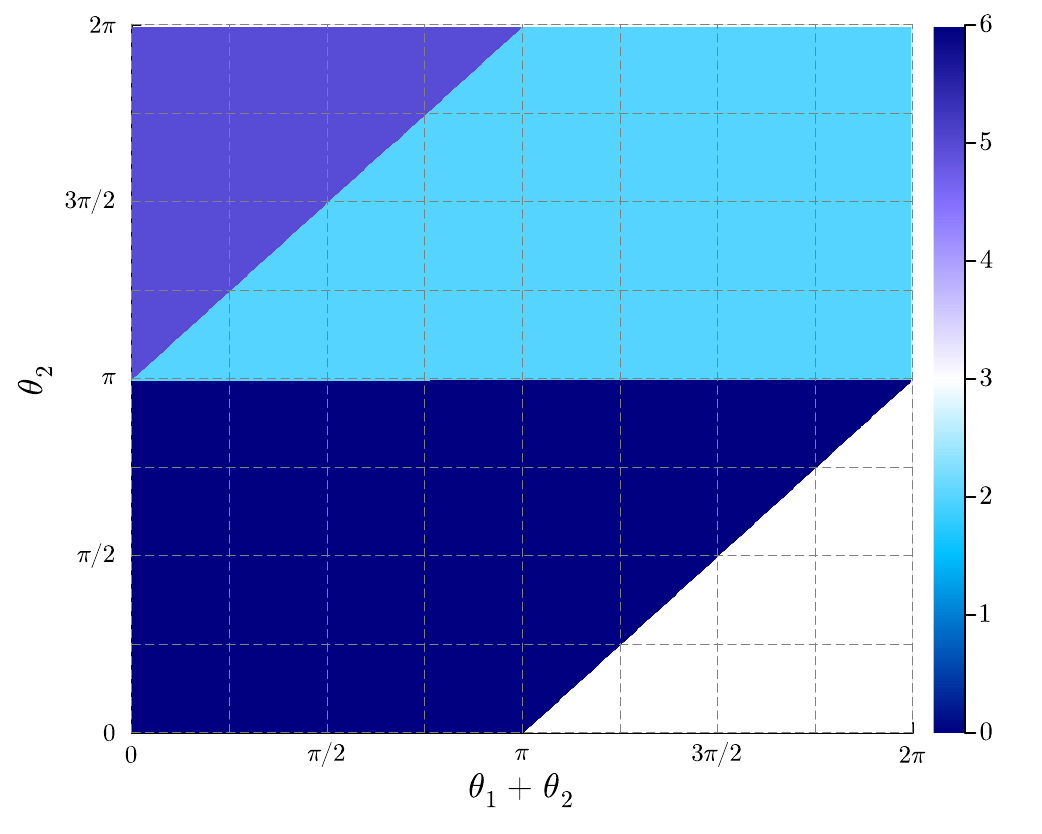}
        \label{fig:2-3_msq=100.0} 
        }
        \vspace{2ex}
    \end{minipage}

    \caption{(\textbf{Top panels})~Phase diagrams of the parent theory, $(N_1,N_2)=(2,5)$, on the $(\theta_1,\theta_2)$ plane with several $m \mu$ in the unit of $2K^{(1)} \rme^{-S_\mathrm{I}^{(1)}/N_1} = 2K^{(2)} \rme^{-S_\mathrm{I}^{(2)}/N_2} = 1$. 
    The topology-changing point is present at $m^*\mu \approx 0.263$, so the phase diagrams (a)-(d) for $m\mu\le m^*\mu$ are expected to be reproduced by the ones of the daughter theory according to the duality conjecture.\\ 
    (\textbf{Bottom panels})~Phase diagrams of the daughter theory, $(N_1, N_2-N_1) = (2, 3)$, on the $(\theta_1'+\theta_2',\theta_2')$ plane, in which the strong scales $2K^{(1')} \rme^{-S_\mathrm{I}^{(1')}/N_1},~ 2K^{(2')} \rme^{-S_\mathrm{I}^{(2')}/(N_2-N_1)}$ are set by the large-$N$ suggestion~\eqref{eq:Kexp_relation_daughter}. 
    The fermion mass $m' \mu$ is also set through \eqref{eq:mass_relation_daughter}, so that the phase diagrams~(f)-(i) are comparable to the above ones~(a)-(d) of the parent theory, respectively.
    }
    \label{fig:duality_comparison}
\end{figure}
\end{landscape}
\clearpage
}

Let us explain the details of Figure~\ref{fig:duality_comparison}. In the top panels, Figures~\ref{fig:2-5_msq=0.02_c}-\ref{fig:2-5_msq=0.3_c}, we show the $(\theta_1, \theta_2)$ phase diagram of the parent theory, $SU(2)\times SU(5)$ QCD(BF), for the range of the mass $0.02\le m \mu \le 0.3$, which includes the topology-changing point $m^*\mu\approx 0.263$ shown in Figure~\ref{fig:2-5_msq_crit}. 
In the bottom panels, Figures~\ref{fig:2-3_mp=0.02}-\ref{fig:2-3_msq=100.0}, we show the $(\theta'_1+\theta'_2,\theta'_2)$ phase diagram of the daughter theory, $SU(2)\times SU(3)$ QCD(BF), for the range of the mass $0.02\lesssim m'\mu\le 100$. We can see that the figures of the parent theory for $m\mu\le m^*\mu$, Figures~\ref{fig:2-5_msq=0.02_c}-\ref{fig:2-5_msq_crit}, and the counterparts of the daughter theory, Figures~\ref{fig:2-3_mp=0.02}-\ref{fig:2-3_msq=100.0}, are showing the same vacuum structures, respectively, and the agreement is remarkable not only qualitatively but also quantitatively:

\begin{itemize}
    \item In Figure~\ref{fig:2-5_msq=0.3_c}, we show the phase diagram of the parent theory for $m\mu=0.3>m^*\mu$. 
    We can see that the topological structure of the phase boundaries is the same with the large-mass limit (left panel of Figure~\ref{fig:two_limits}). 
    
    \item The phase diagram almost at $m\mu=m^*\mu$ of the parent theory is shown in Figure~\ref{fig:2-5_msq_crit}, and it should be compared with the figure just below, Figure~\ref{fig:2-3_msq=100.0} of the daughter theory. 
    In Figure~\ref{fig:2-3_msq=100.0}, we set $m'\mu=100$ as a numerical substitute of the limit $m'\mu\to \infty$, so the phase diagram in the $(\theta'_1,\theta'_2)$ basis looks as the checkerboard. 
    As the duality relates those theta angles to the parent one as $(\theta'_1,\theta'_2)=(\theta_1-\theta_2,\theta_2)$, we need to use the $(\theta'_1+\theta'_2,\theta'_2)$ basis ($=(\theta_1,\theta_2)$) when drawing the figures of the daughter theory. 
    \item We make the fermion mass of the parent theory a little smaller than the topology-changing point, $m\mu=0.2<m^*\mu$, in Figure~\ref{fig:2-5_msq=0.2_c}. 
    On the daughter side, this mass corresponds to $m'\mu=1.2$ according to \eqref{eq:mass_relation_daughter}, and we show its figure in Figure~\ref{fig:2-3_mp=1.2}. 
    We note that $m'\mu=1.2$ of the daughter theory is still quite heavy compared with its strong scales~\eqref{eq:Kexp_relation_daughter}, so Figure~\ref{fig:2-3_mp=1.2} can be understood as the linear transformation of the large-mass limit (left panel of Figure~\ref{fig:two_limits}) to the $(\theta'_1+\theta'_2,\theta'_2)$ basis. Indeed, the direct transition from the $(k_1,k_2)=(0,0)$ state to the $(k_1,k_2)=(1,1)$ state can be seen around $(\theta'_1+\theta'_2,\theta'_2)=(2\pi,\pi)$ in Figure~\ref{fig:2-3_mp=1.2}, while its phase boundary is still tiny.  

    \item As we lower the fermion mass further, the phase diagrams of the parent and daughter theories are deformed in the same manner. 
    At $m\mu=0.06$, the parent theory approaches another topology-changing point as shown in Figure~\ref{fig:2-5_msq=0.06}. 
    On the daughter side, it corresponds to $m'\mu=0.08$ according to \eqref{eq:mass_relation_daughter}, and the phase diagram with the $(\theta'_1+\theta'_2,\theta'_2)$ basis is shown in Figure~\ref{fig:2-3_mp=0.08}. 
    The daughter side is a little far from its topology-changing point, but this quantitative mismatch would be acceptable recalling that our parameter setting is based on the large-$N_{1,2}$ formula.
    \footnote{
    About the parameter setting, we point out the issue coming out of the discrepancy between the large-$N$ result of $m^*\mu|_{\text{large-}N}=0.24$ and the numerical value $m^*\mu|_{\text{numerical}}\approx 0.263$. In our numerical computation of the daughter theory, we use the large-$N$ value for $m^*\mu$ to determine $m'\mu$ in \eqref{eq:mass_relation_daughter} for theoretical consistency. However, this causes a problem in that we cannot determine $m'\mu$ when $m^*\mu|_{\text{large-}N}<m\mu<m^*\mu|_{\text{numerical}}$. We can circumvent this problem if we use $m^*\mu\approx 0.263$ when determining $m'\mu$ in \eqref{eq:mass_relation_daughter}, and then we get a bit lower values of $m'\mu$ compared with current ones. 
    If doing so, Figure~\ref{fig:2-3_mp=0.08} would be a bit closer to the topology-changing point, and the apparent look of phase diagrams may become more similar between the parent and daughter theories. Anyway, we should keep in mind that our mapping of parameters has this kind of ambiguity. } 
    
    \item We also note that the behaviors after this topology changing point are also qualitatively the same between the parent and daughter theories as we can see in Figures~\ref{fig:2-5_msq=0.02_c} and~\ref{fig:2-3_mp=0.02}. 
\end{itemize}

Based on these observations, we are tempted to conclude that the duality map works so nicely even when neither hierarchical nor large-${N}_{1,2}$ limit is strictly taken, and it is useful to understand the topology-changing behaviors for the $(\theta_1,\theta_2)$ phase diagram of the parent theory. 
For example, when we are trying to study the topology-changing point $m\mu=0.06$, Figure~\ref{fig:2-5_msq=0.06}, of the parent theory, it would not be so easy to imagine its behavior since the phase diagram there already experienced the topology-changing phenomenon at $m\mu=m^*\mu\approx 0.263$. 
By switching to the picture of the daughter side, Figure~\ref{fig:2-3_mp=0.08}, this topology-changing phenomenon is its first one, and we can understand it more easily as the deformation from the large-mass limit. 
Indeed, the topology change from Figure~\ref{fig:2-3_mp=0.08} to \ref{fig:2-3_mp=0.02} is basically equivalent to the ones studied in Figure~\ref{fig:app-some-examples}. 
The reason why their topology-changing scales, $m'\mu\approx 0.08$ in Figure~\ref{fig:2-3_mp=0.08} and $m\mu\approx 0.3$ in Figure~\ref{fig:msq03}, are so different can be understood from the fact the duality map~\eqref{eq:Kexp_relation_daughter} sets the strong scales of the daughter side small values. 

In this section, we have examined the validity of the duality for the case when the two strong scales of the parent theory are the same. 
It would be an interesting exercise to check the duality for the case when those strong scales are well separated, $\Lambda_1\ll \Lambda_2$ or $\Lambda_1\gg \Lambda_2$, and those cases are also numerically studied in Appendix~\ref{sec:phase-diag_hierarchy}: The duality works well again.

\begin{figure}[th]
    \centering
    \includegraphics[width=0.5\textwidth]{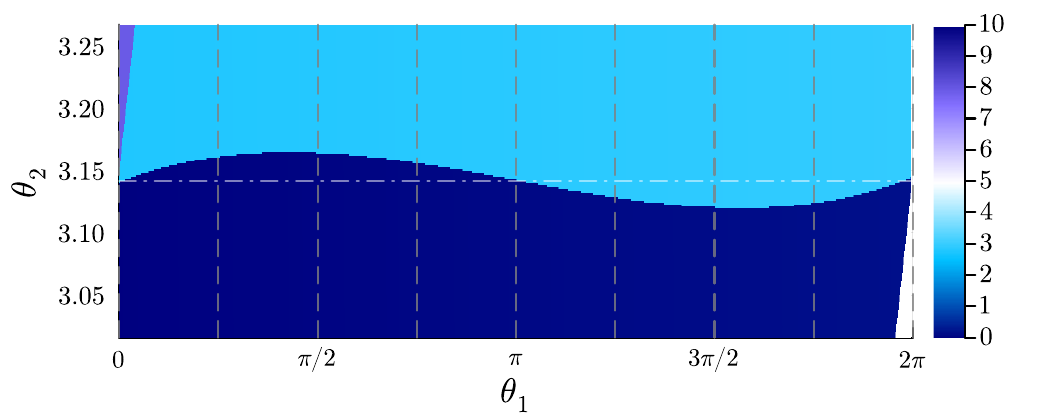}
    \caption{Magnified phase diagram on $(\theta_1,\theta_2)$ plane around $\theta_2 \approx \pi$ with $(N_1,N_2) = (2,5)$ almost at the topology-changing point, $2K^{(1)} \rme^{-S_\mathrm{I}^{(1)}/N_1} = 2K^{(2)} \rme^{-S_\mathrm{I}^{(2)}/N_2}=1$ and $m \mu =0.263$. }
    \label{fig:2-5_msq_top-change}
\end{figure}

Lastly, it might be useful to make a somewhat trivial remark that this duality at finite $N_{1,2}$ is not an exact one. %since the matching of the phase diagrams looks so good. 
As we will elaborate in Appendix~\ref{sec:bdy_is_not_straight}, if the duality relation were exactly true, the phase boundaries at the topology-changing point should consist only of the completely straight lines. 
In Figure~\ref{fig:2-5_msq_top-change}, we enlarge the phase diagram, Figure~\ref{fig:2-5_msq_crit}, of $SU(2) \times SU(5)$ QCD(BF) almost at $m\mu=m^*\mu$ around $\theta_2\approx \pi$. 
The fact that the phase boundary is not a horizontal straight line indicates the non-exactness of the duality relation. 
Conversely, we may say that the violation of the duality relation is so mild, even away from the large-$N$ limit (apart from the subtlety of the parameter mapping), and it is difficult to notice without a magnifying glass. 

\section{Phase diagrams of QCD(BF) for \texorpdfstring{$\operatorname{gcd}(N_1,N_2) \neq 1$}{gcd(N1,N2)>1}}
\label{sec:analysis_gcd!=1}

This section extends our analysis to the cases of $\operatorname{gcd}(N_1,N_2) \neq 1$. 
The discussion is almost parallel to the one in Section~\ref{sec:analysis_gcd1}, but we should keep in mind that the $2$d effective theory has the discrete vacuum labels $(\tilde{k},\tilde{\ell})\in (\mathbb{Z}_{\gcd(N_1,N_2)})^2$ in addition to the extended compact scalar, $\varphi\sim \varphi+2\pi N'_1 N'_2$. 
In Section \ref{sec:limits_gcd!=1}, we consider the massless limit, small-mass limit (in the presence of small mass perturbation), and large-mass limit.
The semiclassical approach produces a phase diagram that aligns with expectations.
In Section \ref{sec:numerical_phase_diagrams_gcd!=1}, we examine the phase diagram of the simplest example, $SU(2) \times SU(4)$ QCD(BF), obtained numerically by the minimization of the effective potential.
We observe the topology-changing phenomenon as the fermion mass is varied, which bridges the small-mass phase diagram (Figure \ref{fig:BFQCDsmallmass}) and the large-mass checkerboard-like phase diagram.
In Section \ref{sec:comment_large_N_gcd!=1}, we point out that the duality at large-$N_{1,2}$  (Section \ref{sec:recovering_large_N_from_semiclassics}) can be easily generalized.
This section is concluded with comments on the domain wall and anomaly inflow, Section \ref{sec:comments_on_DW}.

% \subsection{Small-mass and large-mass limits}
\subsection{Phase diagrams for \texorpdfstring{$\gcd(N_1,N_2)\not=1$}{gcd(N1,N2)!=1} with semiclassics}
\label{sec:limits_gcd!=1}

Let us determine the phase diagrams at the large- and small-mass limits based on the 2d semiclassical description. In Section~\ref{sec:QCDBF_2dEFT}, we have derived the following effective potential,
\begin{align}
    V[\varphi; \Tilde{k},\tilde{\ell}] &= - m \mu \cos (\varphi) - 2K^{(1)} \rme^{-S_\mathrm{I}^{(1)}/N_1} \cos \left( \frac{N_2 \varphi + 2 \pi (N'_2 \tilde{\ell} + M_2 \Tilde{k}) - \theta_1}{N_1} \right) \notag \\
    &~~~~~~~~~ - 2K^{(2)} \rme^{-S_\mathrm{I}^{(2)}/N_2} \cos \left( \frac{N_1 \varphi
    + 2 \pi (N'_1 \tilde{\ell} - M_1 \Tilde{k}) - \theta_2}{N_2} \right), \tag{\ref{eq:potential_2deff}}
\end{align}
where $\varphi$ is the extended compact variable $\varphi \sim \varphi + 2 \pi N_1' N_2'$, 
and we identify $(\varphi;\Tilde{k},\tilde{\ell} + \operatorname{gcd}(N_1,N_2)) \sim (\varphi + 2 \pi; \Tilde{k},\tilde{\ell})$ and $(\varphi; \Tilde{k} + \operatorname{gcd}(N_1,N_2),\tilde{\ell} ) \sim (\varphi ; \Tilde{k},\tilde{\ell})$. 
Our choice of the Bezout coefficient~$(M_1,M_2)$ satisfies $M_1/N_2', M_2/N_1'\in \mathbb{Z}$.

\subsubsection{Large-mass limit \texorpdfstring{$m\to \infty$}{m->infty}}
\label{sec:large-mass_gcd!=1}

Let us briefly look at the phase diagram in the large-mass limit, $\Lambda_1, \Lambda_2 \ll m (\,\ll 1/NL)$.
For this purpose, instead of $(\Tilde{k}, \Tilde{\ell})$, it is convenient to use the original notation for the discrete labels $(k_1,k_2)$, by using (\ref{eq:change_of_discrete_labels}).
With these original labels, the potential is,
\begin{align}
    V [\varphi] &= - m \mu \cos (\varphi) - 2K^{(1)} \rme^{-S_\mathrm{I}^{(1)}/N_1} \cos \left( \frac{N_2 \varphi + 2 \pi k_1 - \theta_1}{N_1} \right) \notag \\
    &~~~~~~~~~ - 2K^{(2)} \rme^{-S_\mathrm{I}^{(2)}/N_2} \cos \left( \frac{N_1 \varphi
    + 2 \pi k_2 - \theta_2}{N_2} \right), \label{eq:eff_potential_with_k1k2}
\end{align}
These labels $(k_1,k_2)$ obeys the identification (\ref{eq:identification_phi_k1_k2}), from which
\begin{align}
    (\varphi ; k_1 + \operatorname{gcd}(N_1, N_2) , k_2) &\sim (\varphi + 2 \pi M_1 ; k_1  , k_2), \notag \\
    (\varphi ; k_1 , k_2+ \operatorname{gcd}(N_1, N_2) ) &\sim (\varphi + 2 \pi M_2 ; k_1  , k_2).
\end{align}
Thus, we can regard $(k_1,k_2)$ as integers in $\{ 0 ,\cdots, \operatorname{gcd}(N_1,N_2) -1 \}$.

In the $m\to \infty$ limit, we should first minimize $-\mu m \cos (\varphi)$, and the candidates of minima are $\varphi =0$ (mod $2\pi$).
As done in Section~\ref{sec:large-mass_gcd=1}, we denote them as $\varphi = 2 \pi (M_1 k_1' + M_2 k_2')$ with $k_1' \in \mathbb{Z}_{N_1'}$ and $k_2' \in \mathbb{Z}_{N_2'}$. 
The potential of order $O(m^0)$ is,
\begin{align}
    V [\varphi] &= - 2K^{(1)} \rme^{-S_\mathrm{I}^{(1)}/N_1} \cos \left( \frac{2 \pi (\operatorname{gcd}(N_1,N_2) k_1'  + k_1) - \theta_1}{N_1} \right) 
    \notag \\
    &~~~
    - 2K^{(2)} \rme^{-S_\mathrm{I}^{(2)}/N_2} \cos \left( \frac{2 \pi (\operatorname{gcd}(N_1,N_2) k_2'  + k_2) - \theta_2}{N_2} \right). 
\end{align}
Then, we can combine the discrete labels $(k_1', k_2', k_1,k_2)$ so that 
$\hat{k}_1 := \operatorname{gcd}(N_1,N_2) k_1'  + k_1 \in \mathbb{Z}_{N_1}$ and $\hat{k}_2 := \operatorname{gcd}(N_1,N_2) k_2'  + k_2 \in \mathbb{Z}_{N_2}$. These labels are nothing but the original $\mathbb{Z}_{N_1} \times \mathbb{Z}_{N_2}$ vacuum labels that arise from the sum over center vortices in Section~\ref{sec:center_vortex_residual_N1N2}.
With these labels $(\hat{k}_1, \hat{k}_2) \in \mathbb{Z}_{N_1} \times \mathbb{Z}_{N_2}$, the vacuum energy $V_{\hat{k}_1, \hat{k}_2} = V [\varphi = 2 \pi (M_1 k_1' + M_2 k_2')]$ reads
\begin{align}
    V_{\hat{k}_1, \hat{k}_2} &=  - 2K^{(1)} \rme^{-S_\mathrm{I}^{(1)}/N_1} \cos \left( \frac{2 \pi \hat{k}_1 - \theta_1}{N_1} \right)- 2K^{(2)} \rme^{-S_\mathrm{I}^{(2)}/N_2} \cos \left( \frac{2 \pi \hat{k}_2 - \theta_2}{N_2} \right).
\end{align}
The $\mathbb{Z}_{N_1} \times \mathbb{Z}_{N_2}$ vacua with the above vacuum energy exactly reproduce the semiclassical picture of the $SU(N_1)\times SU(N_2)$ pure Yang-Mills theory, and the phase boundaries are located on $\theta_1 = \pi ~(\operatorname{mod}2\pi)$ and $\theta_2 = \pi ~(\operatorname{mod}2\pi)$. 
The analysis of the $O(m^{-1})$ contribution at $(\theta_1,\theta_2)=(\pi,\pi)$ is exactly the same as what we have done in Section~\ref{sec:large-mass_gcd=1}, and the large-mass phase diagram is again given by the left panel of Figure~\ref{fig:two_limits}.

We note that the jump of $\varphi$ in the above leading-order discussion occurs only when $\theta_{1,2}$ are shifted by relatively large amounts, such as $\theta_1\to \theta_1+2\pi \gcd(N_1,N_2)$. 
When we restrict our attention to the region $-2\pi<\theta_1<2\pi$ and $-2\pi<\theta_2<2\pi$, the leading-order vacuum configuration is always $\varphi=0$, and the phase transitions at $\theta_1=\pm \pi$, $\theta_2=\pm \pi$ are described by the jump of discrete labels $k_1,k_2$. 
Thus, the discrete labels $k_1,k_2$ (or $\tilde{k}, \tilde{\ell}$) play the primary role in the large-mass regime, and $\varphi$ only affects sub-leading dynamics when $\gcd(N_1,N_2)\not=1$. 

\subsubsection{Massless limit \texorpdfstring{$m\to 0$}{m->0} and small-mass perturbation}

In the massless case, substituting $\varphi+\frac{\theta_2}{N_1}$ into the effective potential yields,
\begin{align}
    V\left[\varphi+\frac{\theta_2}{N_1}; \Tilde{k},\tilde{\ell}\right] 
    &= - 2K^{(1)} \rme^{-S_\mathrm{I}^{(1)}/N_1} \cos \left( \frac{N_2 \varphi + 2 \pi (N'_2 \tilde{\ell} + M_2 \Tilde{k}) - \frac{\theta_{\mathrm{phys}}}{N_1}}{N_1} \right) \notag \\
    &~~~~ - 2K^{(2)} \rme^{-S_\mathrm{I}^{(2)}/N_2} \cos \left( \frac{N_1 \varphi
    + 2 \pi (N'_1 \tilde{\ell} - M_1 \Tilde{k})}{N_2} \right).
\end{align}
Thus, the physics depends only on $\theta_{\mathrm{phys}}=N_1\theta_1-N_2 \theta_2$. 

When $\theta_{\mathrm{phys}}=0$, we can readily find the global minima, 
\begin{equation}
    \varphi=-2\pi \frac{\tilde{\ell}}{\gcd(N_1,N_2)}, \quad \tilde{k}=0,
    \label{eq:chiralbroken_thetaphys=0}
\end{equation}
where $\tilde{\ell}\in \mathbb{Z}_{\gcd(N_1,N_2)}$ due to the identification~\eqref{eq:identification_phi_k_l}.  
These minima represent the spontaneous breakdown of the discrete chiral symmetry (\ref{eq:chiral_transf_semiclassics}), and the vacua have $\gcd(N_1,N_2)$-degeneracy.
From this solution, we can construct the global minima for $\theta_{\mathrm{phys}}=2\pi n$. We note the equivalence between $\theta_{\mathrm{phys}}\to \theta_{\mathrm{phys}}+2\pi n$ and $(\theta_1,\theta_2)\to (\theta_1+2\pi M_2 n, \theta_2-2\pi M_1 n)$, and the latter moves the vacuum label $(k_1,k_2)\to (k_1+M_2n, k_2-M_1n)$. We can translate this result to $(\tilde{k},\tilde{\ell})$ by using \eqref{eq:change_of_discrete_labels}, which gives $\tilde{k}\to \tilde{k}+n$ without changing $\tilde{\ell}$. 
Applying this transformation to \eqref{eq:chiralbroken_thetaphys=0}, we obtain the vacua for $\theta_{\mathrm{phys}}=2\pi n$ as 
\begin{equation}
    \varphi=-2\pi \frac{\tilde{\ell}}{\gcd(N_1,N_2)}, \quad \tilde{k}=n,
    \label{eq:chiralbroken_thetaphys=2pin}
\end{equation}
with $\ell\in \mathbb{Z}_{\mathrm{gcd}(N_1,N_2)}$.

When we adiabatically change $\theta_{\mathrm{phys}}$ as $\theta_{\mathrm{phys}}=2\pi n+\delta \theta_{\mathrm{phys}}$ with $|\delta \theta_{\mathrm{phys}}|<\pi$, the vacuum configurations also continuously change from \eqref{eq:chiralbroken_thetaphys=2pin}:
\begin{equation}
    \varphi=-2\pi \frac{\tilde{\ell}}{\gcd(N_1,N_2)}+\delta\varphi, \quad \tilde{k}=n. 
\end{equation}
We note that the first term of the effective potential is minimized if $\delta\varphi=\frac{\delta \theta_{\mathrm{phys}}}{N_1 N_2}$ and the second term is minimized if $\delta\varphi=0$. 
Thus, the true minima exists in the range $|\delta\varphi|\le \frac{|\delta \theta_{\mathrm{phys}}|}{N_1 N_2}$, so its deviation from \eqref{eq:chiralbroken_thetaphys=2pin} is small. 
When $\delta \theta_{\mathrm{phys}}$ reaches $\pi$, two branches with $\tilde{k}=n$ and $\tilde{k}=n+1$ become degenerate, and the vacuum degeneracy becomes $2\gcd(N_1,N_2)$. 

We now get the result in the massless case. 
There are $\operatorname{gcd}(N_1,N_2)$ degenerate vacua for generic values of $\theta_{\mathrm{phys}}$. 
There are phase transitions described by the jump of $\tilde{k}$, and phase boundaries are located on $\theta_{\mathrm{phys}.} = \pi ~(\operatorname{mod} 2 \pi)$, where the vacuum degeneracy becomes $2 \operatorname{gcd}(N_1,N_2)$.
The phase diagram is consistent with the large-$N$ argument, shown in Figure~\ref{fig:BFQCDmassless}.

% \subsubsection{Small-mass perturbation}
Next, let us consider the small-mass perturbation. Since the change of $\theta_{\mathrm{phys}}$ from $2\pi n$ only gives a tiny effect as we have seen above, let us simply set $\theta_{\mathrm{phys}}=0$. 
Recall that we substituted $\varphi+\frac{\theta_2}{N_1}$ in the above discussion for the massless case, and then the mass perturbation gives the energy splitting for the chiral broken vacua \eqref{eq:chiralbroken_thetaphys=0}  by 
\begin{align}
    \Delta E_{\ell}
    &=-m\mu \cos\left(\frac{\theta_2}{N_1}-2\pi \frac{\tilde{\ell}}{\gcd(N_1,N_2)}\right) \notag\\
    &=-m\mu \cos\left(\frac{\theta_2-2\pi N'_1 \tilde{\ell}}{N_1}\right), 
\end{align}
where $\tilde{\ell}=0,1,\ldots, \gcd(N_1,N_2)-1$. 
When we increase $\theta_2$ from $0$ to $2\pi N'_1$, there is a level crossing at $\theta_2=\pi N'_1$ from $\tilde{\ell}=0$ to $\tilde{\ell}=1$. As we have set $\theta_{\mathrm{phys}}=0$ in this argument, this phase transition point is given by $(\theta_1,\theta_2)=\pi(N'_2, N'_1)$.  
The discrete label $\tilde{\ell}$ has a jump at this point. Thus, this phase transition should form a curve separating the $\tilde{\ell}=0$ and $\tilde{\ell}=1$ states,\footnote{
We can determine this phase transition curve analytically in the small-mass limit: 
The phase boundary between $\Tilde{\ell} = 0$ and $\Tilde{\ell}= 1$ vacua is given by the curve in the $(\theta_1,\theta_2)$ space, on which $\varphi_*=\frac{\pi}{\gcd(N_1,N_2)}$ with $\tilde{\ell}=0, \tilde{k}=0$ is the minimum of the masslesss potential before shifting $\theta_2/N_1$. 
We can see this is actually sufficient by noticing that one of its chiral partners is $\varphi'_*=-\frac{\pi}{\gcd(N_1,N_2)}$ with $\tilde{\ell}'=1$. Then, the mass perturbation gives degenerate energy between them, $-m\mu\cos(\varphi_*)=-m\mu\cos(\varphi'_*)$, which is what we want. 
Substituting $\varphi=\frac{\pi}{\gcd(N_1,N_2)}$ and $\tilde{\ell}=\tilde{k}=0$ to the saddle-point equation of the massless potential, we get 
\begin{align}
\frac{2 N_2 K^{(1)} \rme^{-S_\mathrm{I}^{(1)}/N_1}}{N_1}  
\sin \left( \frac{\pi N_2' - \theta_1}{N_1} \right) 
+ \frac{2 N_1 K^{(2)} \rme^{-S_\mathrm{I}^{(2)}/N_2} }{N_2}  \sin \left( \frac{\pi N_1'  - \theta_2}{N_2} \right) = 0, \notag
\end{align}
and this curve goes through $(\theta_1,\theta_2) = (\pi N_2', \pi N_1')$.
} as expected in the large-$(N_1,N_2)$ argument (Figure~\ref{fig:BFQCDsmallmass}).

% \subsubsection{Numerical investigations}
\subsubsection{Topology change of the phase diagrams as a function of \texorpdfstring{$m$}{fermion mass}}
\label{sec:numerical_phase_diagrams_gcd!=1}

As we have done in Section~\ref{sec:numerical_phase_diagrams_gcd1} for $\gcd(N_1,N_2)=1$, we perform the numerical calculation to see how the large-mass and small-mass phase diagrams are connected for $\gcd(N_1,N_2)\ne 1$. 
Unlike the case of $\gcd(N_1,N_2)=1$, we have the discrete chiral symmetry in the massless limit, and the small-mass phase diagram has extra phase transition lines that exchange chiral broken vacua. 
Moreover, there are nontrivial constraints on the global structure for the phase diagram by anomaly matching conditions when $\gcd(N_1,N_2)\not=1$, so it would be meaningful to observe the topology-changing phenomenon also in this case.

In our computation, we consider the simplest setup: $N_1 = 2$, $N_2 = 4$, and we set their strong scales as $2K^{(1)} \rme^{-S_\mathrm{I}^{(1)}/N_1} = 2K^{(2)} \rme^{-S_\mathrm{I}^{(2)}/N_2}=1$. 
In this setup, $N'_1=1$ and $N'_2=2$, so we can take the Bezout coefficient as $M_1=0$, $M_2=1$, and this satisfies $M_1/N'_2, M_2/N'_1\in \mathbb{Z}$. 
Then, the vacuum labels (\ref{eq:change_of_discrete_labels}) are simply given by $\tilde{k}=k_1-2k_2=k_1$ and $\tilde{\ell}=k_2$ in $\bmod\, 2$, so there is no distinction between $(k_1,k_2)$ and $(\tilde{k},\tilde{\ell})$ as $\mathbb{Z}_{\gcd(2,4)}$ labels. 

% Let us see how the phase diagram changes as the mass parameter $m\mu$ varies.
\begin{figure}[t]
    \centering
    \begin{minipage}[b]{0.4\linewidth}
        \centering \subfloat[$m \mu = 0.05$]{
        \includegraphics[scale=0.275]{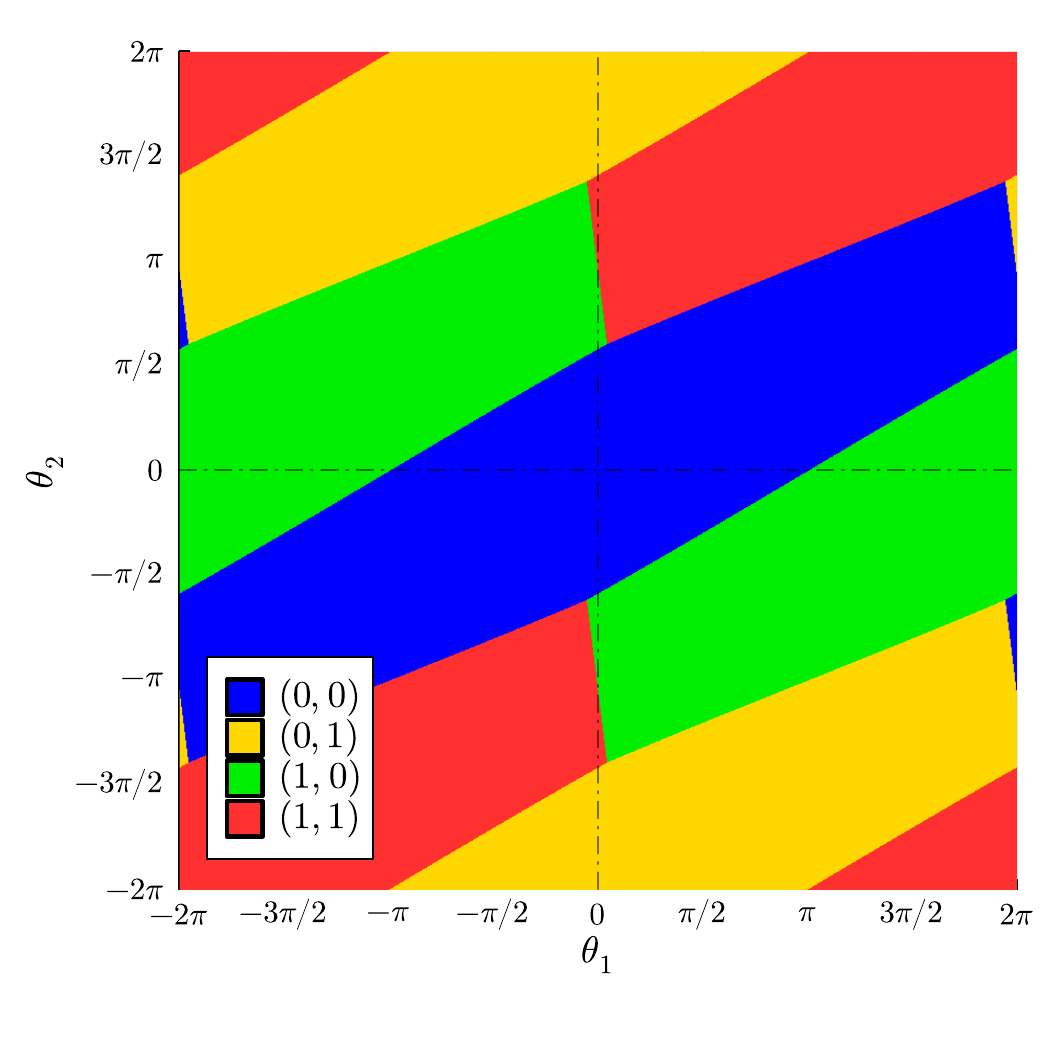}
        \label{fig:2-4_msq=0.05} 
        }
    \end{minipage}
    \begin{minipage}[b]{0.4\linewidth}
        \centering \subfloat[$m \mu = 0.2$]{
        \includegraphics[scale=0.275]{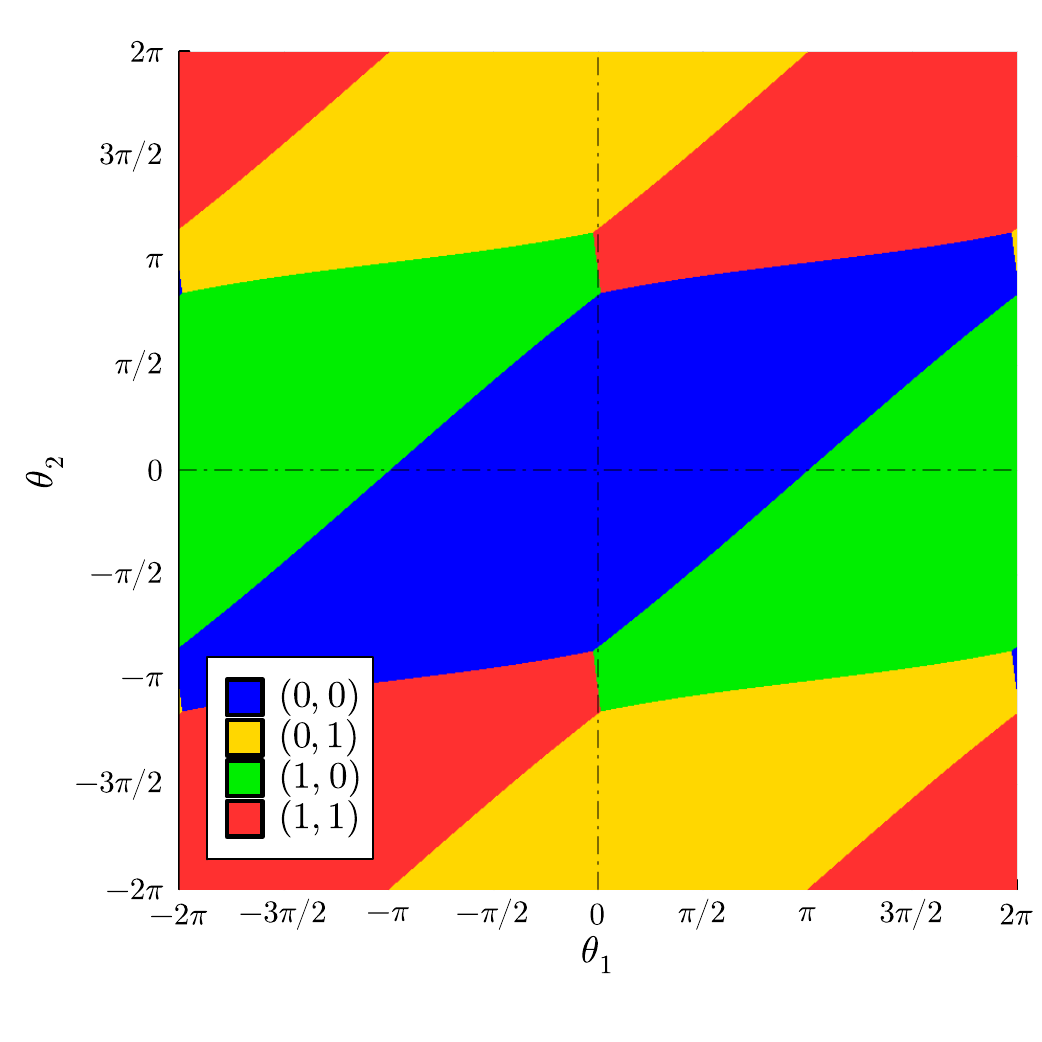}
        \label{fig:2-4_msq=0.2} 
        }
    \end{minipage}
    \\
    \begin{minipage}[b]{0.4\linewidth}
        \centering \subfloat[$m \mu = 0.3$]{
        \includegraphics[scale=0.275]{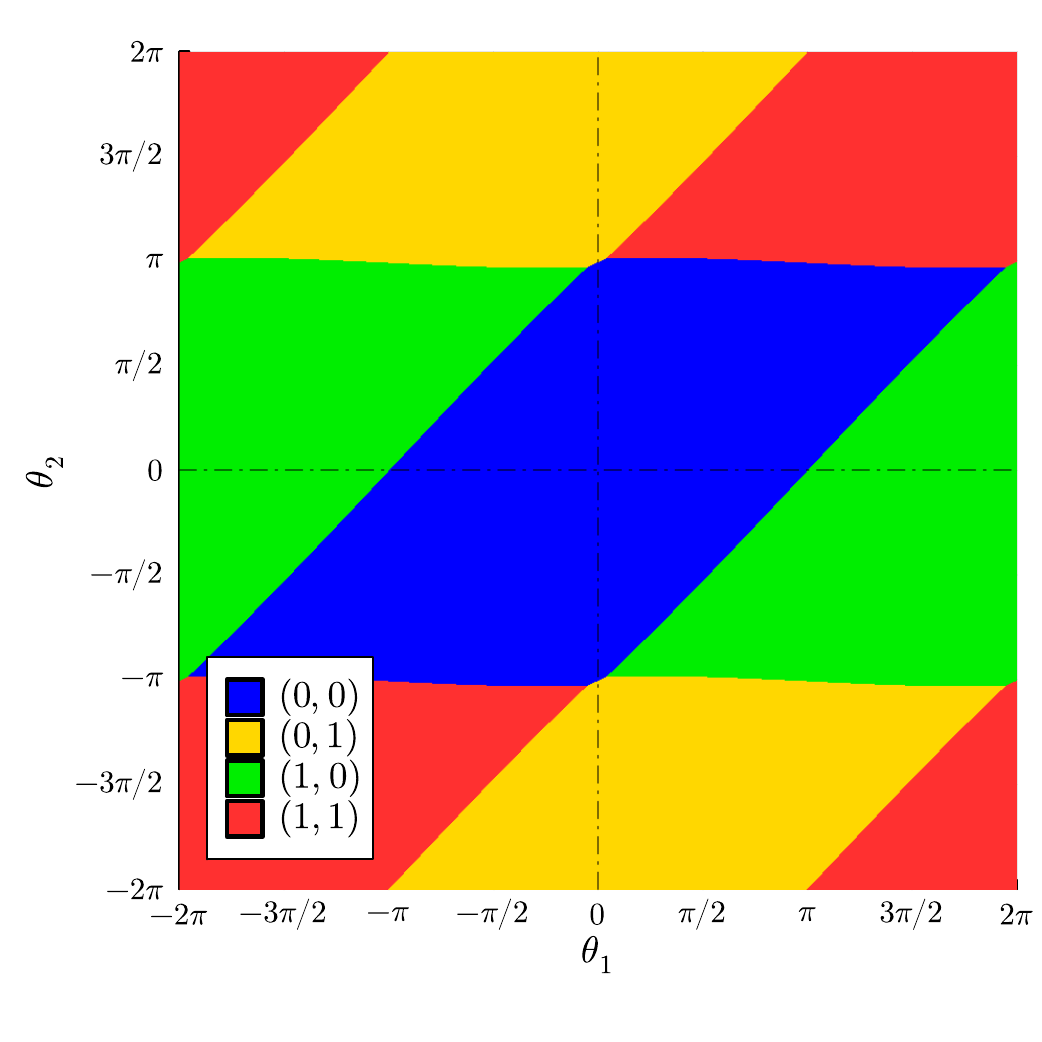}
        \label{fig:2-4_msq=0.3} 
        }
    \end{minipage}
    \begin{minipage}[b]{0.4\linewidth}
        \centering \subfloat[$m \mu = 1.0$]{
        \includegraphics[scale=0.275]{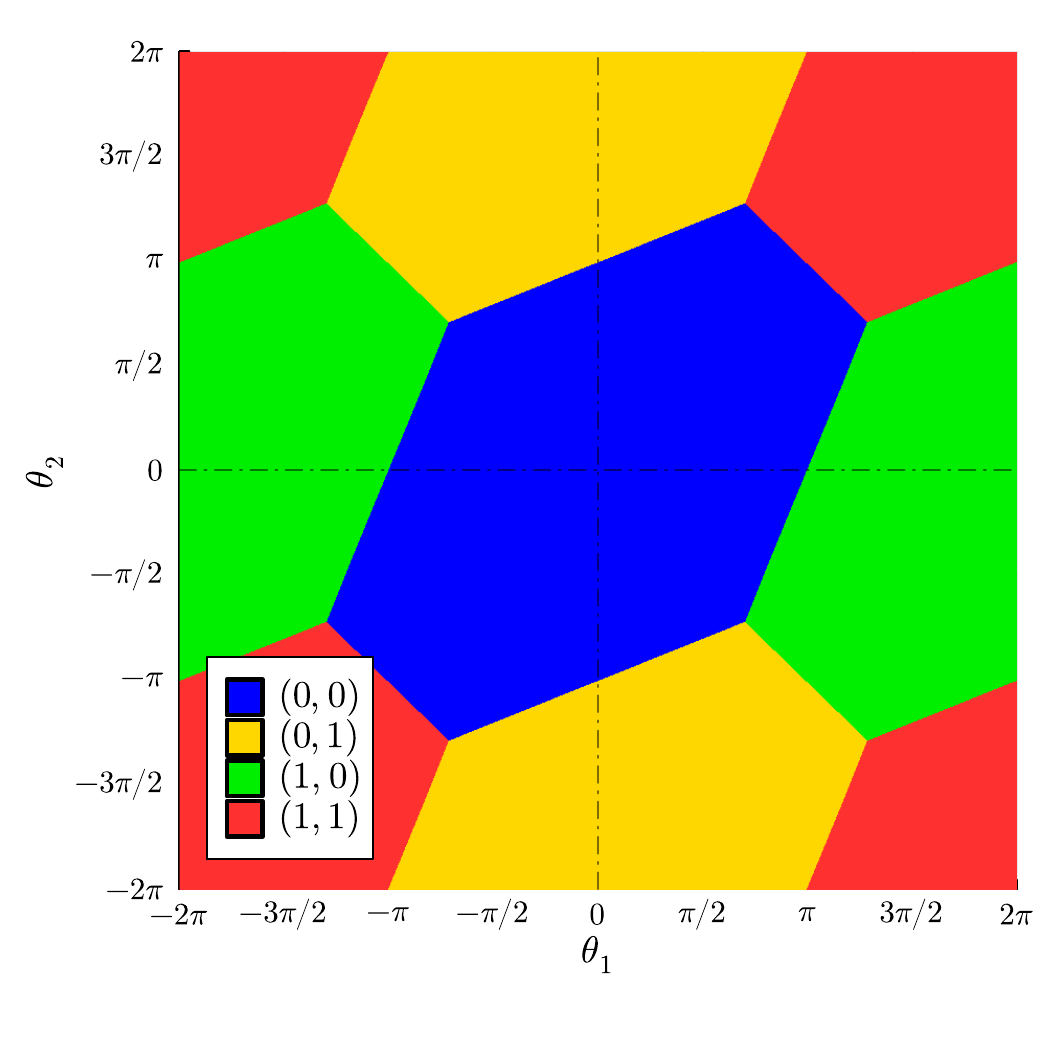}
        \label{fig:2-4_msq=1.0} 
        }
    \end{minipage}
    
    \caption{Phase diagrams on $(\theta_1,\theta_2)$ plane with $(N_1, N_2) = (2, 4)$ and several $m \mu$ in the unit of $2K^{(1)} \rme^{-S_\mathrm{I}^{(1)}/N_1} = 2K^{(2)} \rme^{-S_\mathrm{I}^{(2)}/N_2} = 1$. 
    Upper-right parts correspond to the fundamental domain $0< \theta_1 <2\pi,~0 < \theta_2 < 2\pi$.
    The values of $(k_1, k_2)$ at the global minimum of the potential \eqref{eq:eff_potential_with_k1k2} are depicted by different colors. 
    }
    \label{fig:N1=2_N2=4_K1=1_K2=1}
\end{figure}

In our analysis of the large- and small-mass limits, we found that all the phase transitions are associated with the jump of discrete labels $(\tilde{k},\tilde{\ell})\in (\mathbb{Z}_{\gcd(N_1,N_2)})^2$, and the phase transition lines are separating different $4$d SPT states. 
This is a drastically different point compared with the $\gcd(N_1,N_2)=1$ QCD(BF). 
Of course, as a logical possibility, it would have been possible that some accidental phase transition appears in the intermediate mass range, at which only the continuous field $\varphi$ jumps. 
As far as we numerically studied the case $0.05\le m\mu \le 1.0$, such an accidental transition does not happen.

Figure~\ref{fig:N1=2_N2=4_K1=1_K2=1} depicts the phase diagrams on $(\theta_1,\theta_2)$ plane with $(N_1, N_2) = (2, 4)$ at several $m \mu$.
In each panel, we determine the phases by the discrete labels $(k_1,k_2)$ of the global minimum, and we omit the values of $\varphi$ as its jump turns out to be always associated with the jump of $(k_1,k_2)$.

For a small-mass case (Figure~\ref{fig:2-4_msq=0.05}), the phase diagram slightly alters from the small-mass limit depicted in Figure~\ref{fig:BFQCDsmallmass}. When we change the $(\theta_1,\theta_2)$ along the $\theta_{\mathrm{phys}}$ direction, the phase transition occurs by the jump of $\tilde{k}=k_1$ as expected. When we change the $(\theta_1,\theta_2)$ along the perpendicular direction to $\theta_{\mathrm{phys}}$, the jump of $\tilde{\ell}=k_2$ occurs and the chiral-broken vacua are exchanged. 
When we increase $m$ and it reaches a certain mass, the reconnection (topology change) occurs at $(\theta_1, \theta_2) = (0, \pi), (2 \pi, \pi)$ (see from Figure~\ref{fig:2-4_msq=0.2} to \ref{fig:2-4_msq=0.3}). 
After the reconnection, the phase diagram converges to that of the decoupled Yang-Mills theory (Figure~\ref{fig:2-4_msq=1.0}).
Our semiclassical framework explains the topology-changing behavior speculated in Ref.~\cite{Karasik:2019bxn}.

\subsection{On the duality of semiclassics in the large-\texorpdfstring{$N_{1,2}$}{N1,2} limit}
\label{sec:comment_large_N_gcd!=1}

In this section, we make a brief comment that the duality map of coupling constants~\eqref{eq:duality_semiclssical_parameters} is valid also for the case $\gcd(N_1,N_2)\not =1$ in the large-$N_{1,2}$ limit. 

To see this, we parametrize the $2\pi N'_1N'_2$-periodic scalar as $\varphi = 2 \pi (M_1 k_1' + M_2 k_2') + \delta \varphi$ with $-\pi< \delta \varphi<\pi$ and $(k'_1,k'_2)\in \mathbb{Z}_{N'_1}\times \mathbb{Z}_{N'_2}$. 
Using the extension, $\mathbb{Z}_{N'_i}\to \mathbb{Z}_{N_i}\to \mathbb{Z}_{\gcd(N_1,N_2)}$, the discrete labels $k'_i\in \mathbb{Z}_{N_i}$ and $k_i\in \mathbb{Z}_{\gcd(N_1,N_2)}$ are combined into $(\hat{k}_1, \hat{k}_2) \in \mathbb{Z}_{N_1} \times \mathbb{Z}_{N_2}$, which is exactly what we have done in the discussion for the large-mass limit in Section~\ref{sec:large-mass_gcd!=1}.  
As a result, the potential can be expressed as 
\begin{align}
    V &= - m \mu \cos (\delta \varphi) - 2K^{(1)} \rme^{-S_\mathrm{I}^{(1)}/N_1} \cos \left( \frac{2 \pi \hat{k}_1 - \theta_1}{N_1} + \frac{N_2}{N_1} \delta \varphi  \right) \notag \\
    &~~~~~- 2K^{(2)} \rme^{-S_\mathrm{I}^{(2)}/N_2} \cos \left( \frac{2 \pi \hat{k}_2 - \theta_2}{N_2} + \frac{N_1}{N_2} \delta \varphi  \right).
\end{align}
The rest of the argument is completely the same as the one in Section~\ref{sec:recovering_large_N_from_semiclassics}.
One can find the minimum $\delta \varphi \sim O(1/N_1,1/N_2)$ and have the identical expression for the quadratic form of the energy density with the coefficients (\ref{eq:from_semiclassics_to_large_N}).
We can repeat the same discussion for the daughter side, $SU(N_1)\times SU(N_2-N_1)$ QCD(BF), and we find the concrete form of the duality map~\eqref{eq:duality_semiclssical_parameters}.

Lastly, let us give one cautionary remark. 
Assume that we are studying $SU(N_1)\times SU(N_2)$ QCD(BF) at sufficiently small fermion mass, then we can repeatedly apply the dualities and the gauge-group ranks changes following the Euclidean algorithm, $(N_1,N_2)\to (N_1,N_2-N_1)\to \cdots$. 
For $N=\operatorname{gcd}(N_1,N_2) \neq 1$, the last step of the duality cascade is the duality from $SU(N) \times SU(2N)$ QCD(BF) to $SU(N) \times SU(N)$ QCD(BF), and the latter theory is the equal-rank QCD(BF). 
In this paper, we assumed $N_1\not=N_2$ to construct the $2$d semiclassical theory, so we cannot immediately apply our duality map~\eqref{eq:duality_semiclssical_parameters} to the last step of the cascade, which requires an extra work. 
In our previous paper~\cite{Hayashi:2023wwi}, we construct the $2$d semiclassical description for equal-rank QCD(BF). 
Although there is a difference in the renormalization scheme, the large-$N$ expansion of the equal-rank QCD(BF) gives a parallel expression to (\ref{eq:from_semiclassics_to_large_N}) as the functional form of the energy density.
Thus, it is possible to match these vacuum energies in a similar manner at least formally.
To gain a deep insight into the last step of the duality cascade, we need a more refined understanding of the semiclassical approach including the subtlety of renormalization scheme.

\subsection{Comments on the domain wall and the anomaly inflow}
\label{sec:comments_on_DW}

We here give the brief comment on properties of domain walls at the phase boundaries. 
In Section~\ref{sec:limits_gcd!=1}, we observed that all the phase transitions are characterized by the jump of the discrete vacuum labels $(\tilde{k},\tilde{\ell})$ when $\gcd(N_1,N_2)\not =1$. 
These vacuum labels are identified with the labels $(k,\ell)$ for the $4$d SPT states \eqref{eq:BFQCD-SPT} after a suitable linear transformation~\eqref{eq:newBasis_discreteLabels}, and thus the domain walls in the $4$d setup should support nontrivial dynamics to cancel the anomaly inflow (see e.g. Refs.~\cite{Gaiotto:2017yup, Gaiotto:2017tne, Sulejmanpasic:2016uwq, Anber:2015kea, Komargodski:2017smk, Cox:2019aji}). 
Let us uncover how this is realized in our $T^2$-compactified setup with the baryon-'t~Hooft flux (see also Section~5 of Ref.~\cite{Hayashi:2023wwi}).

We first note that the domain wall in $4$d setup is the dynamical object, which would be spontaneously created if we perturb the $\theta$ parameters to the position-dependent one. 
In the $2$d semiclassical theory, however, the vacua are specified by the discrete labels, which suggests that the relevant degrees of freedom for creating domain walls are heavy and already integrated out in this effective theory. 
Therefore, we need to introduce the loop operator in $2$d that forces the jump of the discrete labels, and this is the counterpart of the $4$d domain wall. 
We can see that this loop-operator insertion satisfies the requirement of anomaly inflow. 

To understand this, it is convenient to consider the decoupling limit $m\to \infty$, and then the vacuum labels $(\tilde{k},\tilde{\ell})$ are reduced to the original ones, $(k_1,k_2)\in \mathbb{Z}_{N_1}\times \mathbb{Z}_{N_2}$, which constrain the total topological charges of the fractional-instanton gas. 
Since $(k_1,k_2)$ specifies the $1$-form symmetry charge of the $SU(N_1)\times SU(N_2)$ pure Yang-Mills theory, their Wilson loops $W_1$ and $W_2$ serve as domain walls connecting vacua with different $(k_1,k_2)$. 
When the dynamics of bifundamental fermion $\Psi$ is turned on with finite mass $m$, the part of $1$-form symmetry loses its information as $\Psi$ caries the same gauge charge of $W_1 W_2^{-1}$. 
The remnant charge of the $1$-form symmetry is specified by the label $k \in \mathbb{Z}_{\gcd(N_1,N_2)}$ in the $4$d SPT action~\eqref{eq:BFQCD-SPT}. 
Importantly, $\Psi$ carries the fractional charge of $U(1)/\mathbb{Z}_{\lcm(N_1,N_2)}$, and this also remains as the robust information described by the label $\ell$ in \eqref{eq:BFQCD-SPT}. 

When two states are discriminated by the $1$-form symmetry charge $k$, then we have to introduce the appropriate Wilson loop to have the domain wall in the $2$d effective theory. 
When two states have the same $1$-form charge $k$ but different levels for $\ell$, then the domain wall can be created either by insertion of the Wilson loops or by excitation of the $\Psi$ quanta. 
Thus, the anomaly inflow to the domain wall in our $T^2$ compactified setup is realized by the fact that the domain wall becomes a non-dynamical object requiring the suitable amount of the loop-operator insertion.

\acknowledgments

The authors thank Zohar Komargodski for discussion and useful comments on the early draft.
The work of Y. T. was supported by Japan Society for the Promotion of Science (JSPS) KAKENHI Grant numbers, 22H01218, 23K22489, and by Center for Gravitational Physics and Quantum Information (CGPQI) at Yukawa Institute for Theoretical Physics.
Y.~H. was supported by JSPS Research Fellowship for Young Scientists Grant No.~23KJ1161.

\appendix

\section{Counting the low-energy fermions for QCD(BF) on \texorpdfstring{$\mathbb{R}^2\times T^2$}{R2xT2}}
\label{app:counting_BFQCD}

In the main text, we used the index theorem (\ref{eq:low_energy_modes_indextheorem}) to count the number of low-energy modes arising from the bifundamental fermion under the $T^2$-compactification.
Here, we present an alternative derivation of the number of low-energy modes by directly solving the zero-mode equation on the torus,
\begin{align}
    \left[ \gamma_3 \partial_3 + \gamma_4 \left( \partial_4 + \im A_{V,4} \right) \right] \Psi(x_3,x_4) = 0,
\end{align}
where the boundary conditions are given by
\begin{align}
    \Psi(x_3+L,x_4) &= \rme^{\im \left( \frac{2 \pi}{N_1 L} - \frac{2 \pi}{N_2 L} \right) x_4} S_1^\dagger \Psi(x_3,x_4) S_2, \\
    \Psi(x_3,x_4+L) &= C_1^\dagger \Psi(x_3,x_4) C_2.  
\end{align}
Here, the $U(1)_V$ background field is 
\begin{align}
    A_V = \left( \frac{2 \pi}{N_1 L^2} - \frac{2 \pi}{N_2 L^2} \right) x_3 dx_4. 
\end{align}

The normalizable solution to the zero-mode equation takes the form of
\begin{align}
    \Psi_+(x_3,x_4) = \rme^{- \frac{\pi}{L^2} \left( \frac{1}{N_1} - \frac{1}{N_2} \right) x_3^2} \Tilde{\Psi}_+(x_3 - \im x_4),
\end{align}
where $\Psi_+$ is a positive eigenmode\footnote{There is no normalizable zero-mode with a negative eigenvalue of $\im \gamma_3 \gamma_4$: $\im \gamma_3 \gamma_4 \Psi_- = -\Psi_-$. Indeed, if $\Psi_-$ were a normalizable zero-mode, we would have
\begin{align}
    0 &= \int_{T^2} \diff^2x~ \left| \left[ \partial_3 + \im \left( \partial_4 + \im A_{V,4} \right) \right] \Psi_-(x_3,x_4) \right|^2 \notag \\
    &= \int_{T^2} \diff^2x~ \left\{ \left|  \partial_3 \Psi_- \right|^2 + \left|  \left( \partial_4 + \im A_{V,4} \right) \Psi_- \right|^2 +  \left( \frac{2 \pi}{N_1 L^2} - \frac{2 \pi}{N_2 L^2} \right) \left| \Psi_- \right|^2\right\},
\end{align}
and this implies $\Psi- = 0$.
} of $\im \gamma_3 \gamma_4$: $\im \gamma_3 \gamma_4 \Psi_+ = +\Psi_+$, and $\Tilde{\Psi}_+(x_3 - \im x_4)$ is a holomorphic function in $x_3 - \im x_4$.
The second boundary condition $\Psi(x_3,x_4+L) = C_1^\dagger \Psi(x_3,x_4) C_2$ can be written as $\Psi_{i,j}(x_3,x_4+L) = \rme^{ - \frac{2 \pi \im}{N_1 L } (i - 1) + \frac{2 \pi \im }{N_2 L } (j - 1) } \Psi_{i,j}(x_3,x_4) $, 
so we can expand the holomorphic function $\Tilde{\Psi}_+(x_3 - \im x_4)$ with Fourier coefficients $c_n^{i,j}$,
\begin{align}
    \Tilde{\Psi}_{+;i,j} (x_3 - \im x_4) = \rme^{ \left[ \frac{2 \pi }{N_1 } (i - 1) - \frac{2 \pi }{N_2 } (j - 1)\right] \left( \frac{x_3 - \im x_4}{L} \right) } \left\{ \sum_{n \in \mathbb{Z}} c_n ^{i,j} \rme^{\frac{2 \pi n}{L} (x_3 - \im x_4) } \right\}.
\end{align}
The first boundary condition $\Psi(x_3+L,x_4) = \rme^{\im \left( \frac{2 \pi}{N_1 L} - \frac{2 \pi}{N_2 L} \right) x_4} S_1^\dagger \Psi(x_3,x_4) S_2$ relates $\Psi_{i,j}$ to $\Psi_{i-1,j-1}$ as follows, 
\begin{align}
   \Tilde{\Psi}_{+;i,j} (x_3 - \im x_4 + L) = \rme^{ \pi \left(\frac{1}{N_1} - \frac{1}{N_2} \right) + \left( \frac{2 \pi}{N_1 L} - \frac{2 \pi}{N_2 L} \right) (x_3 - \im x_4)}  \Tilde{\Psi}_{+;i-1,j-1} (x_3 - \im x_4),
\end{align}
where $i = 0$ (resp.~$j=0$) is understood as $i = N_1$ (resp.~$j=N_2$).
This can be rewritten in terms of the Fourier coefficients $c_n^{i,j}$:
For $1<i \leq N_1$ and $1<j \leq N_2$,
\begin{align}
    c_n^{i,j} =  \rme^{ \pi \left(\frac{1}{N_1} - \frac{1}{N_2} \right)} \rme^{ -\left[ \frac{2 \pi }{N_1 } (i - 1) - \frac{2 \pi }{N_2 } (j - 1)\right]  } \rme^{ - 2 \pi n} c_n^{i-1,j-1}. \label{eq:BC-Fourier-type-1}
\end{align}
The exceptions are listed as follows:
\begin{align}
    c_n^{1,j} &=  \rme^{ \pi \left(\frac{1}{N_1} - \frac{1}{N_2} \right)} \rme^{  - \frac{2 \pi }{N_2 } (j - 1) } \rme^{ - 2 \pi n} c_{n-1}^{N_1,j-1}~~~ (1<j \leq N_2), \label{eq:BC-Fourier-type-2} \\
    c_n^{i,1} &=  \rme^{ \pi \left(\frac{1}{N_1} - \frac{1}{N_2} \right)} \rme^{ -\frac{2 \pi }{N_1 } (i - 1)  } \rme^{ - 2 \pi n} c_{n+1}^{i-1,N_2}~~~ (1<i \leq N_1), \label{eq:BC-Fourier-type-3} \\
    c_n^{1,1} &= \rme^{ \pi \left(\frac{1}{N_1} - \frac{1}{N_2} \right)} \rme^{ - 2 \pi n}  c_n^{N_1,N_2}. \label{eq:BC-Fourier-type-4}
\end{align}

Now, let us count the number of independent solutions.
We first notice that the first boundary condition relates $\Psi_{i,j}$ to $\Psi_{i-1,j-1}$, so we can decompose the $N_1 \times N_2$ matrix into the $\operatorname{gcd}(N_1,N_2)$ classes.
For example, we look at the class including $\Psi_{1,1}$: $\{ \Psi_{1,1}, \Psi_{2,2}, \cdots, \}$.
The periodicity of the sequence $\Psi_{1,1} \rightarrow \Psi_{2,2} \rightarrow \cdots$ is
$\operatorname{lcm}(N_1,N_2)$, as the subscripts can be understood as those of $\mathbb{Z}_{N_1} \times \mathbb{Z}_{N_2}$.
Thus, the $N_1 \times N_2$ matrix $\{ \Psi_{i,j} \}$ can be indeed decomposed into $\operatorname{gcd}(N_1,N_2)$ classes, since each class consists of $\operatorname{lcm}(N_1,N_2) = \frac{N_1 N_2}{\operatorname{gcd}(N_1,N_2)}$ elements.

We then count the number of independent solutions for each class.
For a sequence $\Psi_{i,j} \rightarrow \Psi_{i+1,j+1} \rightarrow \cdots$, 
let us call
\begin{itemize}
    \item jump $\Psi_{N_1,j} \rightarrow \Psi_{1,j+1}$: transition of the 1st index 
    \item jump $\Psi_{i,N_2} \rightarrow \Psi_{i+1,1}$: transition of the 2nd index 
\end{itemize}
Because the number of elements is $\frac{N_1 N_2}{\operatorname{gcd}(N_1,N_2)}$, the sequence $\Psi_{1,1} \rightarrow \Psi_{2,2} \rightarrow \cdots$ undergoes $N_2/\operatorname{gcd}(N_1,N_2)$ transitions of the 1st index as well as $N_1/\operatorname{gcd}(N_1,N_2)$ transitions of the 2nd index.
The boundary conditions (\ref{eq:BC-Fourier-type-1})--(\ref{eq:BC-Fourier-type-4}) relate the Fourier coefficients $c_n^{i,j}$ as,
\begin{align}
    c_n^{i,j} \rightarrow c_n^{i+1,j+1} \rightarrow \cdots \rightarrow c_{n + \frac{N_2 - N_1}{\operatorname{gcd}(N_1,N_2)}}^{i,j}, 
\end{align}
since the transition of 1st index relates $c_n$ to $c_{n+1}$, and that of 2nd index relates $c_n$ to $c_{n-1}$.
Therefore, each class includes $\frac{N_2 - N_1}{\operatorname{gcd}(N_1,N_2)}$ independent solutions.

The $N_1 \times N_2$ matrix  $\{ \Psi_{i,j} \}$ has $\operatorname{gcd}(N_1,N_2)$ classes, and one class has $\frac{N_2 - N_1}{\operatorname{gcd}(N_1,N_2)}$ independent solutions.
In total, the number of zeromodes is,
\begin{align}
    \operatorname{gcd}(N_1,N_2) \times \frac{N_2 - N_1}{\operatorname{gcd}(N_1,N_2)} = N_2 - N_1,
\end{align}
which reproduces the result from the index theorem (\ref{eq:low_energy_modes_indextheorem}).
We therefore get $N_2 - N_1$ low-energy modes before taking the 4-fermi vertex into account.

\section{More on the duality between \texorpdfstring{$SU(2) \times SU(5)$}{SU(2)xSU(5)} and \texorpdfstring{$SU(2) \times SU(3)$}{SU(2)xSU(3)} QCD(BF)}
\label{app:more_phase_diagrams}

\subsection{Quantifying the mild violation of the duality relation}
\label{sec:bdy_is_not_straight}

We have numerically observed the validity of the duality relation in Section~\ref{sec:duality_numerics}, and we found the surprisingly good agreement between the parent theory, $SU(N_1)\times SU(N_2)$ QCD(BF), and the daughter theory, $SU(N_1)\times SU(N_2-N_1)$ QCD(BF). 
In this appendix, we are going to uncover the reason why the duality relation works so well by quantifying the violation of the duality: We are going to demonstrate that the violation is the $O(N_2^{-4})$ effect. 

When we consider the duality map~\eqref{eq:KK_duality_onedirection} at $m=m^*$, we obtain 
\begin{align}
    SU(N_1)&\times SU(N_2) \mathrm{~QCD(BF)~on~} (\theta_1,\theta_2)~\mathrm{at}~m=m^* \notag \\
    &\longrightarrow ~ SU(N_1)\times SU(N_2-N_1) ~\mathrm{pure~YM~on~} (\theta_1-\theta_2,\theta_2). 
\end{align}
We note that the phase boundaries on the daughter side consist only of the straight lines, and, in particular,  they are given by $\theta_1-\theta_2 = \pi~(\operatorname{mod} 2\pi)$ or by $\theta_2 = \pi~(\operatorname{mod} 2\pi)$. 
Therefore, if the duality relation~\eqref{eq:KK_duality_onedirection} were exact, the parent theory would also have the same phase boundaries described by those straight lines. 
By evaluating how the actual phase boundaries of the parent theory deviate from those lines, let us measure the non-exactness of the duality relation.

We focus on the horizontal line $\theta_2=\pi$, which goes through $(\theta_1, \theta_2) = (\pi, \pi)$.
At $(\theta_1, \theta_2) = (\pi, \pi)$, the potential $V[\varphi]$ of the parent theory, given in (\ref{eq:potential_GCD1_BFQCD}), is invariant under the CP symmetry transformation,
\begin{align}
    \varphi \rightarrow -\varphi + 2 \pi (M_1+M_2),
\end{align}
because $N_1 M_2 = 1 ~(\operatorname{mod} N_2)$ and $N_2 M_1= 1 ~(\operatorname{mod} N_1)$.
Recall that the Bezout coefficients~$(M_1,M_2)$ are taken so that $M_1/N_2', M_2/N_1'\in \mathbb{Z}$. %, \eqref{eq:def_K1_K2}.
Thus, we expect two-fold degenerate vacua  at $(\theta_1, \theta_2) = (\pi, \pi)$: $\varphi = \varphi_*, -  \varphi_* + 2 \pi  (M_1+M_2)$.

If the duality were exact, the phase boundary including $(\theta_1, \theta_2) = (\pi, \pi)$ at $m\mu=m^*\mu$ has to be a horizontal line $\theta_2 = \pi$, and we can obtain $\varphi_*$ under this assumption. 
As the two-fold degeneracy should be robust under the perturbation $\theta_1 = \pi + \epsilon$, we must have $V[\varphi_*+O(\ve)]|_{\theta_1=\pi+\ve,\theta_2=\pi}=V[2\pi(M_1+M_2)-\varphi_* + O(\ve)]|_{\theta_1=\pi+\ve,\theta_2=\pi}$. 
Expanding this equality at the lowest order of $\ve$, this requirement constrains $\varphi_*$ as,
\begin{align}
    \sin \left( \frac{N_2 \varphi_* - \pi}{N_1} \right) = 0.
\end{align}
Since one of the vacua is in the same phase as $\varphi = 0$ at $(\theta_1, \theta_2) = (0, 0)$, we deduce $\varphi_* = \pi/N_2$.
To minimize the potential $V[\varphi]$ of (\ref{eq:potential_GCD1_BFQCD}) at $\varphi_* = \pi/N_2$, the following part
\begin{align}
     V_{\mathrm{mass}+SU(N_2)}[\varphi] = - m^* \mu \cos (\varphi)  - 2K^{(2)} \rme^{-S_\mathrm{I}^{(2)}/N_2} \cos \left( \frac{N_1 \varphi - \pi}{N_2} \right) \label{eq:potential_mass_sun2_part}
\end{align}
should have a (local) minimum at $\varphi = \pi/N_2$. In other words, we can estimate the topology-changing point $m^*\mu$ by this requirement, which gives 
\begin{equation}
    m^*\mu = 
    \frac{N_1}{N_2}\frac{\sin\left(\frac{(N_2-N_1)\pi}{N_2^2}\right)}{\sin\left(\frac{\pi}{N_2}\right)}
    \cdot 2K^{(2)}\rme^{-S^{(2)}_{\mathrm{I}}/N_2}. 
\end{equation}
We note that this is consistent with the large-$N$ formula~\eqref{eq:largeN_criticalmass}.

Now, let us check if the doubly-degenerate vacua are preserved for general $\theta_1$ at $\theta_2=\pi$.
First, we look at the vacuum branch that takes $\varphi = \pi/N_2$ at $\theta_1 = \pi$. Let us call this minimum the vacuum A and denote $\varphi = \varphi_*^{(A)}(\theta_1)$.
The potential consists of $V_{\mathrm{mass}+SU(N_2)}[\varphi]$ and $- 2K^{(1)} \rme^{-S_\mathrm{I}^{(1)}/N_1} \cos \left( \frac{N_2 \varphi - \theta_1 }{N_1} \right)$.
The former term has a minimum at $\varphi = \pi/N_2$, and the latter term has a minimum at $\varphi = \theta_1/N_2$.
The vacuum branch is constructed from the superposition of these minima, and $\varphi_*^{(A)}(\theta_1)$ is located between $\varphi = \pi/N_2$ and $\varphi = \theta_1/N_2$ from these two contributions.

Next, we compare this vacuum with the other vacuum, which is $\varphi = - \pi/N_2 + 2 \pi (M_1+M_2)$ at $\theta_1 = \pi$.
We call this minimum the vacuum B and denote $\varphi = \varphi_*^{(B)}(\theta_1)$.
Similarly, the latter vacuum is located between 
$\varphi = - \pi/N_2 + 2 \pi (M_1+M_2)$ (from $V_{\mathrm{mass}+SU(N_2)}[\varphi]$) and $\varphi = (\theta_1 - 2 \pi)/N_2 + 2 \pi (M_1+M_2)$ (from the $SU(N_1)$ part).
For a comparison, let us shift $\varphi$ as $\varphi \rightarrow \varphi + 2\pi/N_2 - 2 \pi (M_1+M_2)$, and we write $\varphi = \tilde{\varphi} + 2\pi/N_2 - 2 \pi (M_1+M_2)$.
With this shift, the vacuum B, $\varphi_*^{(B)}(\theta_1)$, is located between $\tilde{\varphi} = \pi/N_2$ and $\tilde{\varphi} = \theta_1/N_2$, which seems to be the same as the vacuum A $\varphi = \varphi_*^{(A)}(\theta_1)$.
However, this correspondence is not exact.
The vacuum A is the minimum around $\varphi = \pi/N_2$ of the potential
\begin{align}
     V[\varphi]
     &= V_{\mathrm{mass}+SU(N_2)}[\varphi]  - 2 K^{(1)} \rme^{-S_\mathrm{I}^{(1)}/N_1} \cos \left( \frac{N_2 \varphi - \theta_1}{N_1} \right),
\end{align}
On the other hand, the potential in terms of the shifted  $\tilde{\varphi}$ is,
\begin{align}
     V[\varphi &= \tilde{\varphi} + 2\pi/N_2 - 2 \pi (M_1+M_2)] \notag \\
     &= V_{\mathrm{mass}+SU(N_2)}[-\tilde{\varphi} + 2\pi/N_2]  - 2 K^{(1)} \rme^{-S_\mathrm{I}^{(1)}/N_1} \cos \left( \frac{N_2 \tilde{\varphi} - \theta_1}{N_1} \right),
\end{align}
and the vacuum B is the minimum near $\tilde{\varphi} = \pi/N_2$.
The two-fold degeneracy of these vacua, $\varphi = \varphi_*^{(A)}(\theta_1)$ and $ \varphi = \varphi_*^{(B)}(\theta_1)$, is preserved if $V_{\mathrm{mass}+SU(N_2)}[\varphi]$ is symmetric under the reflection over the line $\varphi = \pi/N_2$: $\varphi \rightarrow -\varphi  + 2\pi/N_2$. 
However, this is not true: The potential can be expressed as
\begin{align}
&V_{\mathrm{mass}+SU(N_2)}\left[\frac{\pi}{N_2} + \delta \varphi\right] \notag\\
     &= \left[ - m^* \mu \cos \left( \frac{\pi}{N_2} \right) \cos \left( \delta \varphi \right)  - 2K^{(2)} \rme^{-S_\mathrm{I}^{(2)}/N_2} \cos \left( \frac{(N_1 - N_2)\pi}{N_2^2} \right)  \cos \left( \frac{N_1}{N_2} \delta \varphi \right)   \right] \notag \\
     & +\left[  m^* \mu \sin \left( \frac{\pi}{N_2} \right) \sin \left( \delta \varphi \right)  + 2K^{(2)} \rme^{-S_\mathrm{I}^{(2)}/N_2} \sin \left( \frac{(N_1 - N_2)\pi}{N_2^2} \right)  \sin \left( \frac{N_1}{N_2} \delta \varphi \right)   \right],
\end{align}
where the first line is the $\delta \varphi$-even, and the last line is $\delta \varphi$-odd.
Although we have chosen $m^*\mu$ so that the $O(\delta \varphi)$ term vanishes, the higher-order $\delta \varphi$-odd term $O((\delta \varphi)^3)$ does not.

Thus, the statement at the topology-changing point does not hold for finite $N_1$ and $N_2$ in an exact sense.
However, we note that $\delta \varphi =O(\pi/N_2)$, and thus the amount of the asymmetry in the above potential can be estimated as $O(N_2^{-4})$. 
This smallness would explain why the violation of the duality relation is so mild even for finite parameters.

\subsection{Phase diagrams in the hierarchical limits: \texorpdfstring{$\Lambda_1\ll \Lambda_2$}{Lambda1<<Lambda2} and \texorpdfstring{$\Lambda_1\gg \Lambda_2$}{Lambda1>>Lambda2}}
\label{sec:phase-diag_hierarchy}

In Section~\ref{sec:duality_numerics}, we numerically examined the duality map from $SU(2)\times SU(5)$ QCD(BF) to $SU(2)\times SU(3)$ QCD(BF), when the strong scales of the parent theory satisfies $\Lambda_1 \sim \Lambda_2$. 
In this appendix, we expand our analysis to the case when these two scales of the parent theory are well separated: $\Lambda_1\ll \Lambda_2$ and $\Lambda_1\gg \Lambda_2$. 
In what follows, we restate the key points and show numerical results supporting the duality relation.

\subsubsection*{Case: $\Lambda_1 \ll \Lambda_2$}

\begin{landscape}
\begin{figure}[ht]
    \begin{minipage}[b]{0.196\linewidth}
        \centering \subfloat[$m \mu = 0.5$]{
        \includegraphics[scale=0.265]{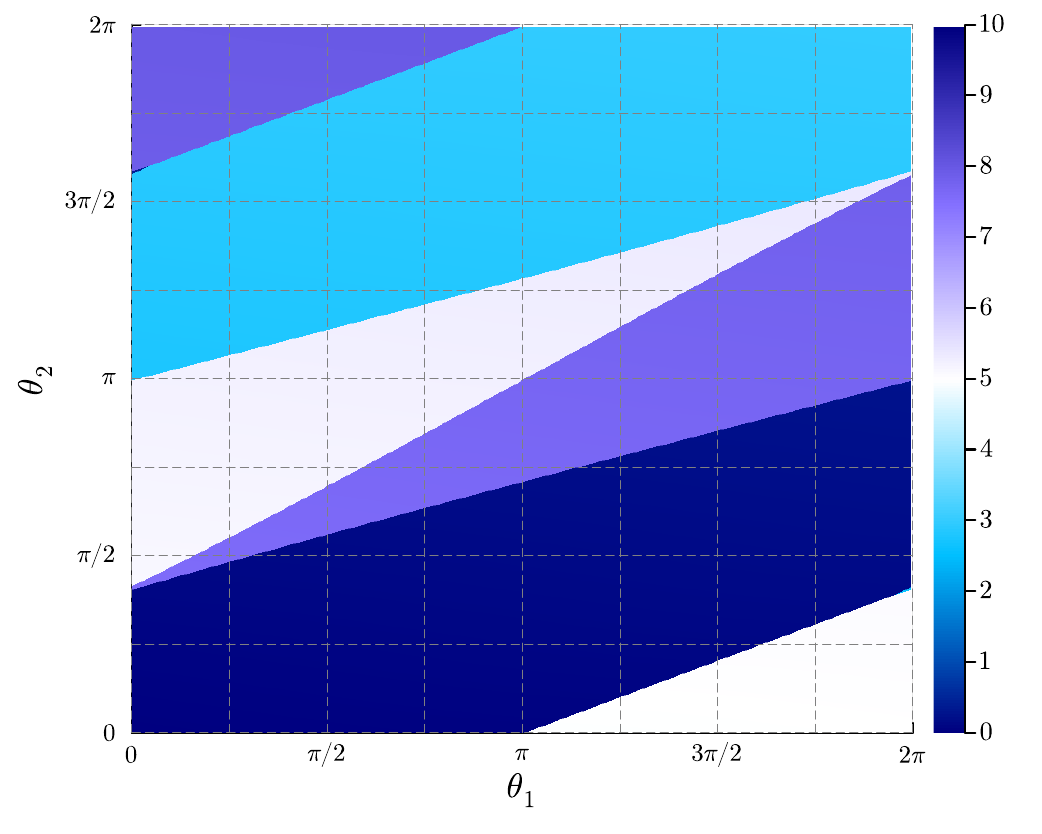}
        \label{fig:N1=2_N2=5_Msq0.5_K1=1_K2=100} 
        }
        \vspace{2ex}
    \end{minipage}
    \begin{minipage}[b]{0.196\linewidth}
        \centering \subfloat[$m \mu = 2.0$]{
        \includegraphics[scale=0.265]{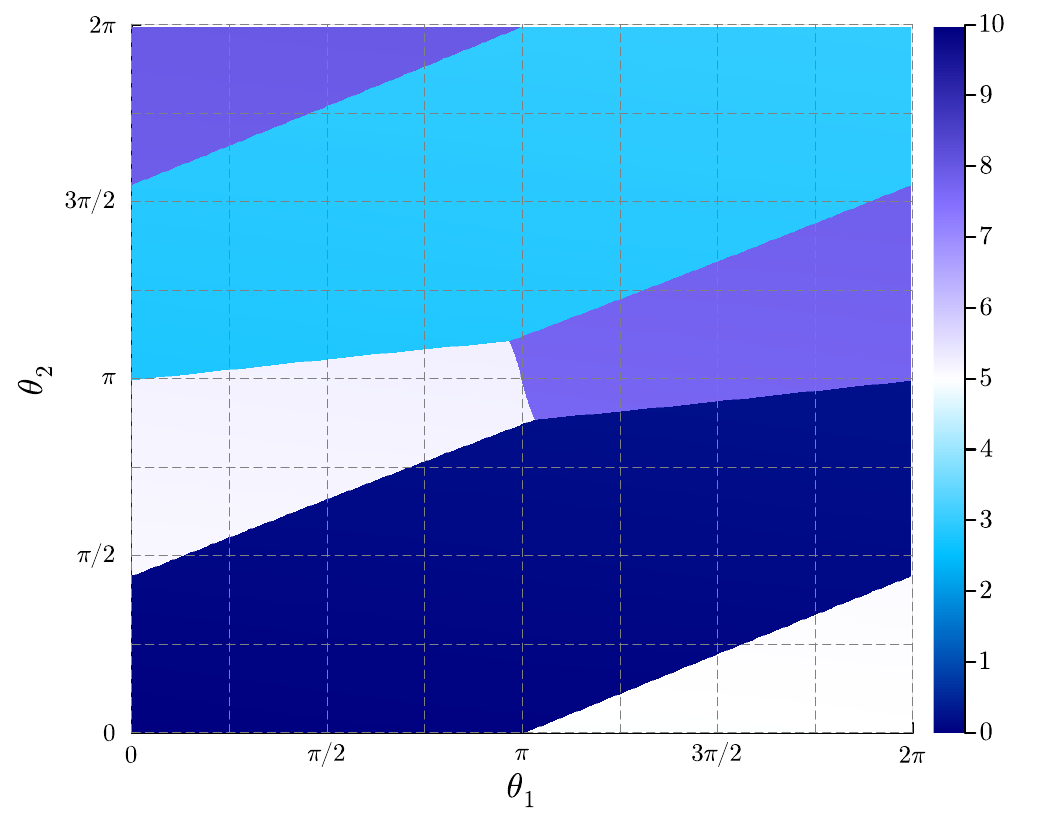}
        \label{fig:N1=2_N2=5_Msq2.0_K1=1_K2=100} 
        }
        \vspace{2ex}
    \end{minipage}
    \begin{minipage}[b]{0.196\linewidth}
        \centering \subfloat[$m \mu = 5.0$]{
        \includegraphics[scale=0.265]{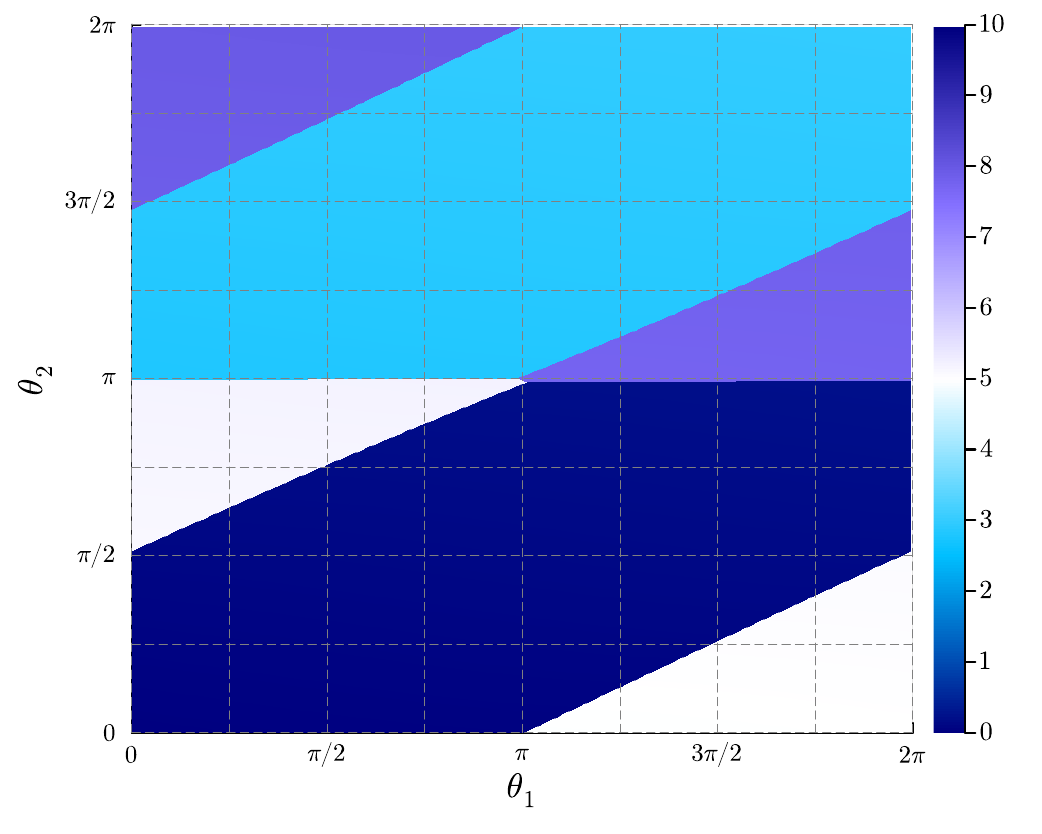}
        \label{fig:N1=2_N2=5_Msq5.0_K1=1_K2=100} 
        }
        \vspace{2ex}
    \end{minipage}
    \begin{minipage}[b]{0.196\linewidth}
        \centering \subfloat[$m \mu = 20.0$]{
        \includegraphics[scale=0.265]{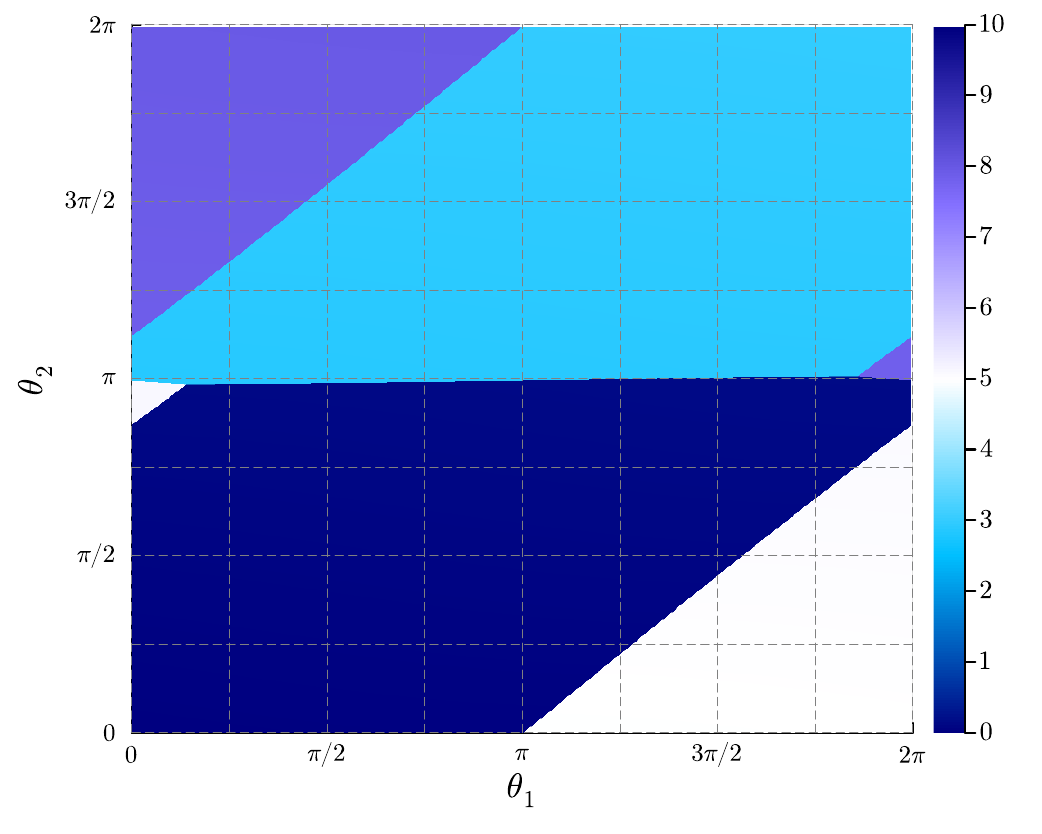}
        \label{fig:N1=2_N2=5_Msq20.0_K1=1_K2=100} 
        }
        \vspace{2ex}
    \end{minipage}
    \begin{minipage}[b]{0.196\linewidth}
        \centering \subfloat[$m \mu = 50.0$]{
        \includegraphics[scale=0.265]{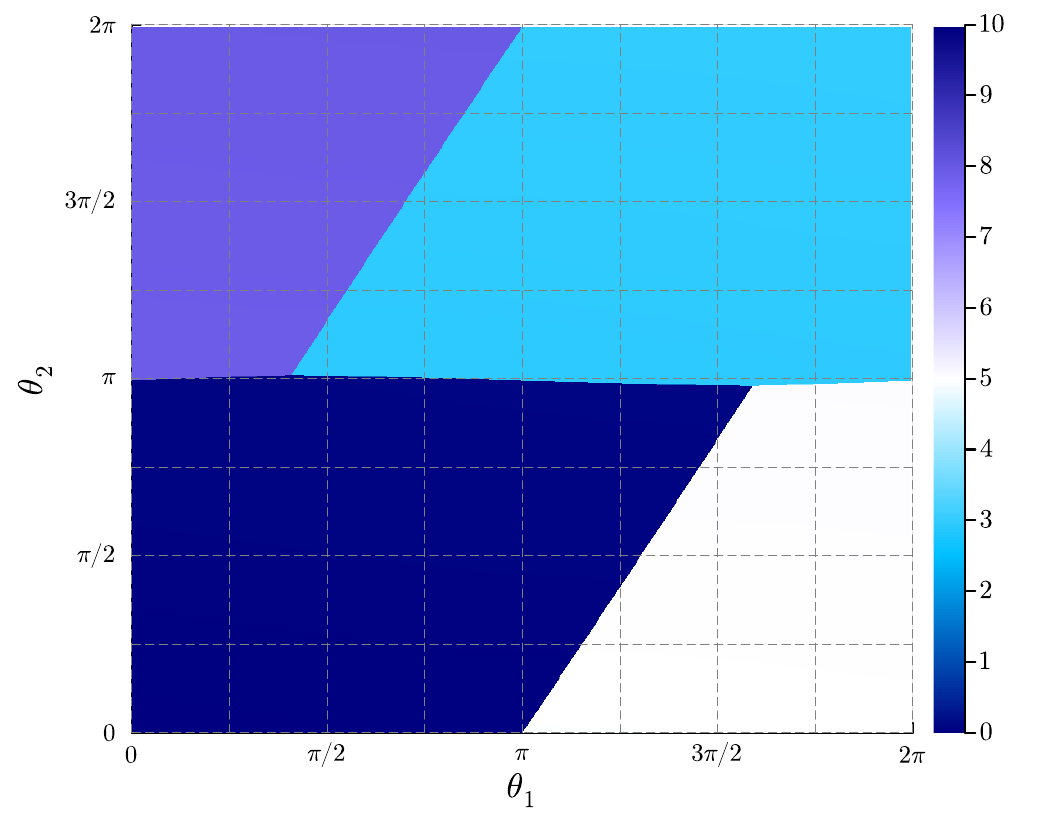}
        \label{fig:N1=2_N2=5_Msq50.0_K1=1_K2=100} 
        }
        \vspace{2ex}
    \end{minipage}
    \\    
    \begin{minipage}[b]{0.196\linewidth}
        \centering \subfloat[$m' \mu \approx 0.5106$]{
        \includegraphics[scale=0.265]{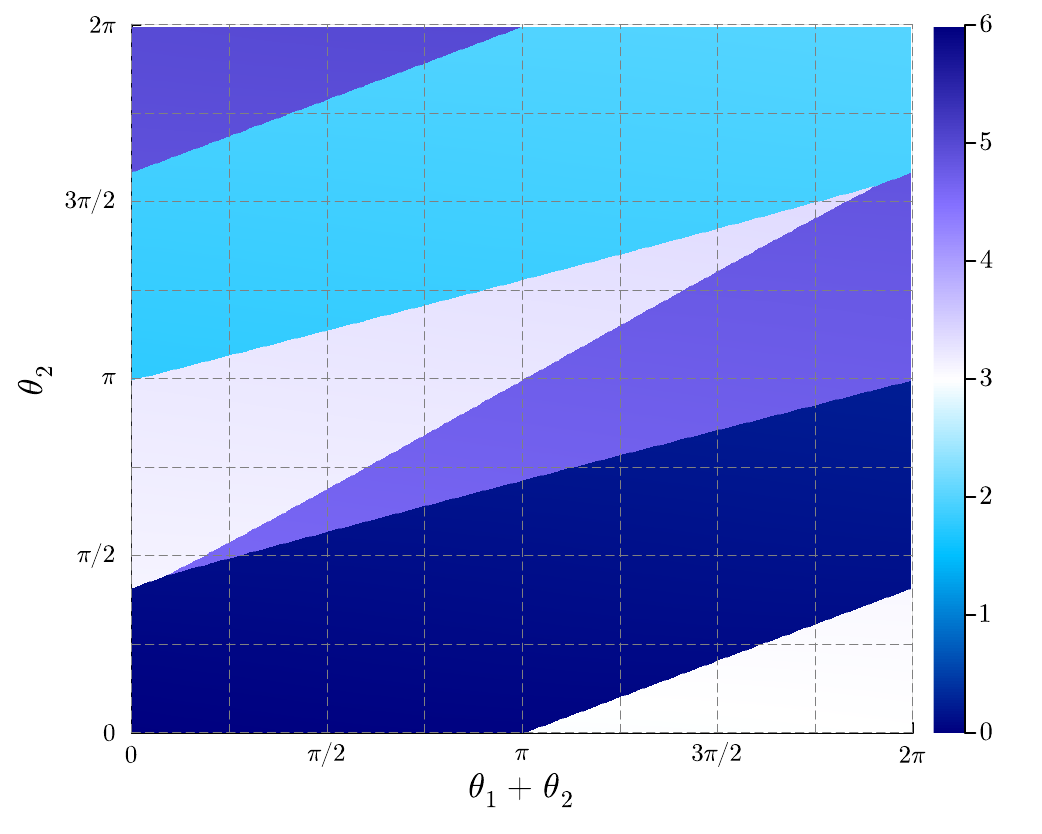}
        \label{fig:N1=2_N2=3_Msq0.5_K1=1_K2=100_redef+}
        }
        \vspace{2ex}
    \end{minipage}
    \begin{minipage}[b]{0.196\linewidth}
        \centering \subfloat[$m' \mu \approx 2.182$]{
        \includegraphics[scale=0.265]{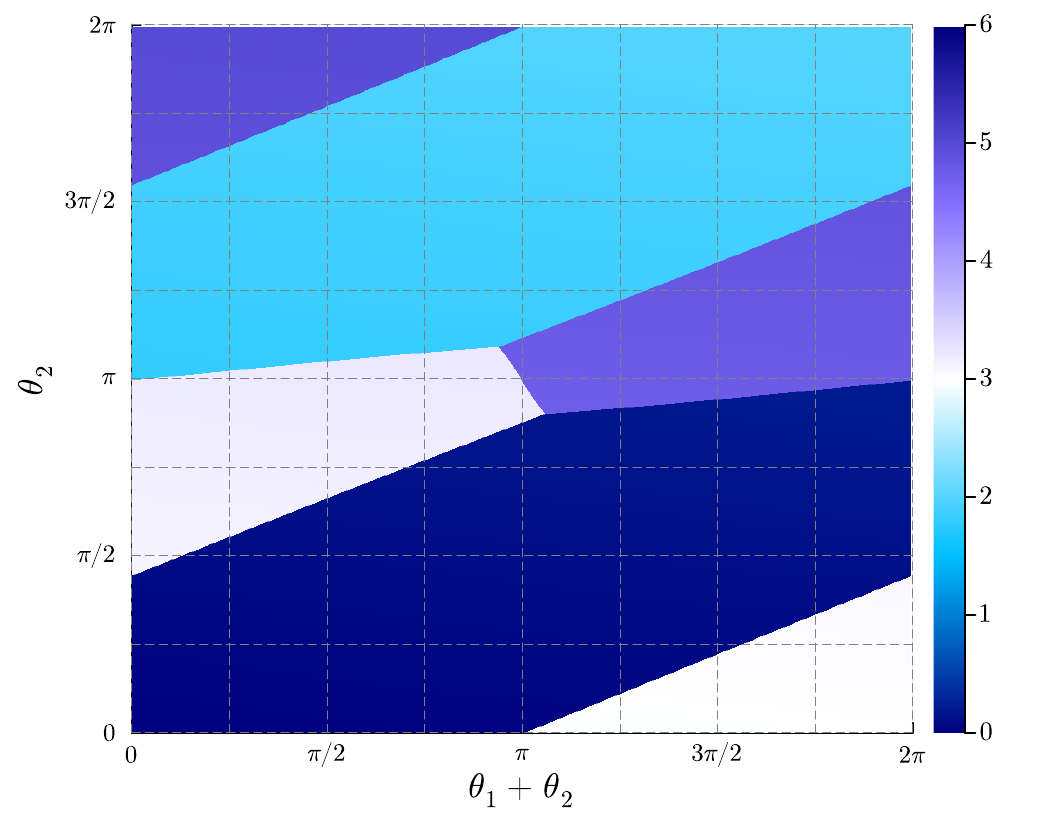}
        \label{fig:N1=2_N2=3_Msq2.0_K1=1_K2=100_redef+}
        }
        \vspace{2ex}
    \end{minipage}
    \begin{minipage}[b]{0.196\linewidth}
        \centering \subfloat[$m' \mu \approx 6.316$]{
        \includegraphics[scale=0.265]{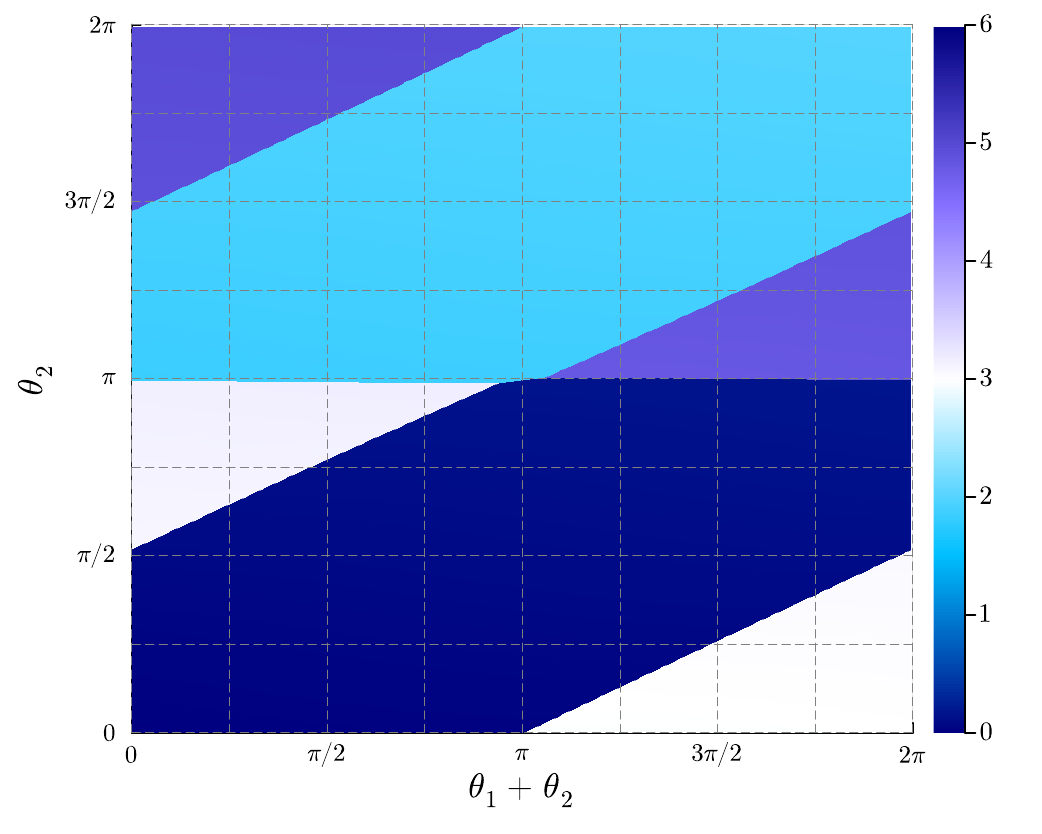}
        \label{fig:N1=2_N2=3_Msq5.0_K1=1_K2=100_redef+}
        }
        \vspace{2ex}
    \end{minipage}
    \begin{minipage}[b]{0.196\linewidth}
        \centering \subfloat[$m' \mu \approx 120.0$]{
        \includegraphics[scale=0.265]{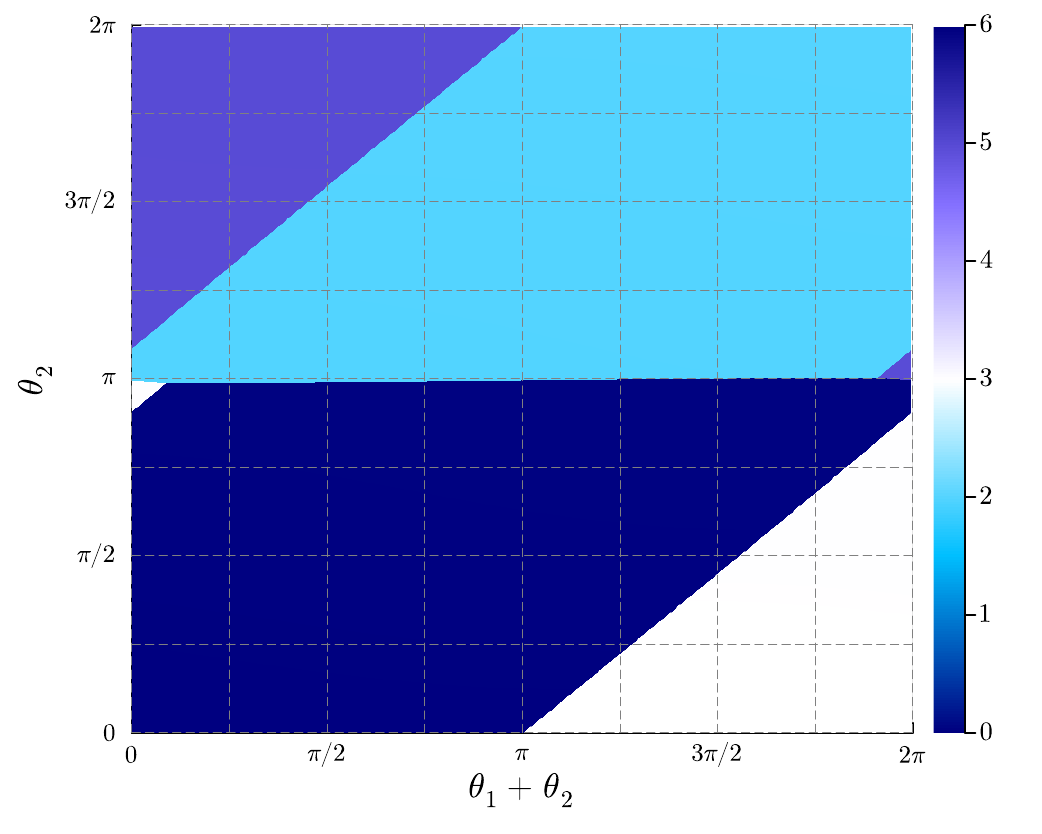}
        \label{fig:N1=2_N2=3_Msq20.0_K1=1_K2=100_redef+}
        }
        \vspace{2ex}
    \end{minipage}

    \caption{(\textbf{Top panels})~Phase diagrams of the parent theory, $(N_1,N_2)=(2,5)$, on the $(\theta_1,\theta_2)$ plane with several $m \mu$ in the unit of $2K^{(1)} \rme^{-S_\mathrm{I}^{(1)}/N_1} = 1,~ 2K^{(2)} \rme^{-S_\mathrm{I}^{(2)}/N_2} = 100$. 
    The topology-changing point is present at $m^*\mu \approx 25.0$, so the phase diagrams (a)-(d) for $m\mu\le m^*\mu$ are expected to be reproduced by the ones of the daughter theory according to the duality conjecture.\\ 
    (\textbf{Bottom panels})~Phase diagrams of the daughter theory, $(N_1, N_2-N_1) = (2, 3)$, on the $(\theta_1'+\theta_2',\theta_2')$ plane, in which the strong scales $2K^{(1')} \rme^{-S_\mathrm{I}^{(1')}/N_1},~ 2K^{(2')} \rme^{-S_\mathrm{I}^{(2')}/(N_2-N_1)}$ are set by the large-$N$ suggestion~\eqref{eq:Kexp_relation_daughter_h2}. 
    The fermion mass $m' \mu$ is also set through \eqref{eq:duality_semiclssical_parameters} (or the left relation of \eqref{eq:mass_relation_daughter}), so that the phase diagrams~(f)-(i) are comparable to the above ones~(a)-(d) of the parent theory, respectively.
    }
    \label{fig:duality_comparison_h2}
\end{figure}
\end{landscape}

Let us put the scale separation $\Lambda_1\ll \Lambda_2$ in the parent theory, $SU(2)\times SU(5)$ QCD(BF). 
We have analytically shown that the duality approximately holds for $m, \Lambda_1 \ll \Lambda_2$ in Section~\ref{sec:duality_hierarchical}, and our numerical analysis here serves also as a consistency check.
We set $2K^{(1)} \rme^{-S_\mathrm{I}^{(1)}/N_1}=1,~ 2K^{(2)} \rme^{-S_\mathrm{I}^{(2)}/N_2}=100$ in the parent theory, and the large-$N$ estimate of the topology-changing point is $m^*\mu=24$. 
Exactly as we have done in Section~\ref{sec:duality_numerics}, we determine the strong scales of the daughter theory by the large-$N$ formula~\eqref{eq:Kexp_relation_daughter}, 
\begin{align}
    2K^{(1')} \rme^{-S_\mathrm{I}^{(1')}/N_1}&=\frac{1}{1+\frac{5^3}{2^3\cdot 100}}\approx 0.865, 
    \notag\\
    2K^{(2')} \rme^{-S_\mathrm{I}^{(2')}/(N_2-N_1)}&=\frac{3^3}{5^3}\cdot 100=21.6.
    \label{eq:Kexp_relation_daughter_h2}
\end{align}
We show the result in Figure~\ref{fig:duality_comparison_h2}: 
The top panels represent the phase diagram of the parent theory on $(\theta_1, \theta_2)$ plane.
The bottom panels represent the phase diagram of the corresponding daughter theory on $(\theta_1'+\theta_2', \theta_2')$ plane through the matching of parameters via the large-$N$ indicated formula.
The fermion masses in the daughter theory are set through \eqref{eq:duality_semiclssical_parameters} to compare the phase diagrams (Figures~\ref{fig:N1=2_N2=3_Msq0.5_K1=1_K2=100_redef+}-\ref{fig:N1=2_N2=3_Msq20.0_K1=1_K2=100_redef+}) to those in the parent theory (Figures~\ref{fig:N1=2_N2=5_Msq0.5_K1=1_K2=100}-\ref{fig:N1=2_N2=5_Msq20.0_K1=1_K2=100}).
Although we have not illustrated the phase diagram at the topology-changing point $m\mu = m^*\mu$, our rough numerical search indicates $m^*\mu \approx 25.0$, which is slightly above but comparable with the large-$N$ estimate.
The fact that the value of $m^*\mu$ is not simply 100 times the value at $2K^{(2)} \rme^{-S_\mathrm{I}^{(2)}/N_2}=1$ reminds us that the topology-changing point in fact depends on both strong scales, $m^*(\Lambda_1,\Lambda_2)$.
Similar to the case in  Figure~\ref{fig:duality_comparison}, we can see the nice agreement between the parent theory for $m\mu\le m^*\mu\approx 25.0$ and the daughter theory as we have expected.

\subsubsection*{Case: $\Lambda_2 \ll \Lambda_1$}
Let us next consider the opposite hierarchy, $\Lambda_1\gg \Lambda_2$, and we are going to set 
\begin{equation}
    2K^{(1)}\rme^{-S^{(1)}/N_1}=100,\quad 
    2K^{(1)}\rme^{-S^{(1)}/N_1}=1, 
\end{equation}
for the parent theory, $SU(2)\times SU(5)$ QCD(BF), in our numerical study. 
The large-$N$ estimate of the topology changing point is $m^*\mu=0.24$. 

Following the large-$N$ duality map~\eqref{eq:Kexp_relation_daughter}, we set the strong scales of the daughter theory, $SU(2)\times SU(3)$ QCD(BF), as follows:
\begin{align}
    2K^{(1')} \rme^{-S_\mathrm{I}^{(1')}/N_1}&=\frac{100}{1+\frac{N_2^3}{N_1^3}100}\approx \frac{N_1^3}{N_2^3}=\frac{8}{125}, \notag\\
    2K^{(2')} \rme^{-S_\mathrm{I}^{(2')}/(N_2-N_1)}&=\frac{(N_2-N_1)^3}{N_2^3}=\frac{27}{125}. 
    \label{eq:Kexp_relation_daughter_h1}
\end{align}
We note that the hierarchy $\Lambda_1\gg \Lambda_2$ does not cause the large hierarchy of the strong scales in the daughter theory. 
However, the above values almost saturate the inequality~\eqref{eq:constraint_KKimage}, which suggests that this daughter theory lives almost at the boundary of the image of the duality map. In this sense, this opposite hierarchical limit provides a nontrivial testing ground for the duality relation~\eqref{eq:KK_duality_onedirection}. 
We relate the fermion mass for $m\mu<m^*\mu$ again by using the large-$N$ indicated formula~\eqref{eq:mass_relation_daughter}. 

\begin{landscape}
\begin{figure}[ht]
    \begin{minipage}[b]{0.196\linewidth}
        \centering \subfloat[$m \mu = 0.02$]{
        \includegraphics[scale=0.265]{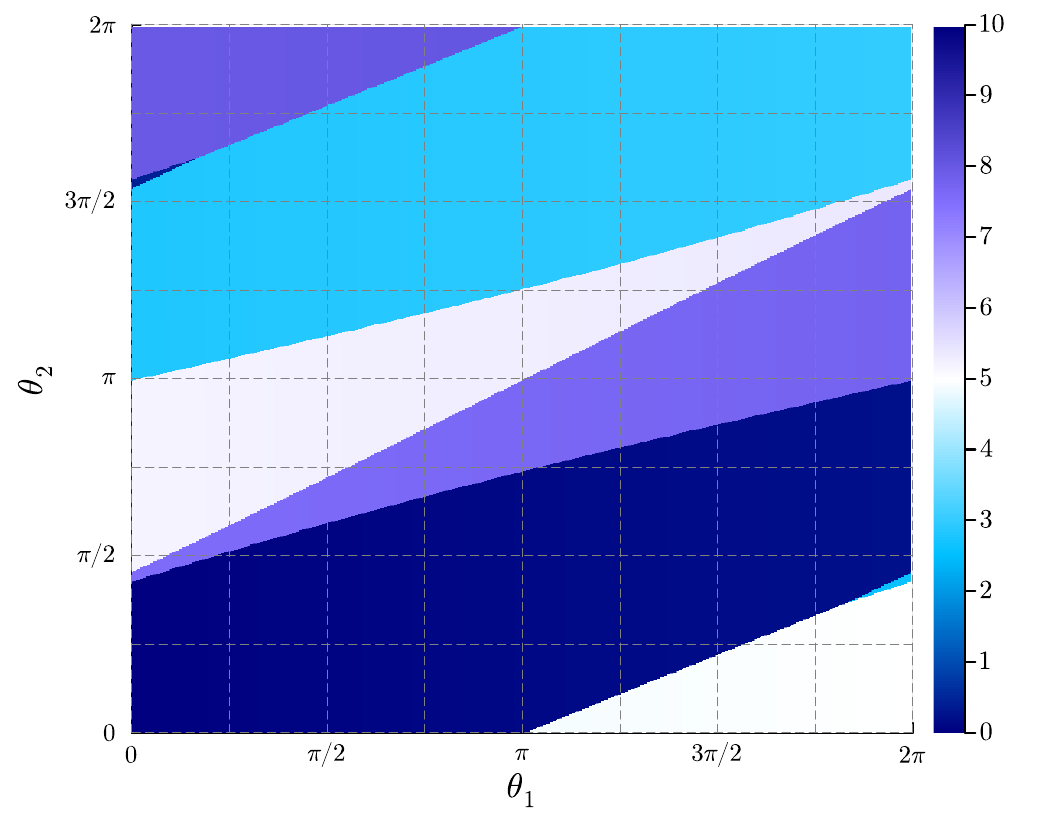}
        \label{fig:N1=2_N2=5_Msq0.02_K1=100_K2=1}
        }
        \vspace{2ex}
    \end{minipage}
    \begin{minipage}[b]{0.196\linewidth}
        \centering \subfloat[$m \mu = 0.06$]{
        \includegraphics[scale=0.265]{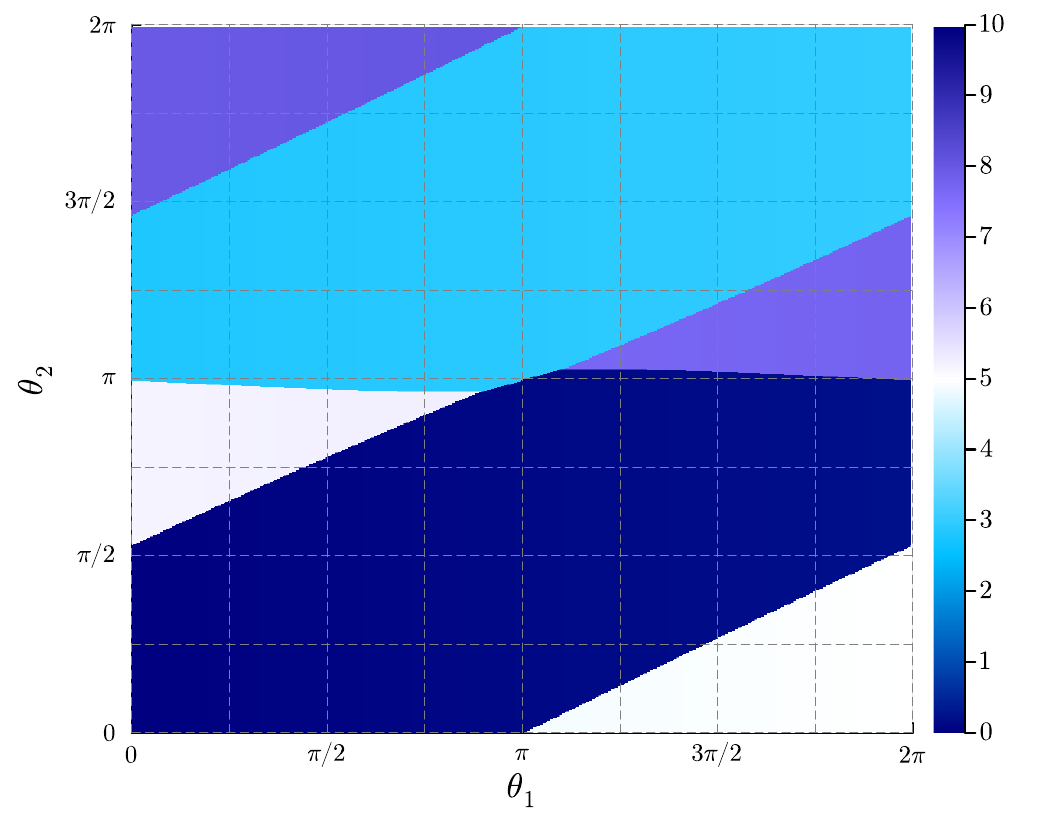}
        \label{fig:N1=2_N2=5_Msq0.06_K1=100_K2=1}
        }
        \vspace{2ex}
    \end{minipage}
    \begin{minipage}[b]{0.196\linewidth}
        \centering \subfloat[$m \mu = 0.2$]{
        \includegraphics[scale=0.265]{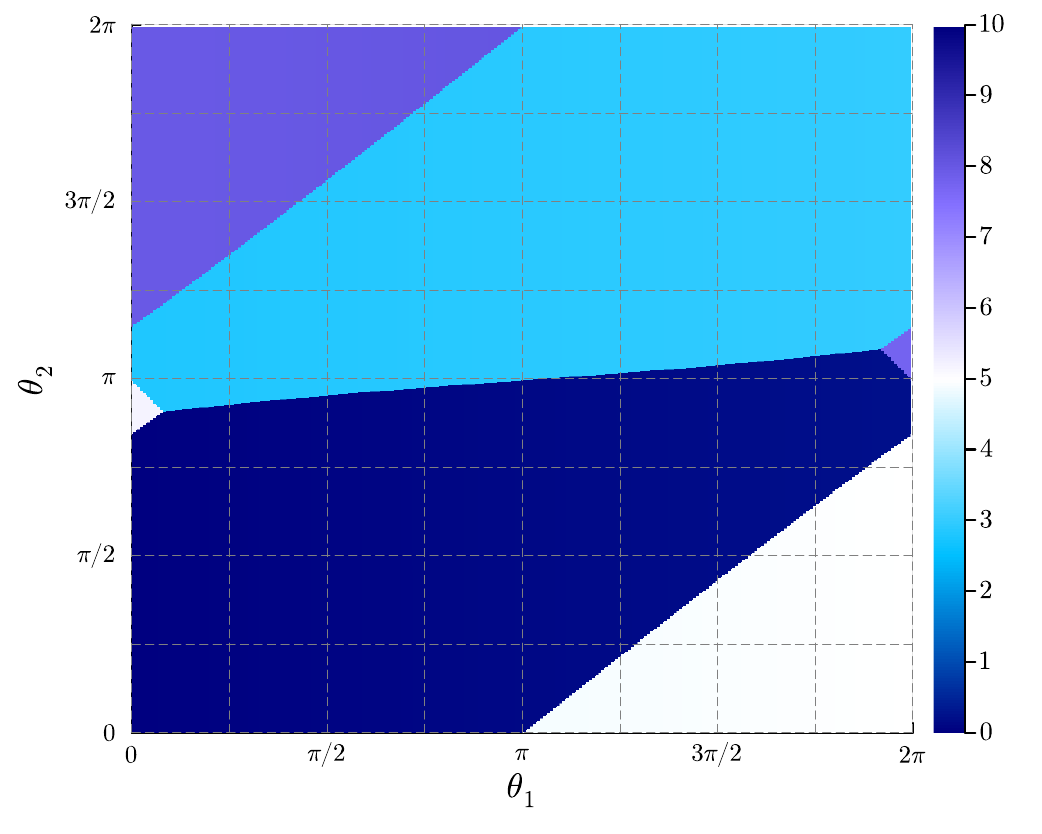}
        \label{fig:N1=2_N2=5_Msq0.2_K1=100_K2=1}
        }
        \vspace{2ex}
    \end{minipage}
    \begin{minipage}[b]{0.196\linewidth}
        \centering \subfloat[$m \mu \approx 0.263 $]{
        \includegraphics[scale=0.265]{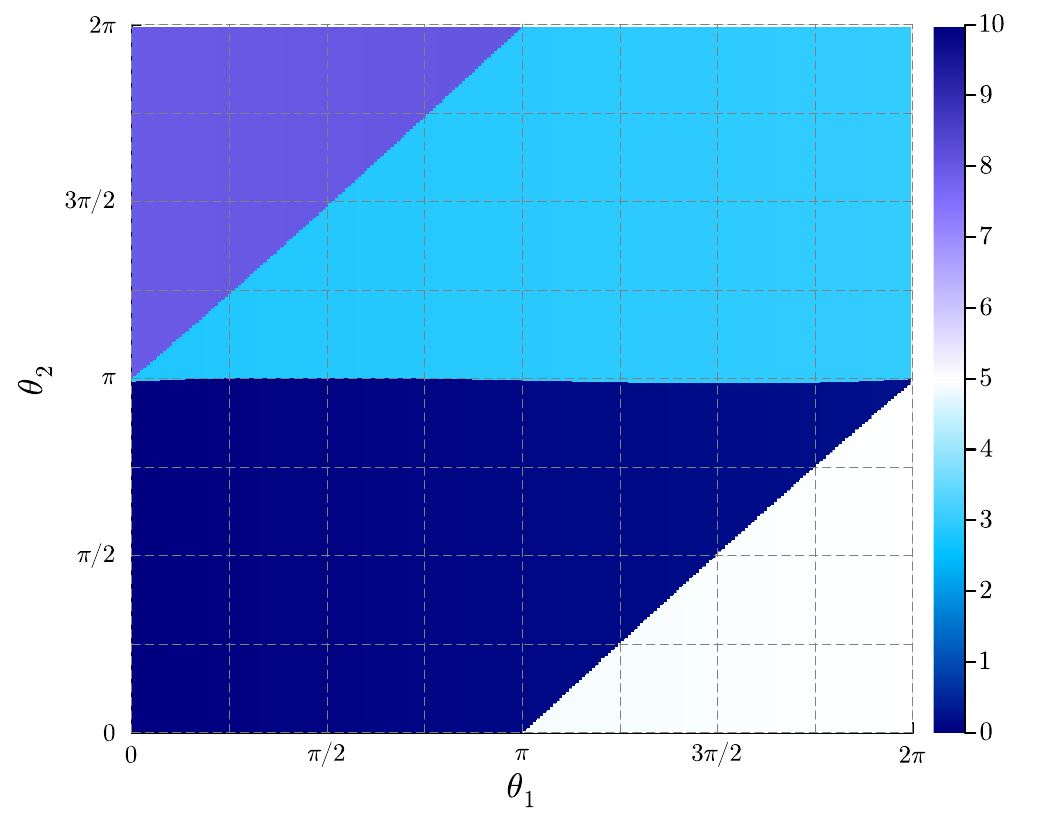}
        \label{fig:N1=2_N2=5_Msq0.263_K1=100_K2=1}
        }
        \vspace{2ex}
    \end{minipage}
    \begin{minipage}[b]{0.196\linewidth}
        \centering \subfloat[$m \mu = 0.3$]{
        \includegraphics[scale=0.265]{Figures/Heatmap_N1=2_N2=5_Msq0.3_K1=1_K2=1.pdf}
        \label{fig:N1=2_N2=5_Msq0.3_K1=100_K2=1}
        }
        \vspace{2ex}
    \end{minipage}
    \\    
    \begin{minipage}[b]{0.196\linewidth}
        \centering \subfloat[$m' \mu = 0.02181$]{
        \includegraphics[scale=0.265]{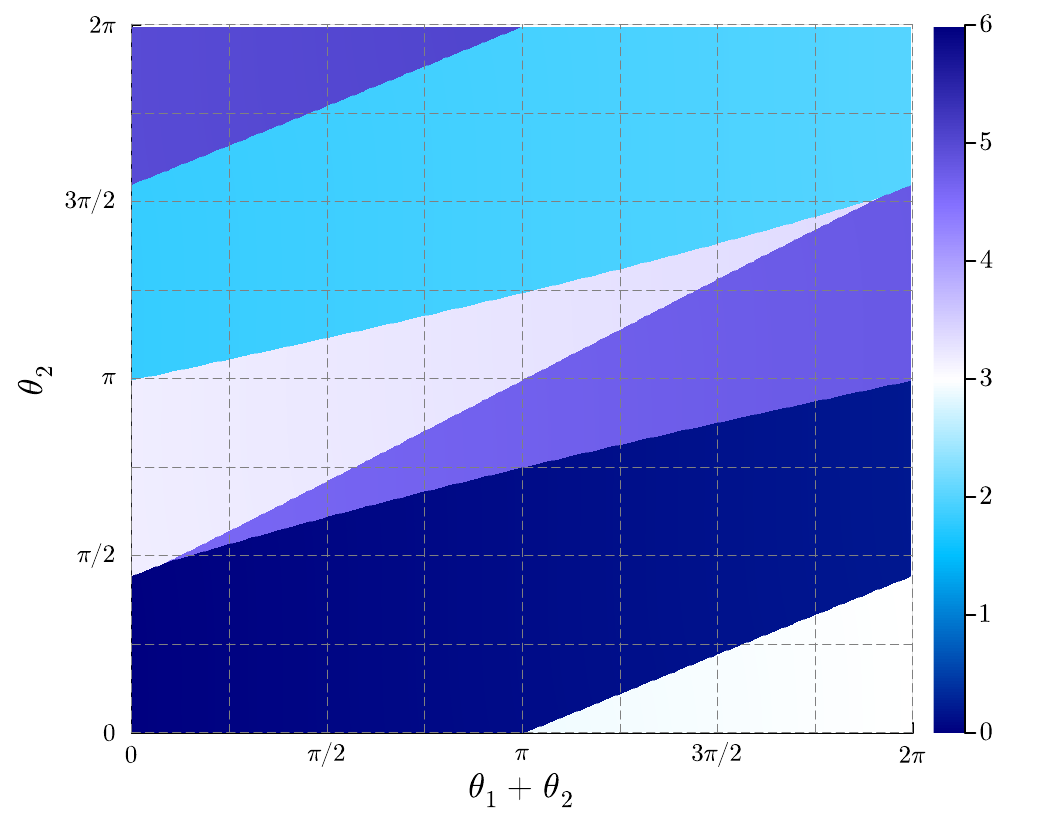}
        \label{fig:N1=2_N2=3_Mp0.02_K1p=100_K2p=1_redef+}
        }
        \vspace{2ex}
    \end{minipage}
    \begin{minipage}[b]{0.196\linewidth}
        \centering \subfloat[$m' \mu = 0.08$]{
        \includegraphics[scale=0.265]{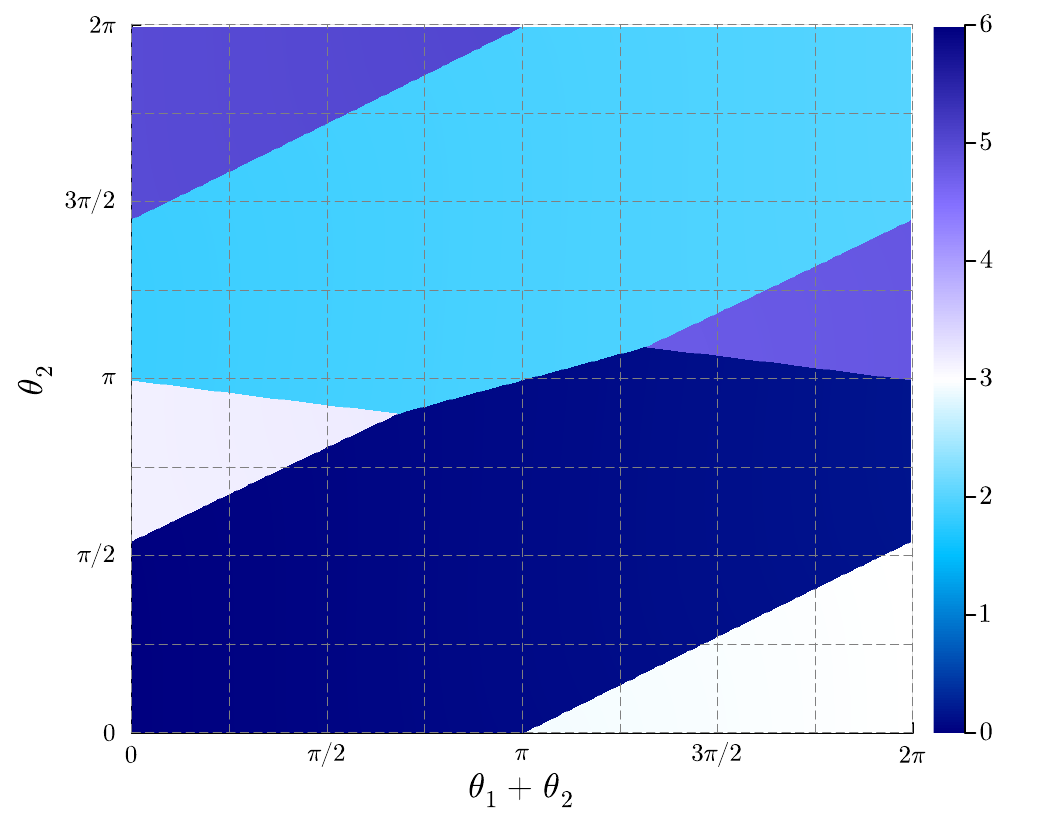}
        \label{fig:N1=2_N2=3_Mp0.06_K1p=100_K2p=1_redef+}
        }
        \vspace{2ex}
    \end{minipage}
    \begin{minipage}[b]{0.196\linewidth}
        \centering \subfloat[$m' \mu = 1.2$]{
        \includegraphics[scale=0.265]{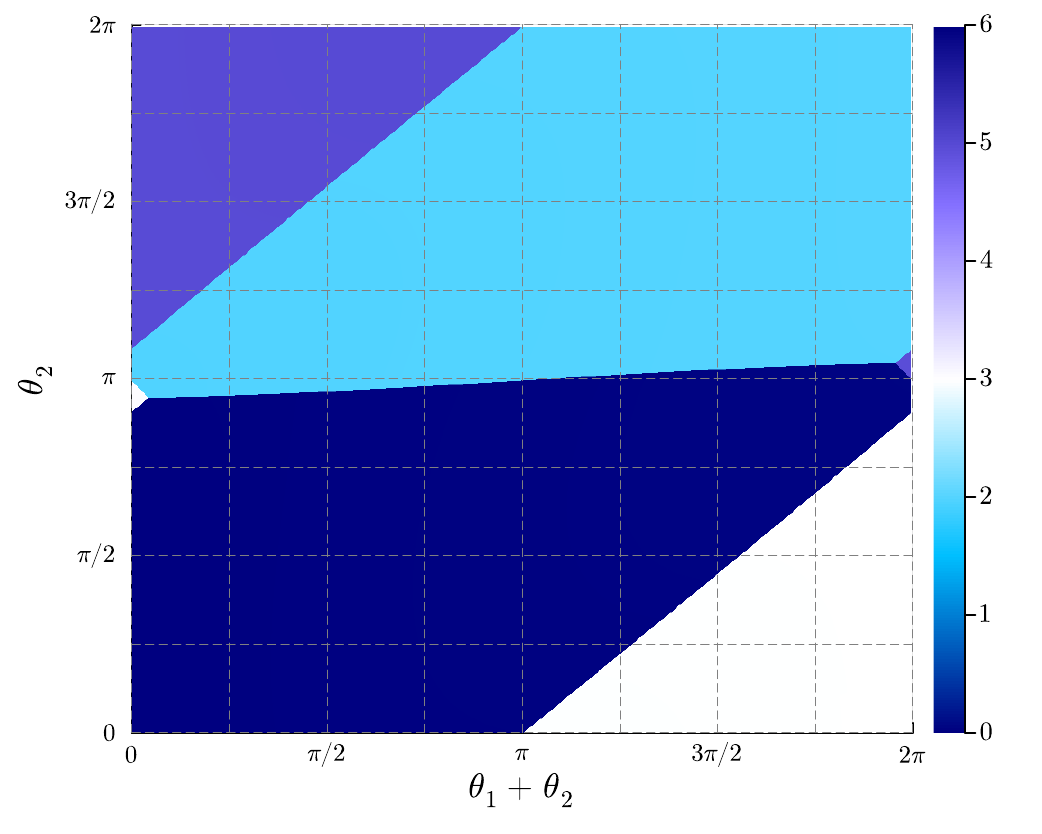}
        \label{fig:N1=2_N2=3_Mp0.2_K1p=100_K2p=1_redef+}
        }
        \vspace{2ex}
    \end{minipage}
    \begin{minipage}[b]{0.196\linewidth}
        \centering \subfloat[$m' \mu = 100$]{
        \includegraphics[scale=0.265]{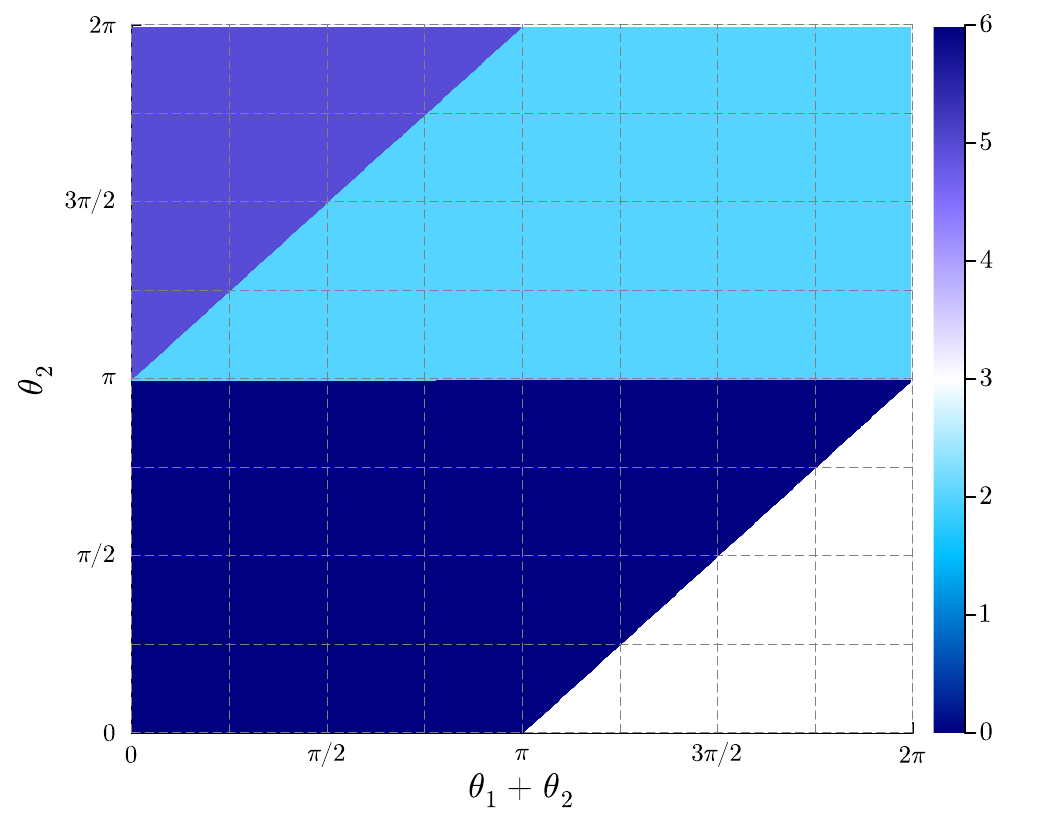}
        \label{fig:N1=2_N2=3_Msq100.0_K1p=100_K2p=1_redef+}
        }
        \vspace{2ex}
    \end{minipage}

    \caption{(\textbf{Top panels})~Phase diagrams of the parent theory, $(N_1,N_2)=(2,5)$, on the $(\theta_1,\theta_2)$ plane with several $m \mu$ in the unit of $2K^{(1)} \rme^{-S_\mathrm{I}^{(1)}/N_1} = 100,~ 2K^{(2)} \rme^{-S_\mathrm{I}^{(2)}/N_2} = 1$. 
    The topology-changing point is present at $m^*\mu \approx 0.263$, so the phase diagrams (a)-(d) for $m\mu\le m^*\mu$ are expected to be reproduced by the ones of the daughter theory according to the duality conjecture.\\ 
    (\textbf{Bottom panels})~Phase diagrams of the daughter theory, $(N_1, N_2-N_1) = (2, 3)$, on the $(\theta_1'+\theta_2',\theta_2')$ plane, in which the strong scales $2K^{(1')} \rme^{-S_\mathrm{I}^{(1')}/N_1},~ 2K^{(2')} \rme^{-S_\mathrm{I}^{(2')}/(N_2-N_1)}$ are set by the large-$N$ suggestion~\eqref{eq:Kexp_relation_daughter_h1}. 
    The fermion mass $m' \mu$ is also set through \eqref{eq:duality_semiclssical_parameters} (or the left relation of \eqref{eq:mass_relation_daughter}), so that the phase diagrams~(f)-(i) are comparable to the above ones~(a)-(d) of the parent theory, respectively.
    }
    \label{fig:duality_comparison_h1}
\end{figure}
\end{landscape}

We show the result in Figure~\ref{fig:duality_comparison_h1}, and the duality works again nicely. 
Moreover, we can immediately notice that those figures are almost identical to those of Figure~\ref{fig:duality_comparison}, including the location of the topology-changing point $m^*\mu\approx 0.263$. 
As a result, the duality works so well also in the opposite hierarchical limit within our semiclassical framework. 
Let us consider why this is so. 
When $m\mu<m^*\mu$, we have the duality map to the daughter theory. When $\Lambda_1\gtrsim \Lambda_2$, the $SU(N_1)$ strong scale of the daughter theory does not change so much according to the formula~\eqref{eq:Kexp_relation_daughter}. 
Even though the two setups $\Lambda_1 \sim \Lambda_2$ and $\Lambda_1\gg \Lambda_2$ seem to be very different in the parent theory, the corresponding daughter theories have roughly the same parameters, and thus the phase diagrams in the small-mass range, $m\mu\lesssim m^*\mu$, are insensitive to the values of $\Lambda_1$. 
It seems not to be straightforward to understand this feature of the parent theory without relying on the duality relation.

\clearpage

\bibliographystyle{JHEP}
\bibliography{./QFT.bib, ./refs.bib} 

\end{document}